\documentclass[useAMS,usenatbib]{mn2e} 
\usepackage{amsmath,graphics,color,subfigure,epsfig,amssymb}
\usepackage{mathptmx}
\usepackage{epsfig}
\usepackage{booktabs}
\usepackage{rotating,rotfloat}
\usepackage{longtable}
\usepackage{graphicx}
\usepackage{url} 
\usepackage{listings}
\usepackage{color}
\usepackage{pslatex} 
\usepackage{booktabs}
\usepackage{dcolumn}
\usepackage[hyperindex,breaklinks=true, colorlinks, citecolor=blue]{hyperref}

\let\la=\lesssim    
\def\reff@jnl#1{{\rm#1\/}}
\def\aj{\reff@jnl{AJ}}                  
\def\araa{\reff@jnl{ARA\&A}}            
\def\actaa{\reff@jnl{Acta. Astron}}     
\def\apj{\reff@jnl{ApJ}}                
\def\apjl{\reff@jnl{ApJ}}               
\def\apjs{\reff@jnl{ApJS}}              
\def\ao{\reff@jnl{Appl.Optics}}         
\def\apss{\reff@jnl{Ap\&SS}}            
\def\aap{\reff@jnl{A\&A}}               
\def\aapr{\reff@jnl{A\&A~Rev.}}         
\def\aaps{\reff@jnl{A\&AS}}             
\def\azh{\reff@jnl{AZh}}                
\def\baas{\reff@jnl{BAAS}}              
\def\jrasc{\reff@jnl{JRASC}}            
\def\memras{\reff@jnl{MmRAS}}           
\def\mnras{\reff@jnl{MNRAS}}            
\def\pra{\reff@jnl{Phys.Rev.A}}         
\def\prb{\reff@jnl{Phys.Rev.B}}         
\def\prc{\reff@jnl{Phys.Rev.C}}         
\def\prd{\reff@jnl{Phys.Rev.D}}         
\def\prl{\reff@jnl{Phys.Rev.Lett}}      
\def\pasp{\reff@jnl{PASP}}              
\def\pasj{\reff@jnl{PASJ}}              
\def\qjras{\reff@jnl{QJRAS}}            
\def\skytel{\reff@jnl{S\&T}}            
\def\solphys{\reff@jnl{Solar~Phys.}}    
\def\sovast{\reff@jnl{Soviet~Ast.}}     
\def\ssr{\reff@jnl{Space~Sci.Rev.}}     
\def\zap{\reff@jnl{ZAp}}                
\def\nat{\reff@jnl{Nature}}             
\def\pasa{\reff@jnl{Publ. Astron. Soc. Aust.}}            
\def\jcap{\reff@jnl{Journal of Cosmology and Astroparticle Physics}}     
\def\jqsrt{\reff@jnl{Journal of Quantitative Spectroscopy \& Radiative Transfer}}
\def\nar{\reff@jnl{New Astronomy Reviews}}

\newcolumntype{K}{D{.}{.}{2,2}}
\newcolumntype{H}{D{,}{\pm}{4.4}}

\newcommand{\wmap}{{\em WMAP~}}
\newcommand{\planck}{{\em Planck~}}

\newcommand{\dg}{$^{\circ}$~}
\newcommand{\beq}{\begin{equation}}
\newcommand{\eeq}{\end{equation}}

\newcommand{\kband}{K--band}
\newcommand{\kaband}{Ka--band}
\newcommand{\qband}{Q--band}
\newcommand{\vband}{V--band}
\newcommand{\wband}{W--band}
\newcommand{\Q}{{\em Q}}
\newcommand{\U}{{\em U}}
\newcommand{\I}{{\em I}}

\newcommand{\mdg}{^{\circ}}
\newcommand{\mdgp}{^{\circ}\!\!.}

\newcommand{\kms}{\,km\,s$^{-1}$}
\newcommand{\mparcm}{'\!\!.}

\title[Polarised radio filaments outside the Galactic plane]{Polarised
  radio filaments outside the Galactic plane}

\author[Vidal et al.]{Matias~Vidal\thanks{E-mail: matias.vidal@manchester.ac.uk (MV)}, C.~Dickinson, R.~D.~Davies and J.~P.~Leahy
  \\
  Jodrell Bank Centre for Astrophysics, Alan Turing Building,
  School of Physics and Astronomy, The University of Manchester,\\
  Oxford Road, Manchester M13 9PL, UK. \\
}

\begin{document}
   

\pagerange{\pageref{firstpage}--\pageref{lastpage}} \pubyear{2002}

\maketitle

\label{firstpage}

\begin{abstract}

We used data from the \wmap satellite at 23, 33 and 41 GHz to study
the diffuse polarised emission over the entire sky. The emission
originates mostly from filamentary structures with well-ordered
magnetic fields. Some of these structures have been known for decades
in radio continuum maps. Their origin is not clear and there are many
filaments that are visible for the first time. We have identified and
studied 11 filaments.  The polarisation fraction of some of them can
be as high as 40\%, which is a signature of a well ordered magnetic
field. The polarisation spectral indices, averaged over 18 regions in
the sky is $\beta = -3.06 \pm 0.02$, consistent with synchrotron
radiation. There are significant variations in $\beta$ over the sky
($\Delta\beta\approx0.2$).

We explore the link between the large-scale filaments and the local
ISM, using the model of an expanding shell in the solar vicinity. We
compared observed polarisation angles with the predictions from the
model and found good agreement. This strongly suggests that many large
scale filaments and loops are nearby structures. This is important in
the context of the Galactic magnetic field as these structures are
normally included in global models, neglecting the fact that they
might be local. We also studied the level of contamination added by
the diffuse filaments to the CMB polarisation power spectra.  We
conclude that, even though these filaments present low radio
brightness, a careful removal will be necessary for future all-sky CMB
polarisation analysis.

\end{abstract}

\section{Introduction}

The ``radio loops'' are some of the largest structures on the sky and
have been studied for more than 50 years.  Some of these structures,
are also visible in microwaves, X-rays and gamma-rays.
\citet{brown:60} review for the first time possible theories about the
origin of the most obvious of them, Loop I, also called the North
Polar Spur (NPS). Loop II \citep{large:62}, Loop III
\citep{quigley:65} and Loop IV \citep{large:66} where discovered
shortly after. The emission from all these features is non-thermal
with temperature spectral indices ($T_b \propto \nu^{\beta}$) $\beta$
around 1\,GHz ranging between $-2.7$ and $-2.9$
\citep{berkhuijsen:73,borka:07}. These loops are also visible in
polarisation, both in starlight \citep{mathewson:70} and radio
\citep{page:07}.

There are different interpretations for the origin of these loops and
their true nature is not completely clear. The lack of precise
measurements of their distance is the main limitation in their study.
Four possible explanations for their origin have been proposed:
\begin{itemize}
\item Old and nearby supernova remnants
  \citep{berkhuijsen:71b,spoelstra:73}
\item Outflow from the Galactic Centre
  \citep{sofue:77,bland-hawthorn:03}
\item Bubbles/shells powered by OB associations
  \citep{egger:95b,wolleben:07}. 
\item Magnetic field loops illuminated by relativistic electrons
  \citep{heiles:98}.
\end{itemize}

The spectral indices of these structures indicate a synchrotron origin
for the radio emission. At frequencies below $\sim$1\,GHz, synchrotron
dominates the sky brightness.  The synchrotron intensity depends on
the relativistic electron density ($n_e$), their energy distributions
of and the strength of the magnetic field. The energy distribution of
cosmic ray electrons (CRE) is observed to follow a power-law, 
$N(E) \propto E^{-p}$ for energies $E>10$\,GeV
\citep[e.g.][]{pamela:11,ackermann:12}. Gamma-ray data suggests that
this distribution is fairly smooth, both spectrally and spatially over
the sky \citep{strong:04,strong:07}. Typical values measured for the
spectral index $p$ are close to $p=3$, and therefore, this value is
usually adopted in synchrotron and magnetic field modelling
\protect\citep[e.g.][]{miville:08,jaffe:10}.

Synchrotron emission is intrinsically linearly polarised.  For a
uniform magnetic field, the fractional polarisation is $\Pi =
(p+1)/(p+7/3) $. For a typical value of cosmic ray spectral index
($p\approx 3$) the polarisation faction can be as high as 75\%. The
observed value is smaller than this due to superposition of different
field directions along the line-of-sight, beam averaging and, at
cm--wavelengths, intrinsic Faraday rotation.

The synchrotron spectral index can vary spatially across the sky
\citep[e.g.][]{lawson:87,reich:88a} and also with frequency \citep[see
  e.g.][]{banday:91a,giardino:02,platania:03}. The loss of energy of
the CRE through synchrotron radiation is larger at higher energies
($\propto E^2$), so over time, the synchrotron spectral index becomes
steeper (more negative).  $\beta$ is also expected to be steeper with
higher Galactic latitude \citep{strong:07}. This is because most
cosmic rays are produced in supernova (SN) explosions, close to the
Galactic plane. As the CRE diffuse away from the plane, to higher
Galactic latitudes, the ageing effect will produce a steepening in the
synchrotron spectral index.

Averaged values for the measured spectral indices vary from
$\beta=-2.55$ between 45\,MHz and 408\,MHz \citep{guzman:11},
$\beta=-2.71$ between 408\,MHz and 2.3\,GHz
\citep{platania:03,giardino:02} and $\beta=-3.01$ between 23\,GHz and
33\,GHz \citep{davies:06,dunkley:09,dickinson:09b}. \citet{macellari:11} measured
the synchrotron spectral index over the entire sky using polarisation
and intensity data from \wmap between 23 and 33\,GHz, and found it to
be $\beta=-3.32\pm0.12$ for intensity and $\beta=-3.01\pm0.03$ for
polarised intensity. More recently \citet{fuskeland:14} have measured
small variations in the synchrotron polarisation spectral index using
\wmap polarisation data at 23 and 33\,GHz.

In this paper we aim to characterise the filamentary structure visible
in the polarised sky of \wmap\!\!. The relatively high frequency of
the \wmap bands is useful to study the polarised Galactic synchrotron
because Faraday rotation is negligible at this frequency range. This
allows us to study the emission from regions on the Galactic plane in
the inner Galaxy, which are washed out by Faraday rotation in
low-frequency polarisation surveys.

The paper is organised as follows: Section \ref{sec:analysis}
describes the data and the processing applied to it, such as smoothing
and bias correction. In Section \ref{sec:fil_angles} we define a
number of filaments that are visible in polarised intensity and
measure how close the observed magnetic field vectors are to the
orientation of the filament. A spectral index analysis in polarisation
is described in Section \ref{sec:spectral_indices} while in Section
\ref{sec:pol_fraction} we study the polarisation fraction at
\kband. In Section \ref{sec:discussion} we discuss the different
results and Section \ref{sec:conclusions} gives conclusions.

\section{Data processing}
\label{sec:analysis}
\subsection{\wmap data}

In this work, we use the 9-year \wmap maps \citep{bennett:13}
available from the Legacy Archive for Microwave Background Data
Analysis (LAMBDA) website\footnote{http://lambda.gsfc.nasa.gov}. The
data consist of 5 full sky maps of the Stokes \I, \Q\, and \U\,
parameters at central frequencies of 22.7\,GHz (\kband), 32.9.\,GHz
(\kaband), 40.6\,GHz (\qband), 60.5\,GHz (\vband) and 93.0\,GHz
(\wband) with the corresponding beam profiles and noise models. We
focus on K, Ka and Q--bands where the signal-to-noise ratio (SNR) is
highest and synchrotron can be detected at least along bright
filaments. Besides, dust contamination at these frequencies is
negligible. The maps are provided in the
HEALPix\footnote{http://healpix.sourceforge.net/} pixelisation scheme
\citep{gorsky:05} with $N_{\text{side}}=512$, which corresponds to a
pixel size of $\sim 7'$.


\subsection{Smoothing}

To increase the SNR of the maps, and to minimise any systematic effect
due to beam-asymmetries in polarisation, we smoothed the maps to
common resolutions of 1\dg and 3\dg full-width-half-maximum
(FWHM). The filaments that we study are typically diffuse at scales
larger than 1\dg\!\!. The smoothing was done by first deconvolving in
harmonic space the azimuthally symmetrized effective beam and then
convolving with a Gaussian beam.  \citet{wehus:13} showed that the
beam asymmetries in \wmap K-- and Ka--bands give rise to unstable
spectral index measurements in polarisation. Here we will use an
effective resolution of 3\dg\!\!\!  for measuring the spectral indices
so the beam asymmetries have minimal impact. We also generated
smoothed maps with an angular resolution of 1\dg\!\!\! FWHM to study
the polarisation angle direction along the filamentary
structures. After smoothing, the 3\dg\!\!\! resolution maps were
downgraded to a HEALPix resolution of $N_{\text{side}}=32$ (pixel size
$\sim 1.8^{\circ}$) while the 1\dg\!\!\! resolution maps were
downgraded to $N_{\text{side}}=256$ (pixel size $\sim 13\mparcm6$).

In order to assess the noise level after the smoothing of the maps, we
ran Monte Carlo simulations to calculate the uncertainties after the
smoothing and pixel downgrading. We generated 500 normally distributed
noise realisations of Stokes \I, \Q~and \U~for each frequency band
based on the covariance matrix from the \wmap data.  We calculated the
statistics of this assembly and we used the dispersion on each pixel
(for the \I, \Q, and \U~maps) as the statistical error on the 3\dg
smoothed maps. We also calculated the covariance term between \Q~and
\U~for the ensemble ($\sigma^2_{QU}$). By this route, we have smoothed
maps for Stokes \I, \Q~and \U~with their respective uncertainties and
$(Q,U)$ correlations for each pixel. We note that this is an
underestimate of the uncertainty due to the presence of correlated
noise in large scales.  For all the noise values estimated, we need to
add in quadrature the 0.2\% \wmap absolute calibration error
\citep{bennett:13}.

\begin{figure*} 
  \centering %
  \newcommand{\widthfig}{0.49}
  \includegraphics[angle=90,width=\widthfig\textwidth]{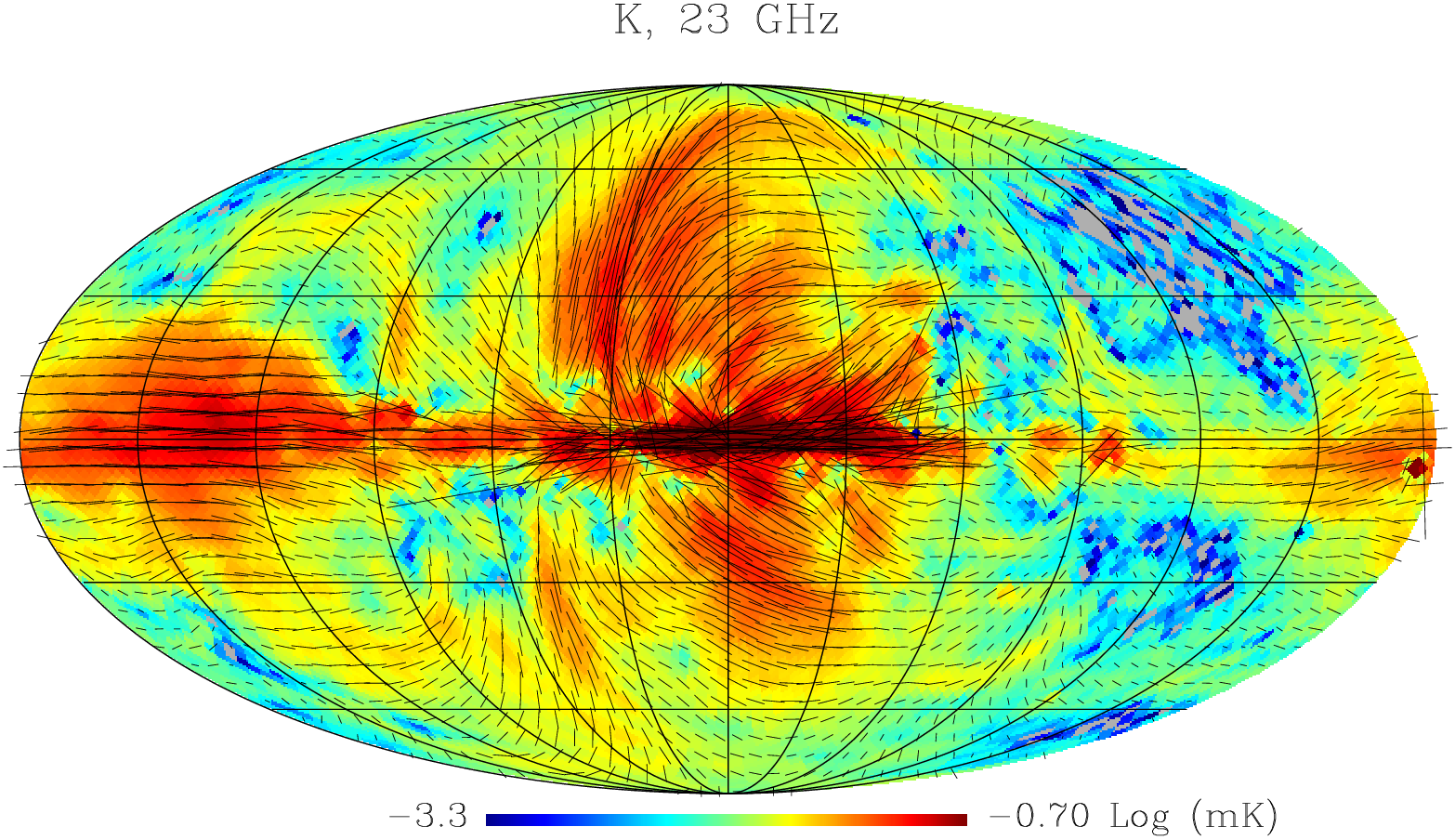}
  \includegraphics[angle=90,width=\widthfig\textwidth]{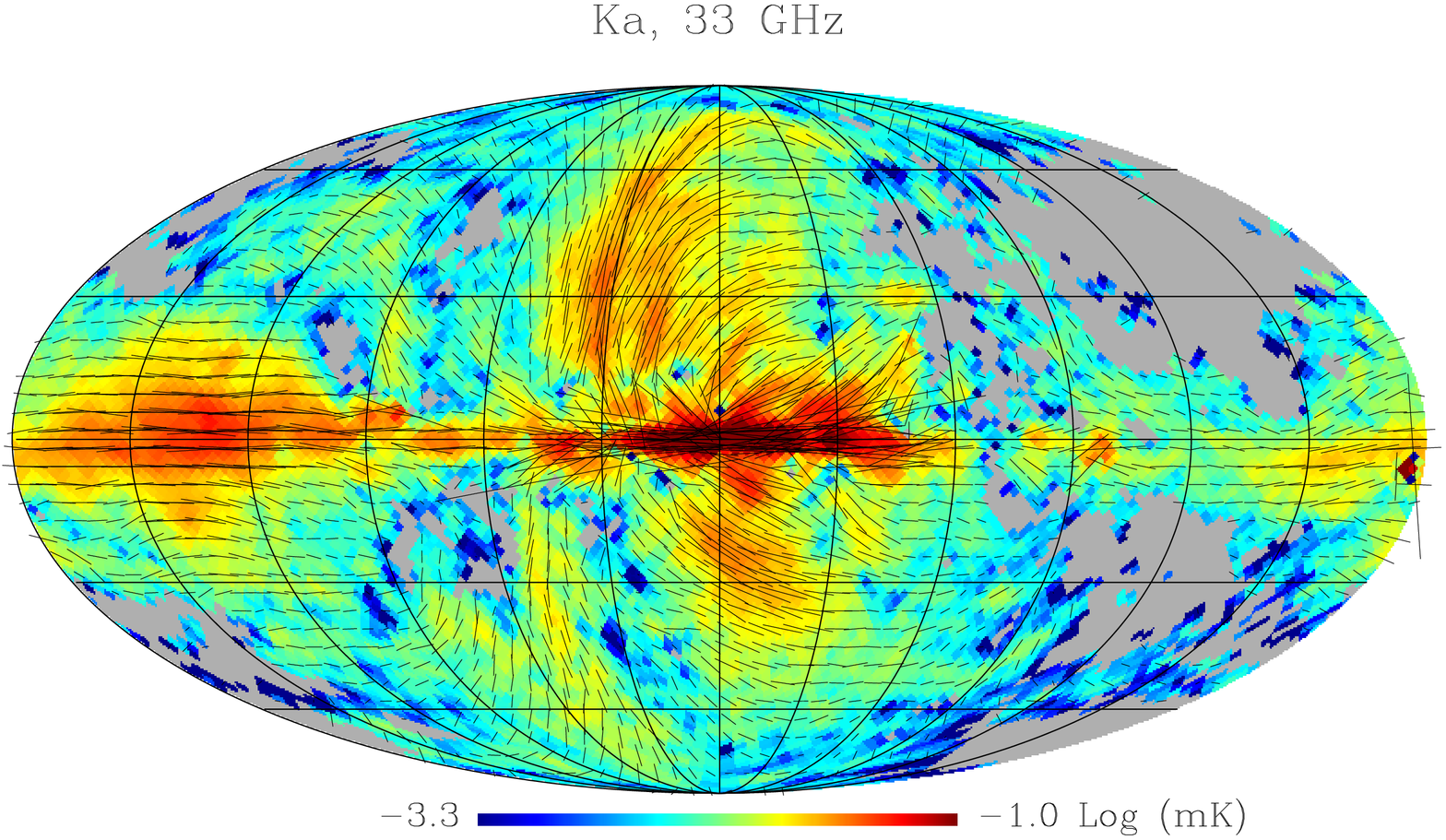}
  \includegraphics[angle=90,width=\widthfig\textwidth]{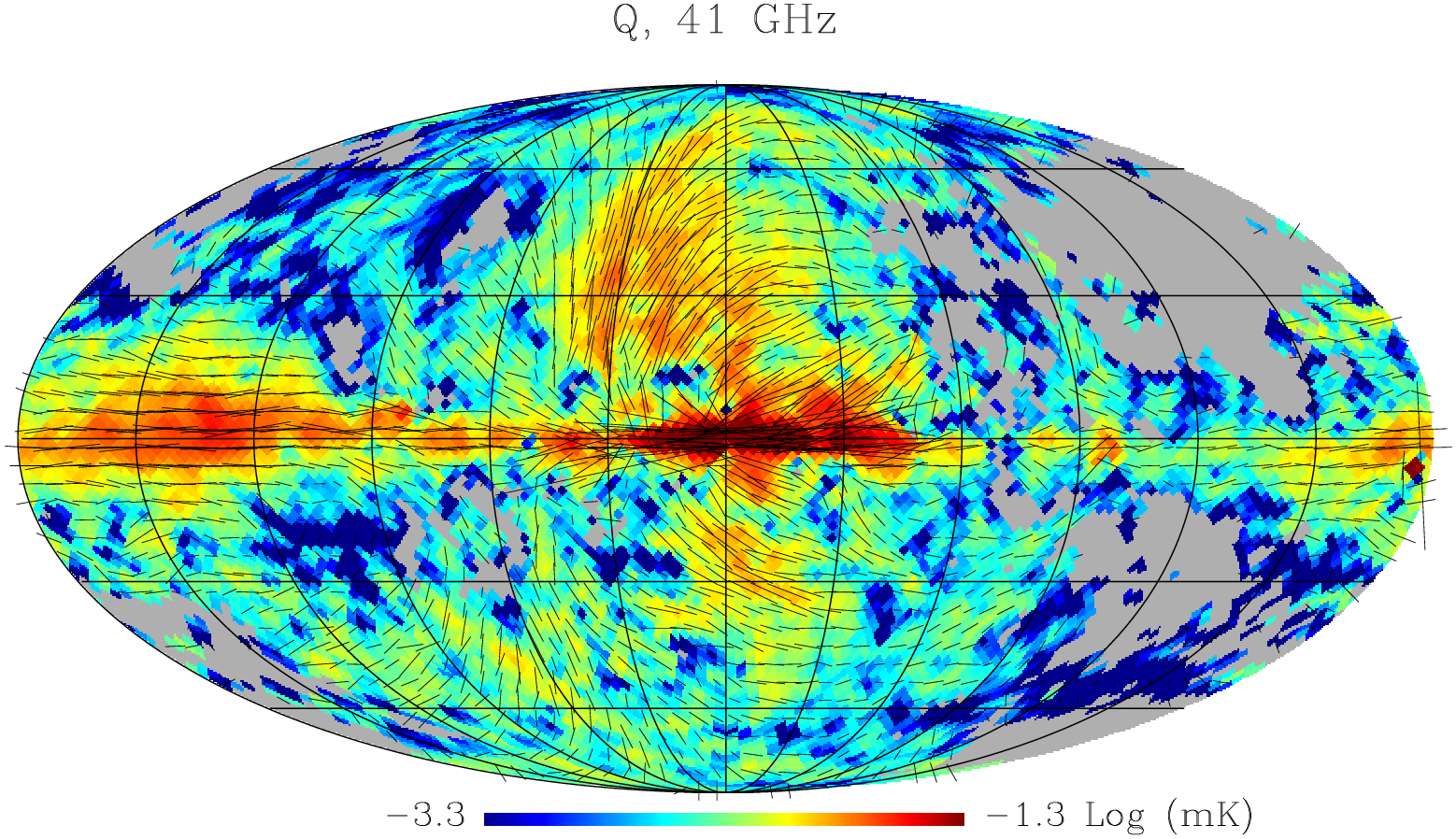}
  \includegraphics[angle=90,width=\widthfig\textwidth]{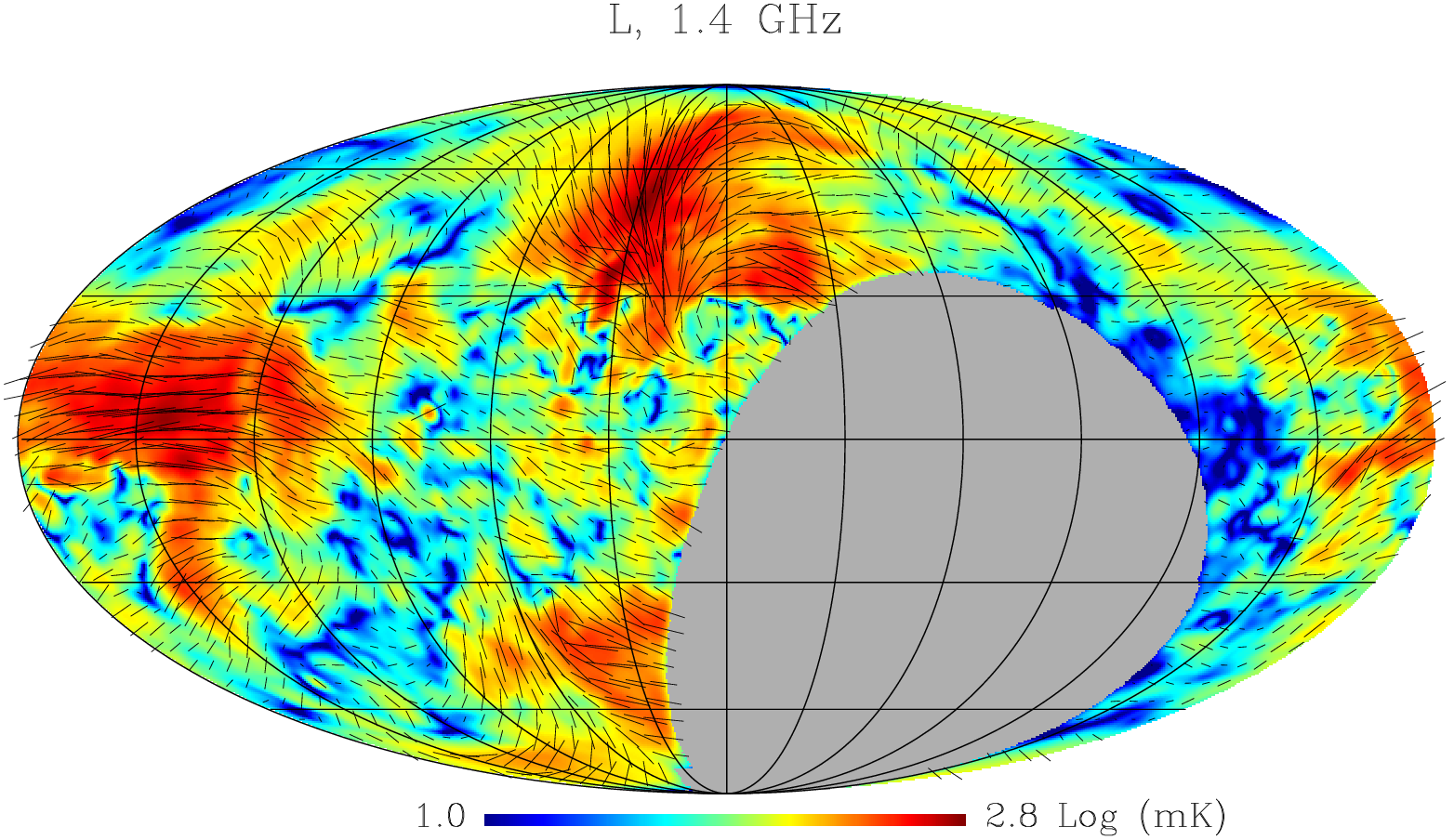}
  \caption[De-biased \wmap polarisation maps]{Polarisation intensity
    maps of \wmap K- ({\it top left}), Ka- ({\it top right}) and
    Q-band ({\it bottom left}) and the \citet{wolleben:06} 1.4\,GHz
    map ({\it bottom right}). The vectors indicate the magnetic field
    direction and their length is proportional to the polarisation
    intensity.. All maps have been smoothed to an angular resolution
    of 3\dg FWHM. The \wmap maps have been corrected for polarisation
    bias.  The masked areas (grey pixels) in the Ka and Q-band maps
    correspond to the regions where the SNR in the polarisation angle
    at K-band is lower than 5.  The masked area in the 1.4\,GHz map is
    the region that was not observed during the survey. The Galactic
    coordinate grid has a separation of $30^{\circ} \times
    30^{\circ}$. 
  }
  \label{fig:wmap_3deg_all}
\end{figure*}

\subsection{Polarisation bias}
\label{sec:pol_bias}
It has long been known that observations of linear polarisation are
subject to bias \citep{serkowski:58}. Given the positive nature of
$P=\sqrt{Q^2+U^2}$, even if the true Stokes parameters are zero, $P$
will yield a positive estimate in the presence of noise. The effect is
particularly important in the low SNR regime.

Ways to correct for the bias have been studied in detail
\citep{wardle:74, simmons:85,vaillancourt:06,quinn:12} for the special
case where the uncertainties for $(Q,U)$ are equal and normally
distributed around their true value $(Q_0,U_0)$. \citet{wardle:74}
proposed an estimator, which is widely used both for its simplicity
and for its good performance. The debiased polarisation intensity,
$P_{wk}$, is calculated as
\begin{equation}
P_{wk} = \sqrt{P' - \sigma_P^2},
\label{eq:pwk}
\end{equation}
where $P'=\sqrt{Q^2 + U^2}$ is the observed linear polarisation
intensity and $\sigma_P=\sigma_Q=\sigma_U$.

The case with asymmetric uncertainties ($\sigma_Q \neq \sigma_U$) on
the other hand is interesting as many polarisation data sets have this
characteristic. In particular, the correlations between the $(Q,U)$
uncertainties in \wmap data are mainly due to non-uniform azimuthal
coverage for each pixel in the sky. In this case, the bias will depend
on the polarisation angle $\chi$ as well as on $p$.  A generalisation
of the Wardle \& Kronberg estimator can be written for this case. It
has the form:
\begin{equation}  
  \hat{p}_{wk}=\sqrt{P'^2 - (\sigma_Q^2\sin^22\chi'
    +\sigma_U^2\cos^22\chi'-2\sigma_{QU}\cos2\chi'\sin2\chi' ) },
\label{eq:pwk_chi}
\end{equation}
where $\sigma_{QU}$ is the correlation term of the uncertainty in
$Q,U$. This estimator reduces to Eq. \ref{eq:pwk} when the errors are
isotropic.  Here, the observed polarisation angle, $\chi'$ is used as
a surrogate for the true polarisation angle $\chi$. At the high SNR
regime (e.g. the bright regions in \wmap \kband), the approximation
$\chi'\approx\chi$ is excellent. For instance, for a SNR of 15, the
uncertainty in the polarisation angle $\sigma_{\chi} \approx 1\mdgp9$.

The estimator in Eq. \ref{eq:pwk_chi} can be used to de-bias the
polarisation maps at \kband, where there is a reasonable SNR ($\gtrsim
5\sigma$) along large areas of the sky. For the rest of the bands,
where the SNR is lower we can derive a new estimator. For this, we use
the fact that we can obtain a good estimation of the {\em true}
polarisation angle $\chi$ from \wmap \kband. Moreover, at \wmap
frequencies, Faraday rotation is negligible over most of the sky (see
Section \ref{sec:far_rot}) so we can assume that the polarisation
angle $\chi$ observed at \kband, will be the same at the higher
frequency bands. The additional information about the polarisation
angle in the higher frequency bands will help to reduce the total bias
in the pixels with low SNR. The new estimator requires the observed
$Q', U'$ values, the uncertainties $\sigma_Q, \sigma_U, \sigma_{QU}$
and an independent measurement of the polarisation angle $\chi$.  The
derivation for this known-angle estimator and the test of its
performances is described in \cite{Vidal2014b}. The estimator takes
the following form:

\begin{align}
  \hat{p}_{\chi} = \frac{ \sigma^2_U Q'\cos 2\chi - \sigma_{QU}(Q'\sin
    2\chi +U'\cos{2\chi}) + \sigma^2_Q U' \sin 2\chi } {
    \sigma^2_U\cos^22\chi - 2\sigma_{QU}\sin 2\chi \cos 2\chi +
    \sigma^2_Q\sin^22\chi } .
  \label{eq:pdeb_chi}
\end{align}

and its uncertainty,

\begin{equation}
  \sigma^2_{\hat{p}_{\chi}}=\frac{ \sigma^2_Q\sigma^2_U -
    \sigma^2_{QU} } {\sigma^2_U\cos^22\chi - 2\sigma_{QU}\sin 2\chi
    \cos 2\chi + \sigma^2_Q\sin^22\chi }.
\end{equation}

In \cite{Vidal2014b}, simulations are used to test the effectiveness
of this new estimator and to quantify any residual bias due to noise
in $\chi$.

In summary, we use the generalised Wardle \& Kronberg estimator from
Eq. \ref{eq:pwk_chi} to correct the polarisation intensity map at
K-band and the known-angle estimator to correct the Ka and Q-band
maps.

\subsection{Bias-corrected polarisation maps}
\label{sec:pol_maps}
We prepared bias-corrected polarisation intensity maps of \wmap K, Ka
and Q-bands maps, smoothed to a common resolution of 3\dg\!\!.  At
K-band, some pixels show a noise value (terms inside the bracket in
Eq. \ref{eq:pwk_chi}) that is larger than the observed polarisation
intensity. This will produce some pixels with imaginary values after
the bias correction (see Eq. \ref{eq:pwk_chi}). These pixels are
masked-out for the analysis. Given the nature of the known-angle
de-biasing method that we use for Ka-- and Q--bands --the knowledge of
the true polarisation angle-- we have masked out the pixels that show
an uncertainty in the polarisation angle at K-band larger than
5$^{\circ}$\!\!.8, which are the ones that have a SNR in polarisation
intensity, SNR$_{\rm P} <5$. This corresponds to a total masked
sky-area of 17.3\%. This SNR cutoff was chosen to be sure that any
residual bias will be smaller than 5\%, as shown in
\citet{Vidal2014b}.  In Fig. \ref{fig:wmap_3deg_all} we show the
maps, with the masked areas shown as grey. Also plotted on each map
are the polarisation vectors, aligned parallel to the magnetic field
direction (we have rotated the vectors by 90\dg\!\!). We also include
the 1.4\,GHz polarisation map from \cite{wolleben:06}.

By looking at \wmap \kband~in Fig. \ref{fig:wmap_3deg_all}, it can be
seen that most of the emission outside the Galactic plane comes from a
number of individual large scale features. Regions we can easily
identify are the Galactic plane, the fan region centred at $l\approx
140^{\circ}$, and a number of loops and filaments than run mostly
perpendicular to the Galactic plane. The biggest of these filaments is
Loop I, running perpendicular to the plane at $l \sim 30^{\circ}$.  We
note also that the magnetic field vectors are coherent with the
direction of the filamentary structures; we will quantify this
observation in the next section.

The most obvious difference between the \wmap maps and the Wolleben
map at 1.4\,GHz is the inner Galaxy between $b \la
|30^{\circ}|$. Faraday rotation produces a depolarisation region in
the 1.4\,GHz map, which cancels out most of the polarised emission
close to the Galactic plane, but as noted above, this is negligible at
\wmap frequencies, except towards the Galactic Centre (see
Section~\ref{sec:far_rot}).


\newcommand{\mytoprule}{\specialrule{1.5pt}{0em}{0em}}
\renewcommand{\arraystretch}{1.1}
\begin{table*}
  \centering
  \caption{Circular loops and filaments parameters. The {\it top}
    section lists the previously identified continuum loops. The {\it
      middle} section lists the loops and circular arcs that are
    visible only in polarisation.  The {\it bottom} section lists the
    filaments that are not circular arcs.  The 14 loops, arcs and
    filaments that are visible in polarisation are shown in the {\it
      bottom} panel of Fig. \ref{fig:unsharp_haslam_kpol}.  All the
    numerical values in the table are in degrees.  }
  \begin{tabular}{@{}lcccccccl@{}}
    \mytoprule
    Loop  &   \multicolumn{3}{c}{Continuum}  &   & \multicolumn{3}{c}{Polarisation} &   Comments\,/\,Reference  \\ 
    Name & $l$ & $b$ & $r$ && $l$ & $b$ & $r$    &\\
    \midrule    
    I    & 329.0 & 17.5  & 58     && $  332.6$ & $   20.7$ & $ 54.3$ & NPS, \citet{brown:60}\\ 
    II   & 100.0 & $-32.5$ & 45.5 &&   $--$    & $--$      &  $--$     & \citet{large:62}\\
    III  & 124.0 & 15.5  & 32.5   && $  118.8$ & $   13.2$ & $   31.6$ & \citet{quigley:65}\\
    IV   & 315.0 & 48.5  & 19.8   && $  315.8$ & $   48.1$ & $   19.3$ & \citet{large:66} \\
    V    & 127.5 & 18.0  & 67.2   &&  $--$     &  $--$     &  $--$     & \citet{milogradov-turin:97}, same as loop III.  \\
    VI   & 120.5 & 30.8  & 72.4   &&  $--$     &  $--$     &  $--$     & \citet{milogradov-turin:97}, not visible in \wmap polarisation. \\
    \midrule        
    GCS  & $--$ & $--$ &  $--$    && 344.0     & 4.8   & 18.5          & Galactic centre spur from \citet{sofue:89}. \\
    III{\sc s}$^*$   & $--$ & $--$ &  $--$    && 106       & $-22$ & 50.0          & Below the plane at the "fan" region. \\
    VIIa & $--$ & $--$ &  $--$    && $  345.0$ & $0.0$     &   $ 65$   & From \citet{wolleben:07} using the 1.4\,GHz map. \\
    VIIb$^*$ &      &      &          && $  0.7$   & $-23.3$   & $ 45.9$   & Values found in this work for  VIIa. \\
    IX$^*$   & $--$ & $--$ &  $--$    && 332.0     & 16    & 46.5          & Inside the NPS.\\
    X$^*$ & $--$ & $--$ &  $--$    && 30        & 35    & 67.0          & Tangential to the plane.\\
    XI$^*$ & $--$ & $--$ &  $--$    && 227       & 38    & 81.0          & Large arc tangential to the plane.\\
    XII  & $--$ & $--$ &  $--$    && 300       & 0.7    & 27.6          & Close to the LMC.\\
    \midrule
    VIII  & $--$ & $--$ &  $--$    && $--$   & $--$    & $--$          & Southern part of the GCS? \\
    XIII  & $--$ & $--$ &  $--$    && $--$   & $--$    & $--$          & Weak continuum emission at 0.408\,GHz.  \\
    XIV   & $--$ & $--$ &  $--$    && $--$   & $--$    & $--$          & Possible H{\sc I} counterpart (see Sec. \ref{sec:hi}). Weak continuum emission at 0.408\,GHz. \\
    \bottomrule
  \end{tabular}
   \begin{flushleft}
     $^*$: These arcs are clearly visible only in polarisation. \\
   \end{flushleft}
  \label{tbl:circ_loops}
\end{table*}

\section{Polarised large-scale features.}
\label{sec:fil_angles}

\begin{figure}
  \centering %
  \includegraphics[angle=90,width=0.49\textwidth]{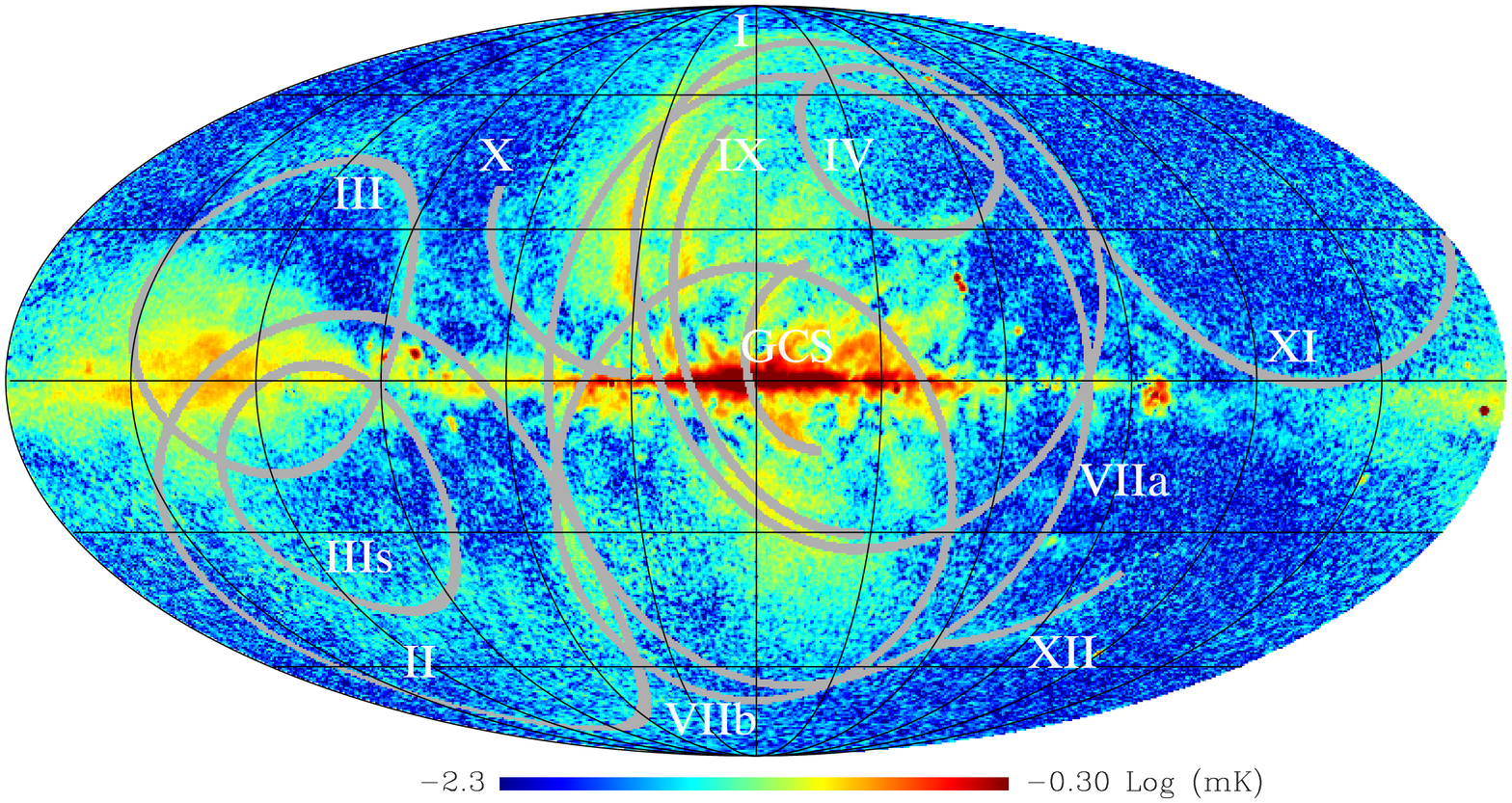}
  \includegraphics[angle=90,width=0.49\textwidth]{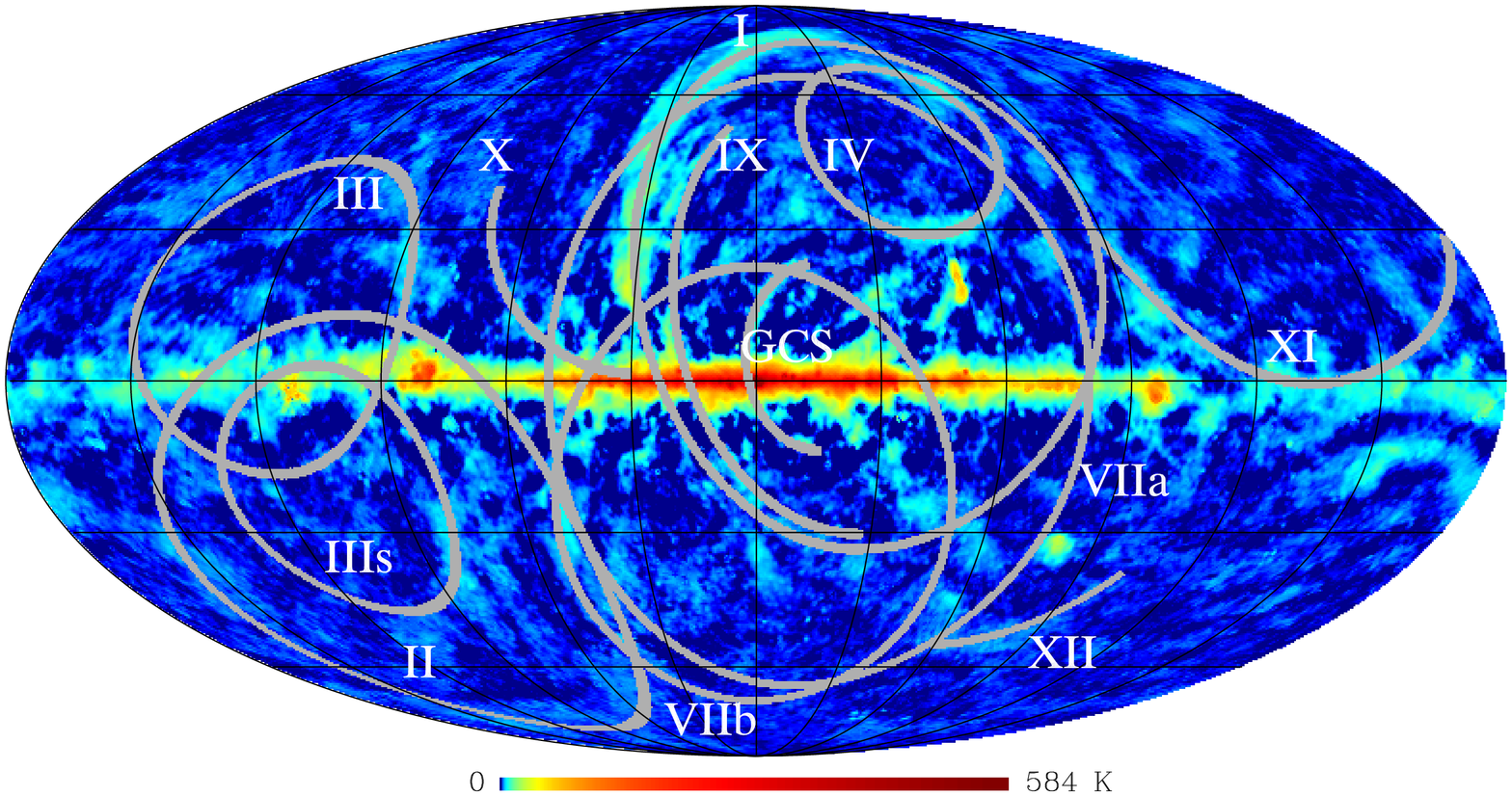}
  \includegraphics[angle=90,width=0.49\textwidth]{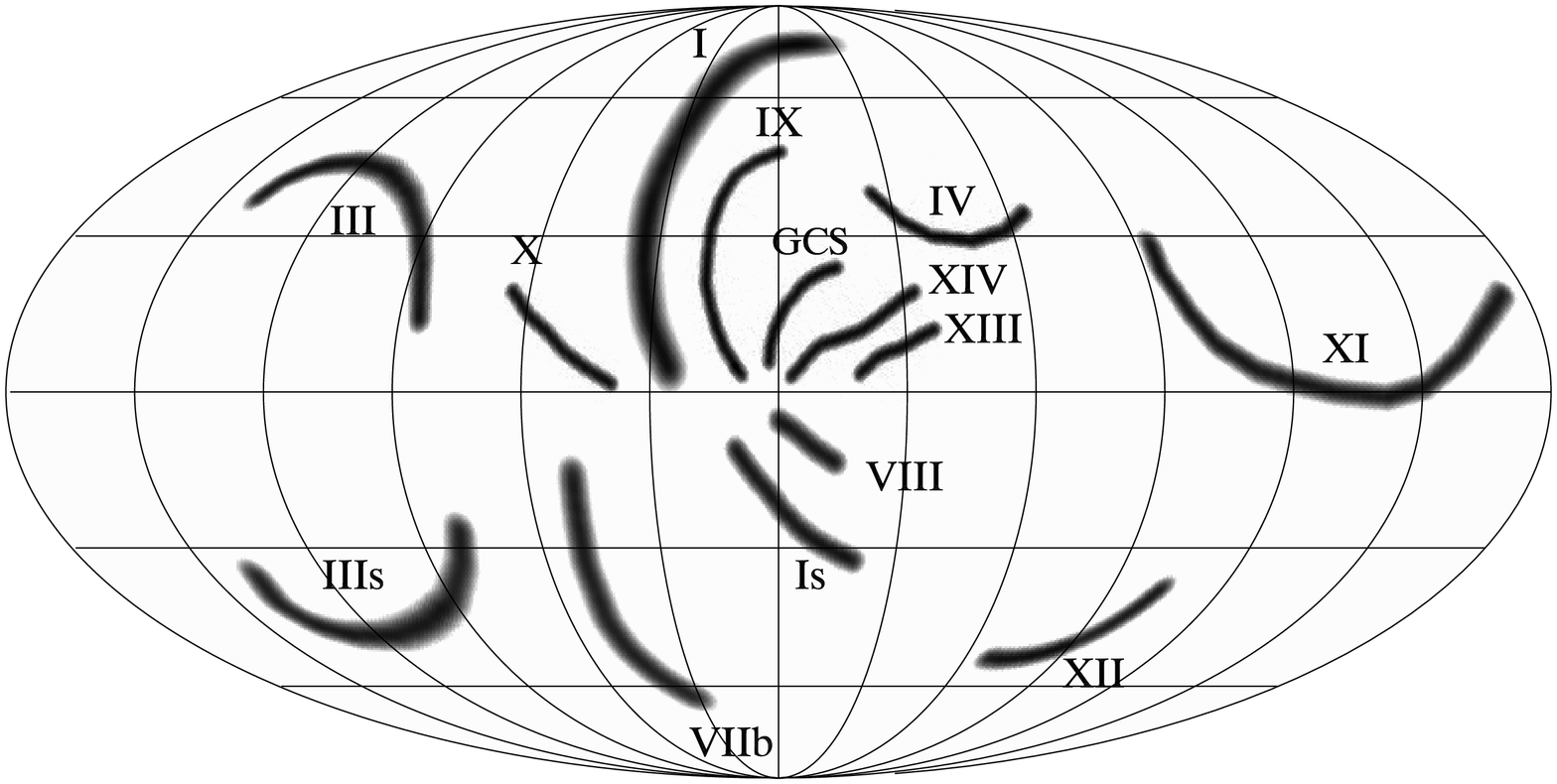}
  \caption{{\it Top: } \wmap--K band polarisation intensity map.  {\it
      Middle:} Un-sharp mask version of the Haslam et al. map.  The
    angular resolution of the unfiltered maps is 1\dg and the filter
    beam has a size of 10\dg FWHM.  The location and sizes of the 13
    loops and arcs are listed in Table \ref{tbl:circ_loops}.  {\it
      Bottom:} Template showing the filaments visible on the {\it top}
    panel. They are defined following maximum brightness points along
    each filament.  }
  \label{fig:unsharp_haslam_kpol}
\end{figure}

A number of radio loops have been described in the literature, Loop I or NPS being
the best known.  Most have been observed in low frequency ($\nu
<1$\,GHz) continuum radio surveys. Six loops have been described in
the literature. These loops are well fitted by small circles on the
sky \citep[see e.g.][]{berkhuijsen:71b} in the low frequency continuum
data. Table \ref{tbl:circ_loops} lists some properties of the loops
along with comments and references. We identify the loops in
polarisation based on the 1\dg de-biased polarised intensity map at
\kband. Additionally, we prepared a high-pass filtered version of the
0.408\,GHz map from \citet{haslam:82} to highlight the filamentary
structure in the continuum map. This was done using an unsharp-mask
\citep{sofue:89} which filters out the emission at angular scales
larger than 10\dg\!\!.  In Fig. \ref{fig:unsharp_haslam_kpol} we show
the \wmap K--band polarisation intensity map and the filtered version
of the 0.408\,GHz Haslam et al. map.

A number of filaments are easily recognisable over the sky; most of
them lie in the inner Galaxy, with Galactic longitude in the range
$-90\mdg<l<90\mdg$. Here we concentrate on the filaments that have a
``circular'' arc shape. The continuum loops I, III and IV are visible
in polarisation although the more diffuse Loop II is not obvious.

We fitted small-circle arcs to the coordinates of the pixels that
trace the peak of the circular features in
Fig. \ref{fig:unsharp_haslam_kpol}. Table \ref{tbl:circ_loops} lists
the parameters found for the three previously known continuum loops
that are visible in the polarisation data, as well as the parameters
published on the literature from continuum fits to the data. We did
not attempt to fit for Loop II (Cetus Arc) due to the very low
emission observed in the \wmap polarisation maps. There is a good
agreement for the radii and centres of the loops between the low
frequency continuum data and the \wmap polarisation data.  Loop III is
the one that shows the larger discrepancy between the geometry
measured in continuum with respect to polarisation, where the emission
in polarisation peaks 5\dg away from the continuum. This shift between
the total intensity and the polarisation morphology was previously
noted by \citet{spoelstra:72}.

We have identified five new arcs that are visible only in these
polarisation data (marked with an asterisk in Table
\ref{tbl:circ_loops}) that can be fitted by small circle arcs.
\citet{wolleben:07} identifies a ``New Loop'' (loop VIIa in our
nomenclature) using the 1.4\,GHz polarisation map from the DRAO survey
\citep{wolleben:06}. The 1.4\,GHz maps present a depolarisation band
at Galactic latitudes $|b|<30\mdg$ due to Faraday
depolarisation. Because of this, they only use data at $b<-35\mdg$ to
define the location and size if this new loop. Here we include data
closer to the Galactic plane that belongs to the arc in order to
define its geometry. This filament is brighter closer to the Galactic
plane and is listed as VIIb.


  \newcommand{\widtha}{0.37}
  \newcommand{\widthb}{0.57}  

\begin{figure*}
  \renewcommand{\widtha}{0.27} \renewcommand{\widthb}{0.27} \centering
  \includegraphics[angle=0,width=\widtha\textwidth]{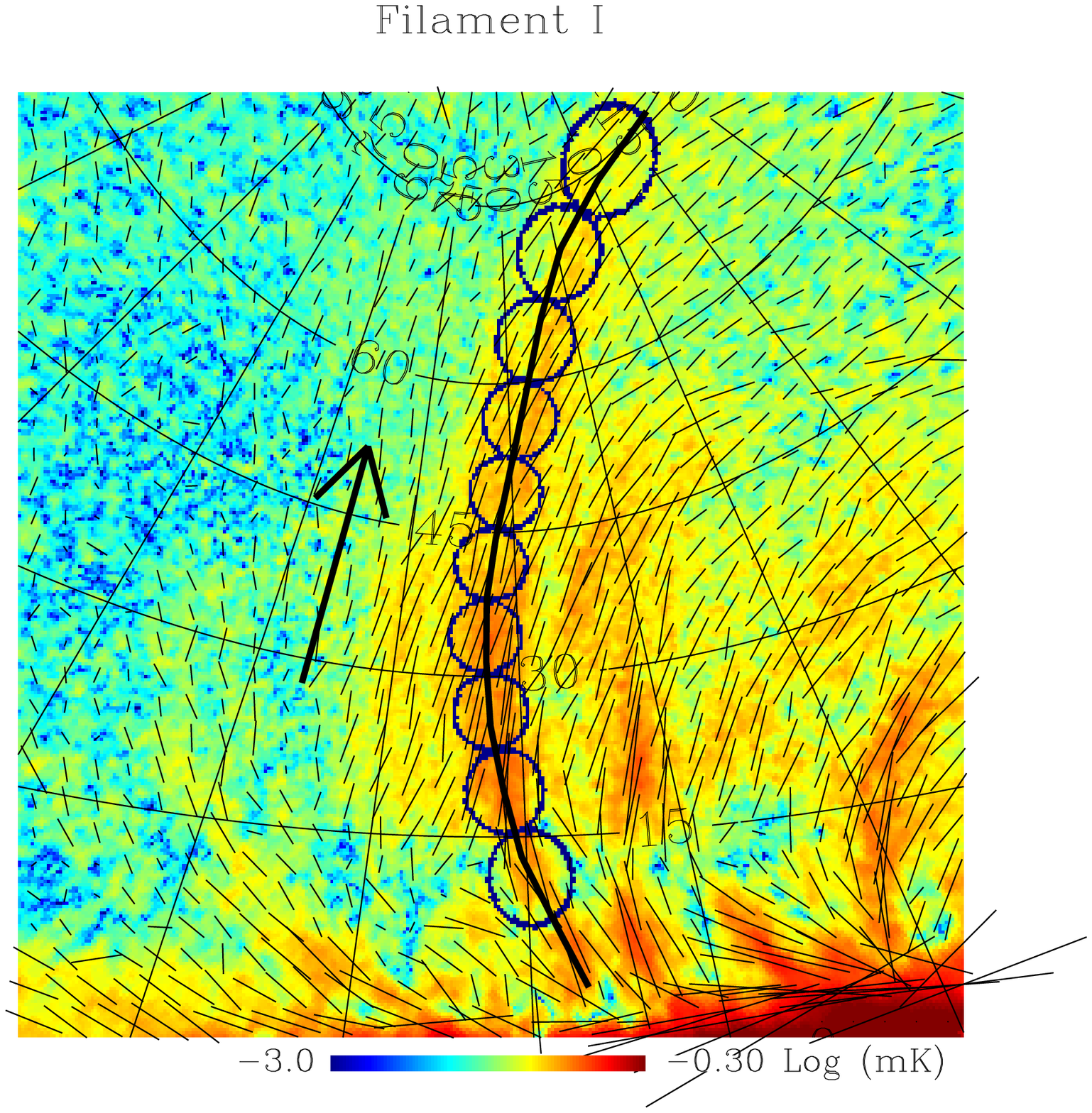}
  \includegraphics[angle=0,width=\widtha\textwidth]{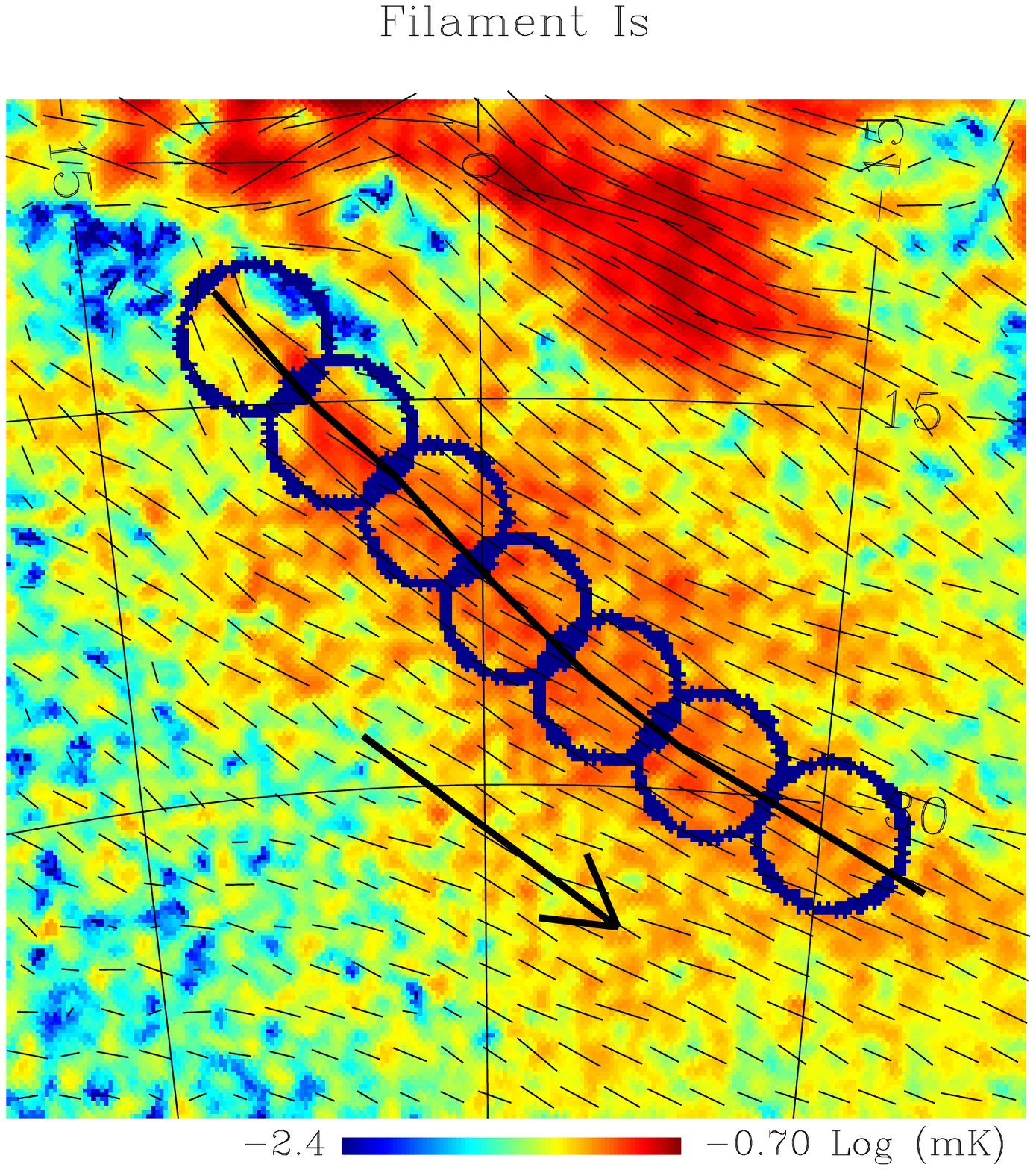}
  \includegraphics[angle=0,width=\widtha\textwidth]{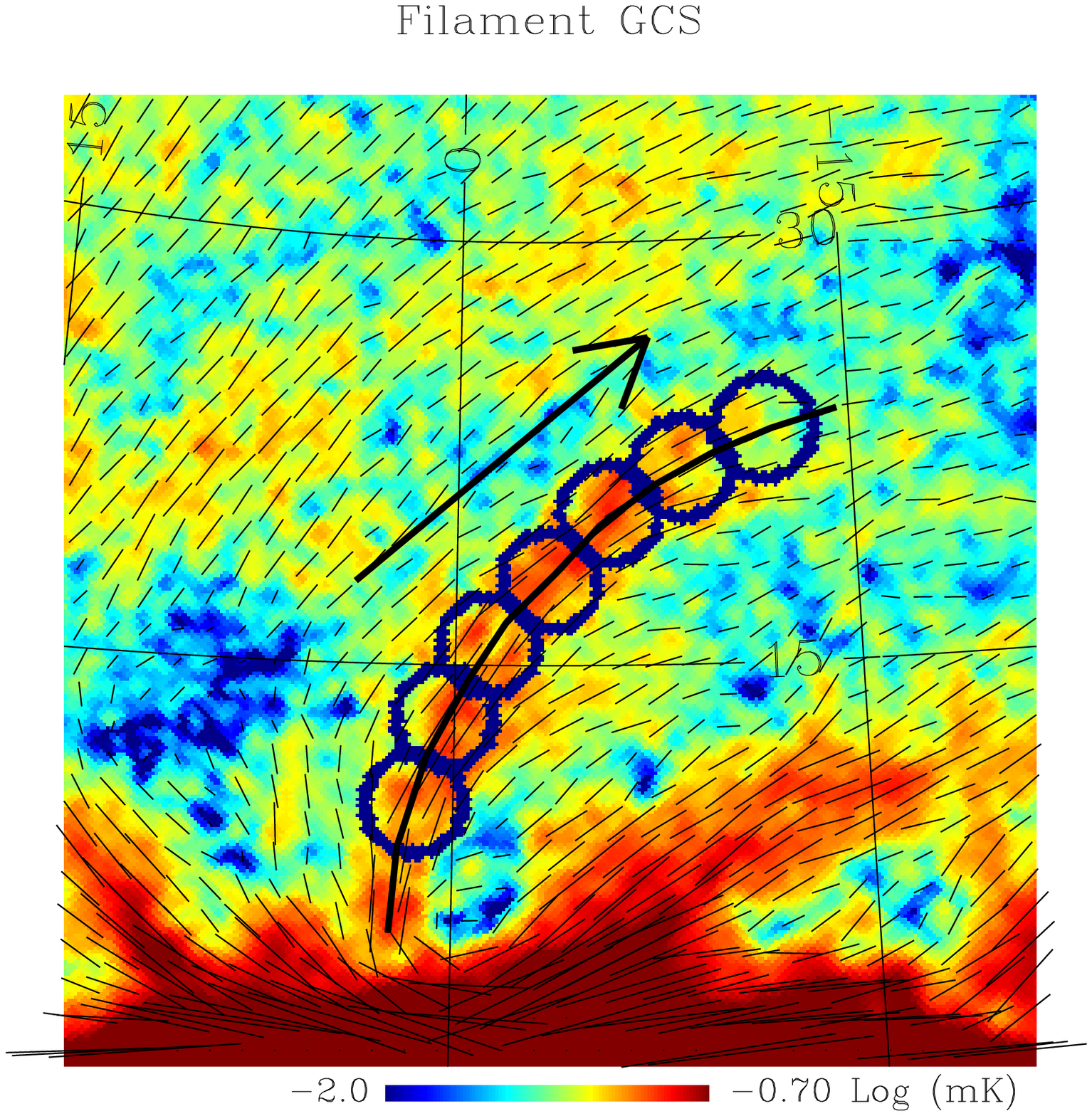}   
  \includegraphics[angle=0,width=\widthb\textwidth]{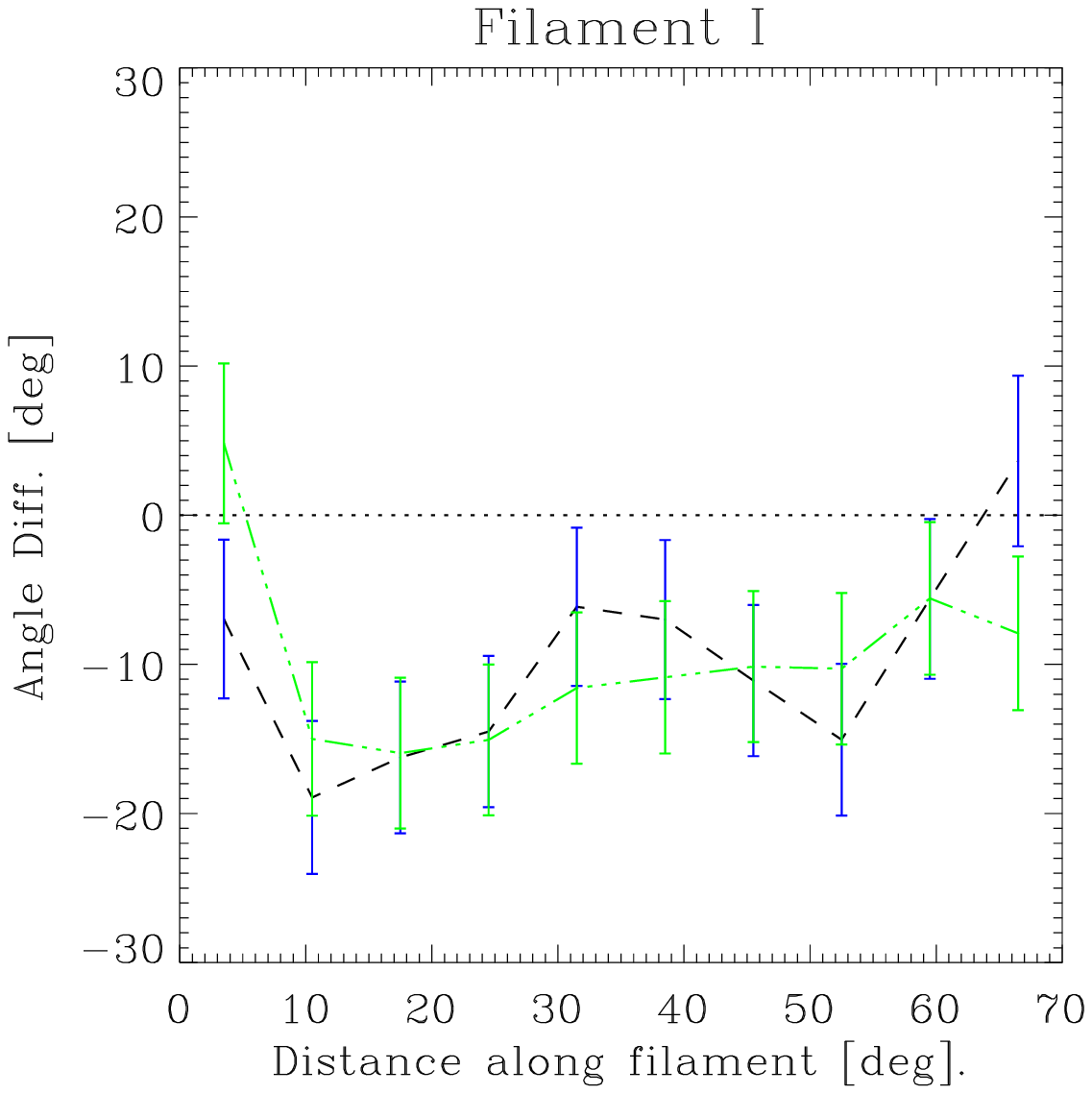}
  \includegraphics[angle=0,width=\widthb\textwidth]{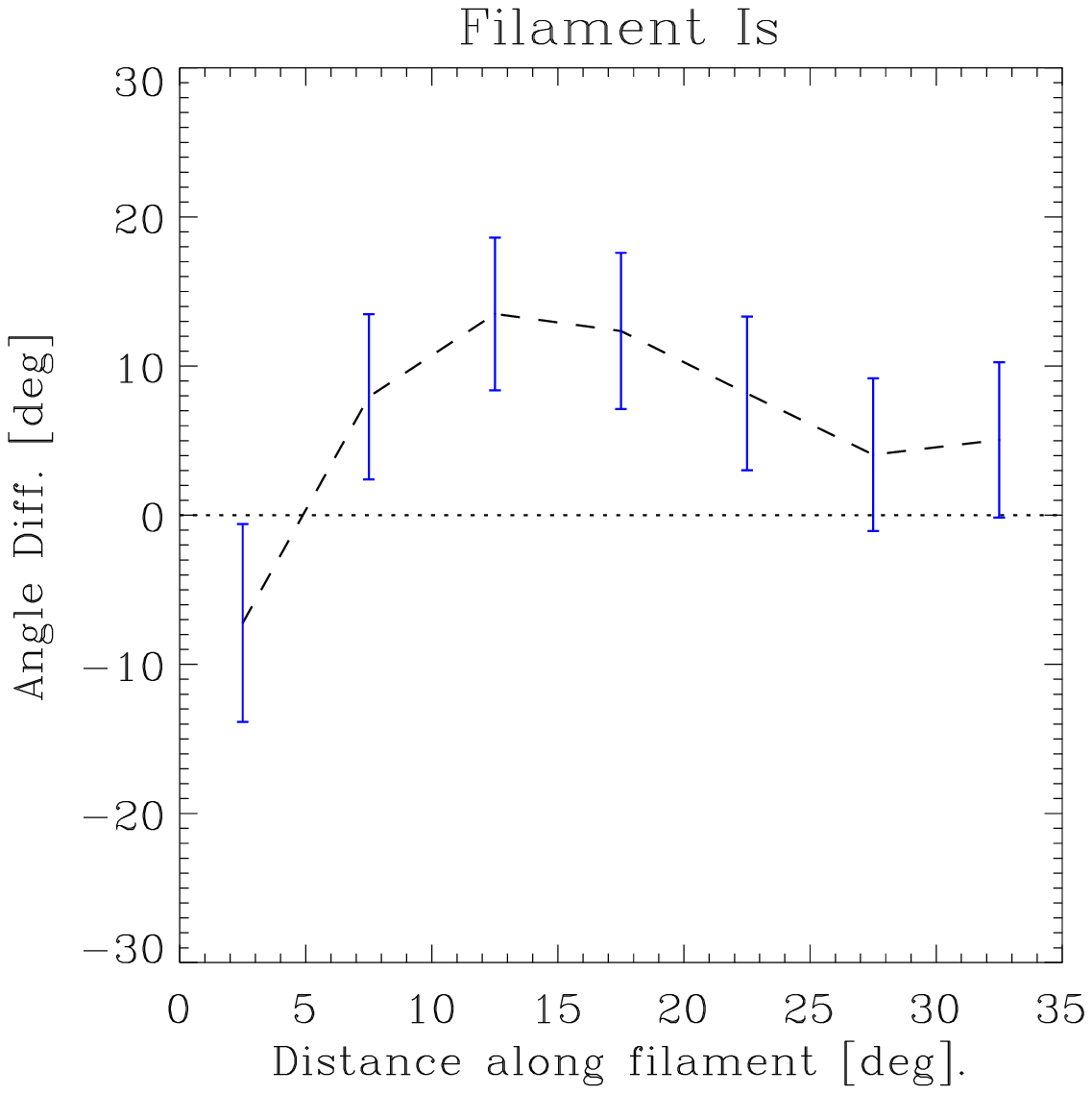}
  \includegraphics[angle=0,width=\widthb\textwidth]{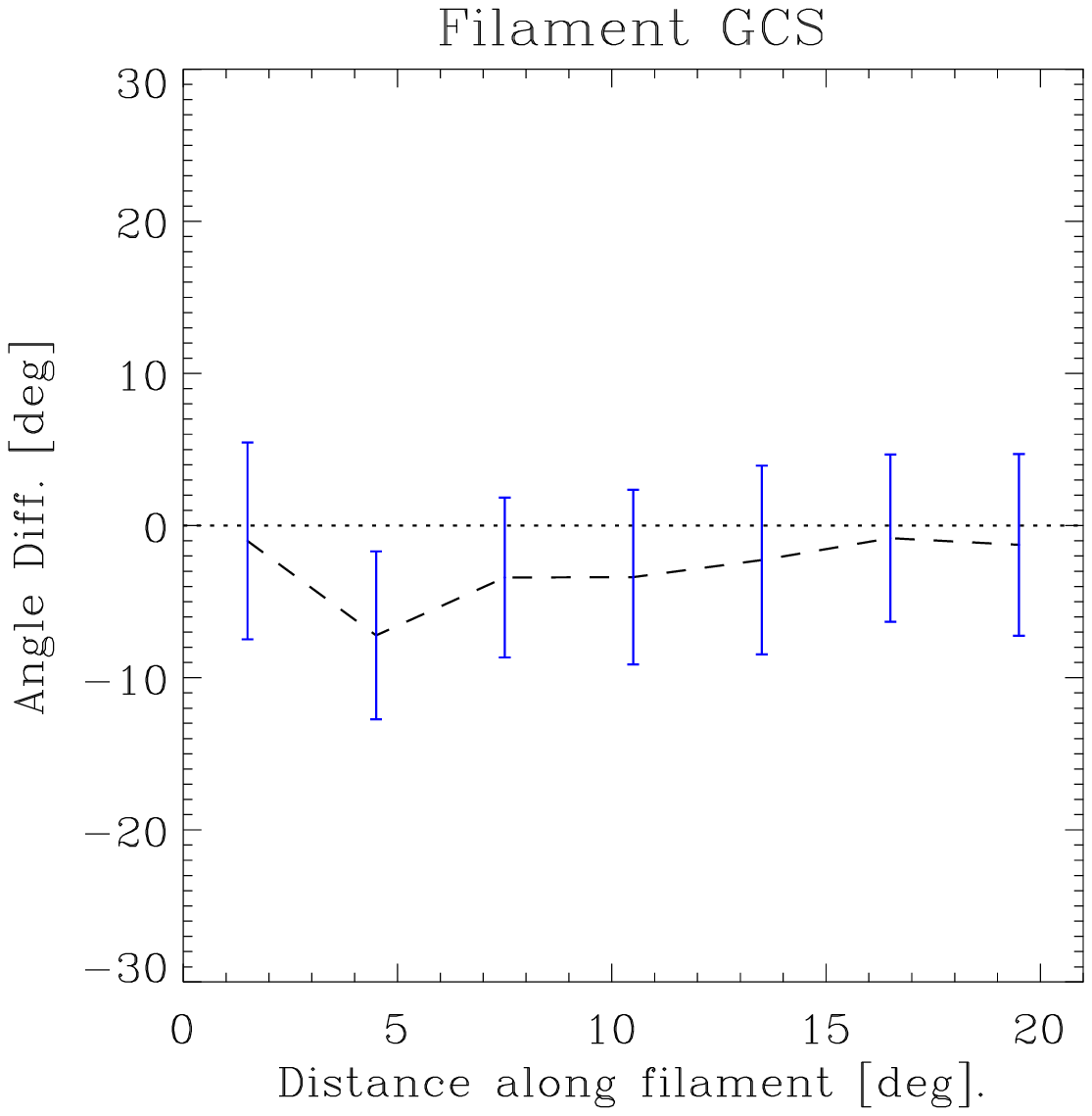}
  \includegraphics[angle=0,width=\widtha\textwidth]{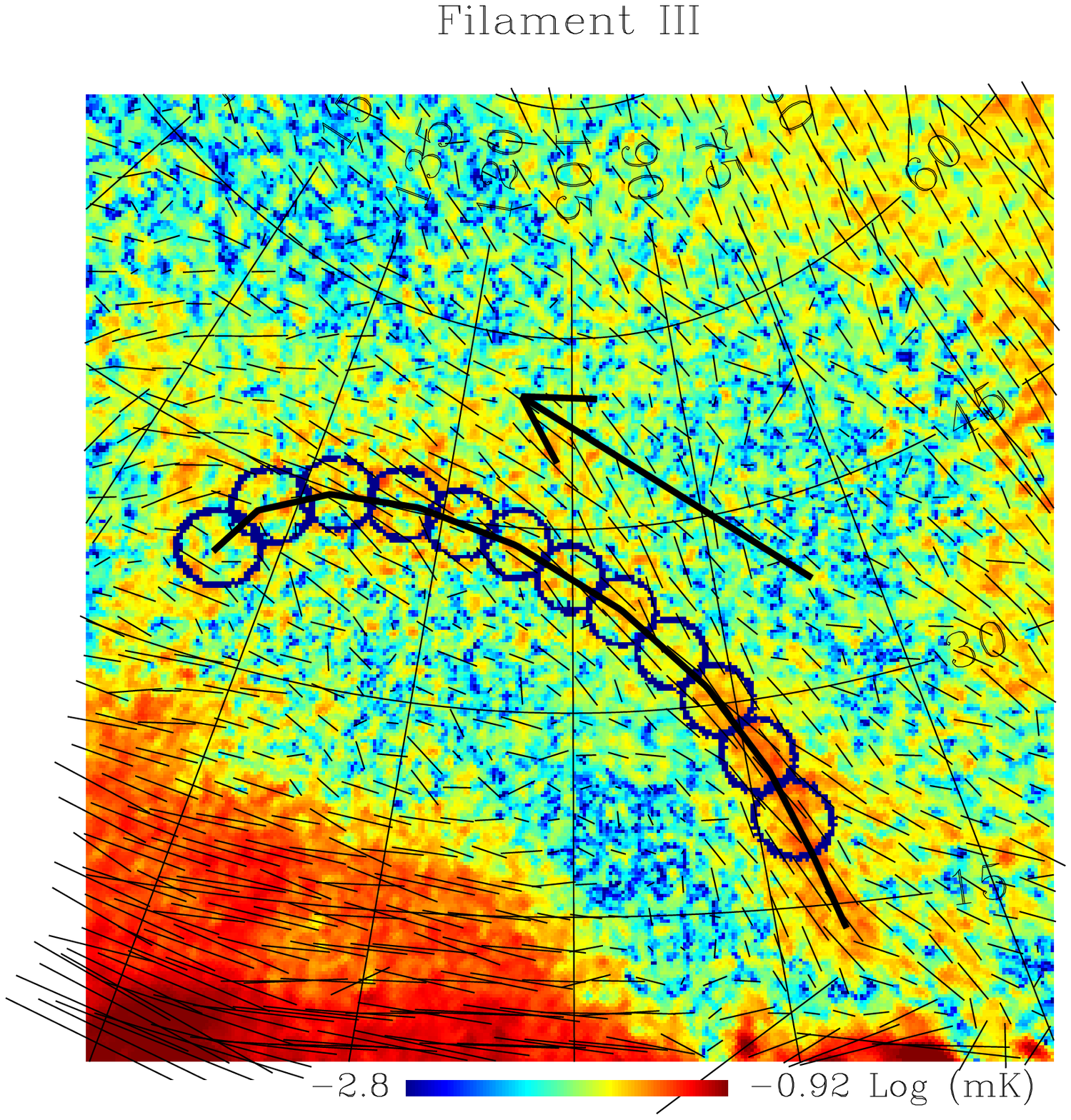}
  \includegraphics[angle=0,width=\widtha\textwidth]{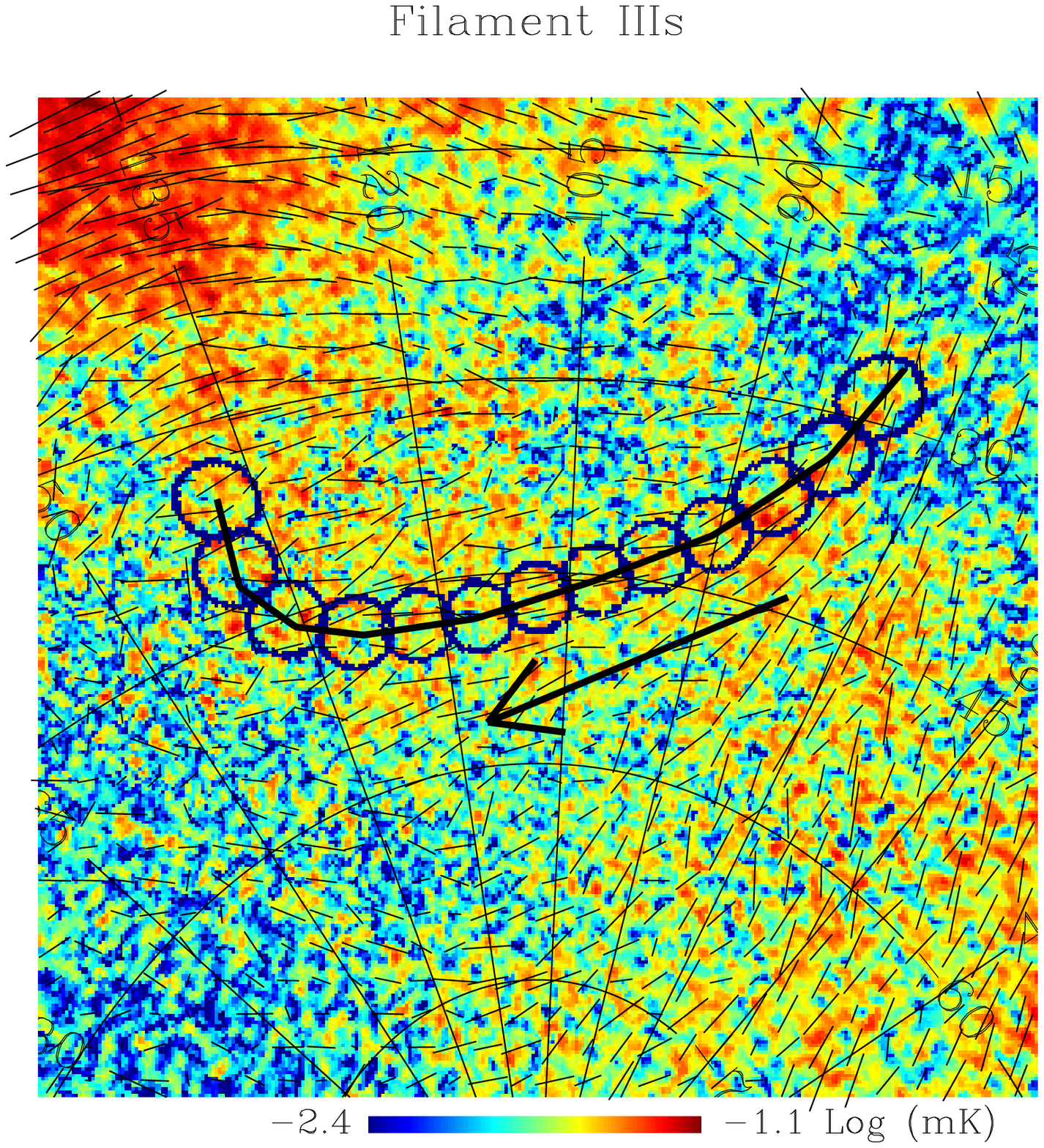}
  \includegraphics[angle=0,width=\widtha\textwidth]{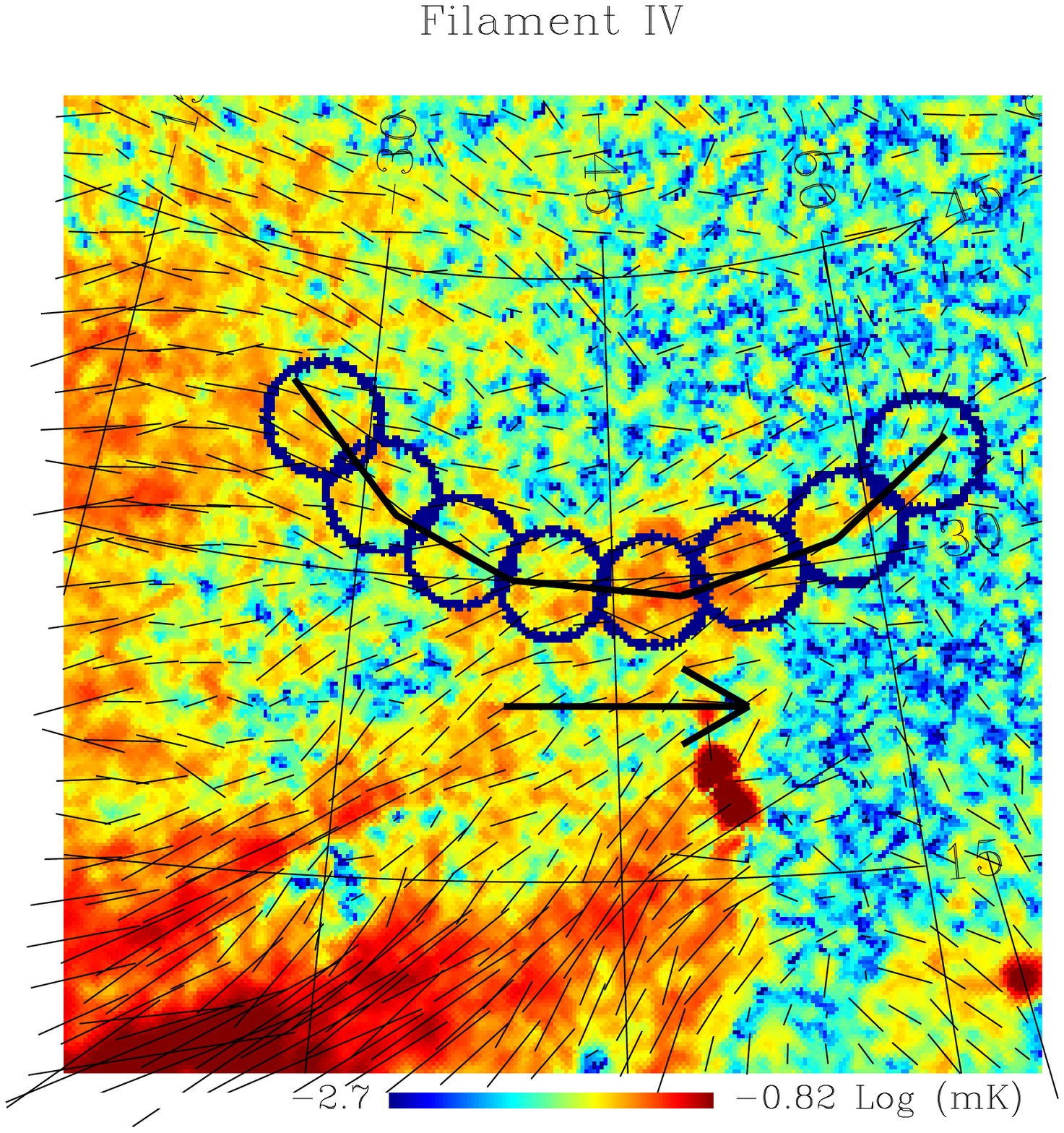}
  \includegraphics[angle=0,width=\widthb\textwidth]{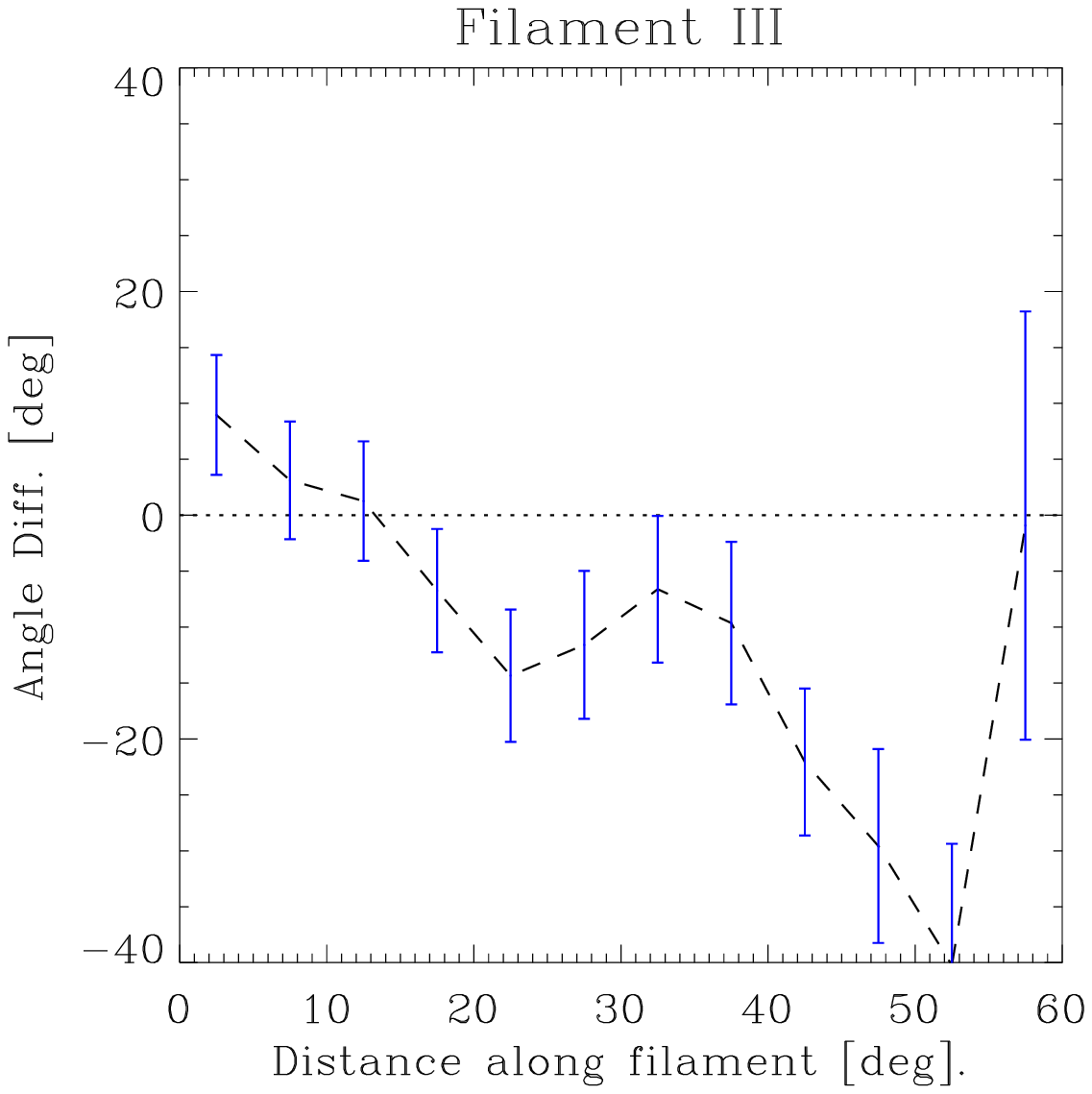}
  \includegraphics[angle=0,width=\widthb\textwidth]{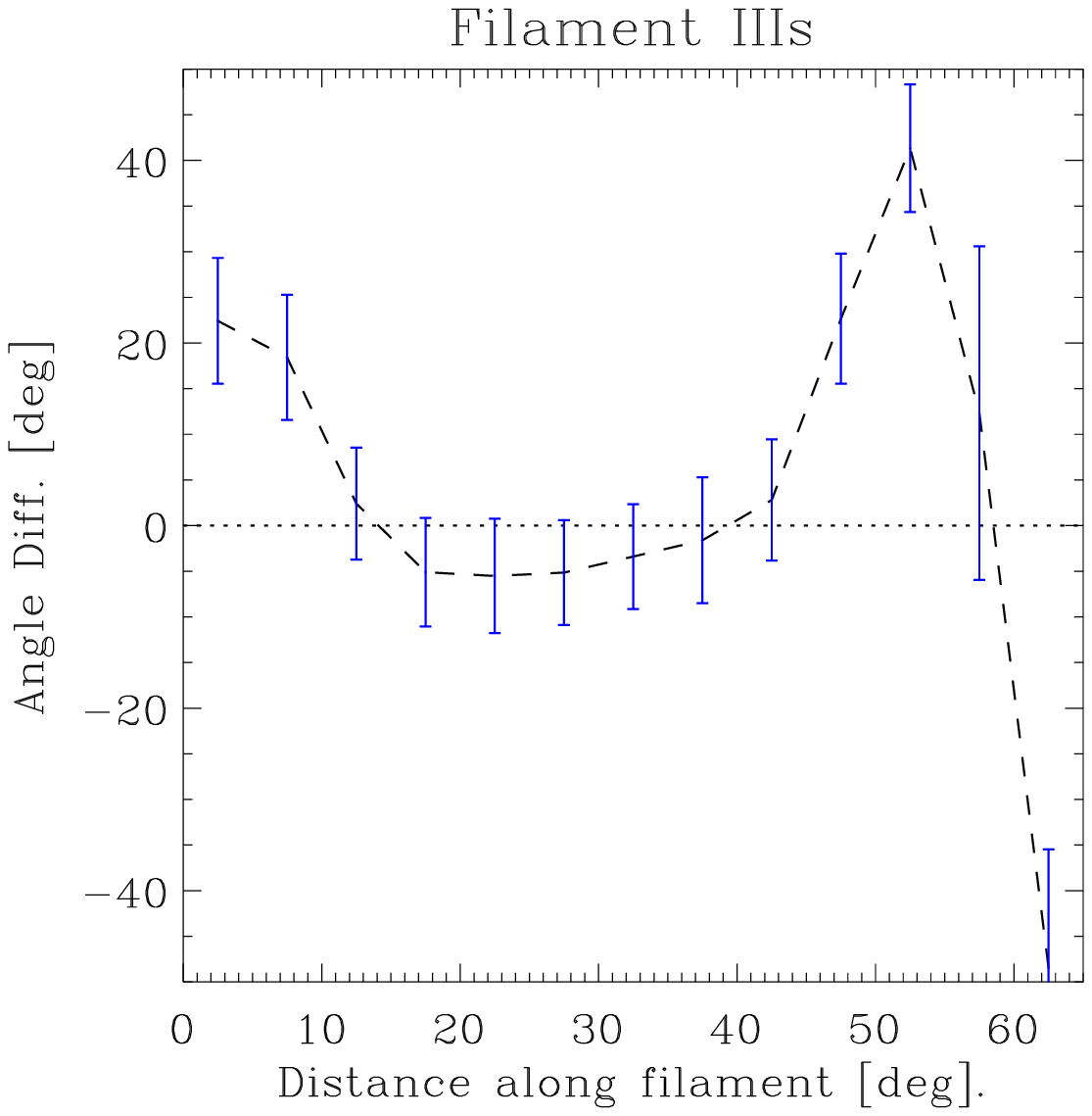}
  \includegraphics[angle=0,width=\widthb\textwidth]{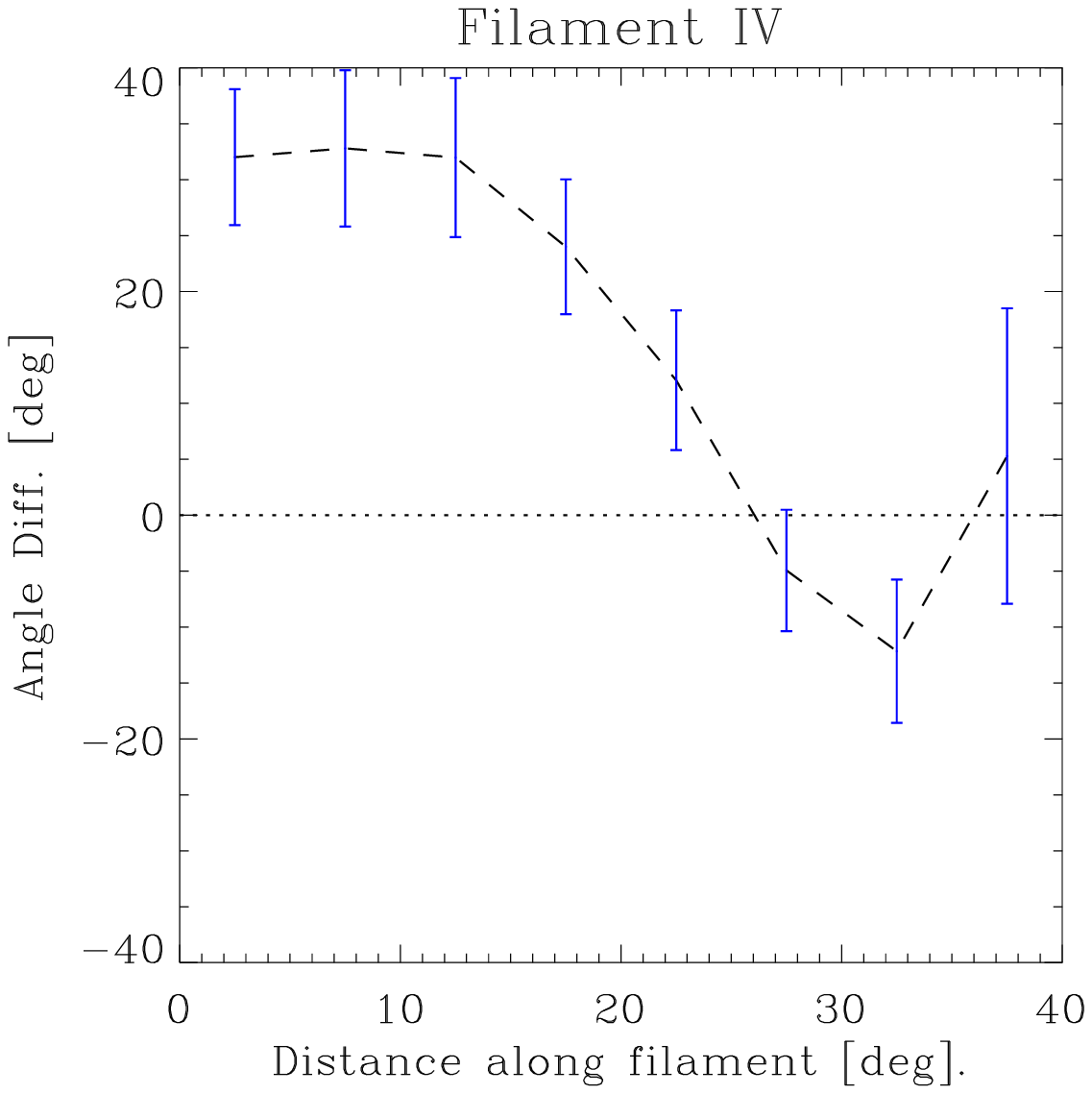}
  \caption[Comparison between the polarisation angle $\chi$ along the
    filaments and the direction $\alpha$ defined by its extension]
          {The panels on the {\it top} show maps of each filament as
            defined in the {\it bottom} panel of Fig
            \ref{fig:unsharp_haslam_kpol}. The adjacent circles in
            each filament show the apertures that we use to measure
            the difference between the polarisation angle (represented
            by the small black vectors) and the direction of the
            filament (defined by thin grey line running along the
            filament). The panels at the {\it bottom} show the
            difference between the polarisation angle, $\chi$, and the
            angle defined by the direction of each filament, $\alpha$,
            along the filament. In these plots, the distance along the
            filament is measured in the direction indicated by the
            black arrow. The error bars represent the random
            fluctuation in the polarisation angle, including an
            additional 5\degr uncertainty assigned as a conservative
            systematic error in the definition of the direction of the
            filament. }
  \label{fig:diff_angle_1}
\end{figure*}

\begin{figure*}
  \renewcommand{\widtha}{0.27}
  \renewcommand{\widthb}{0.27}
  \centering
  \includegraphics[angle=0,width=\widtha\textwidth]{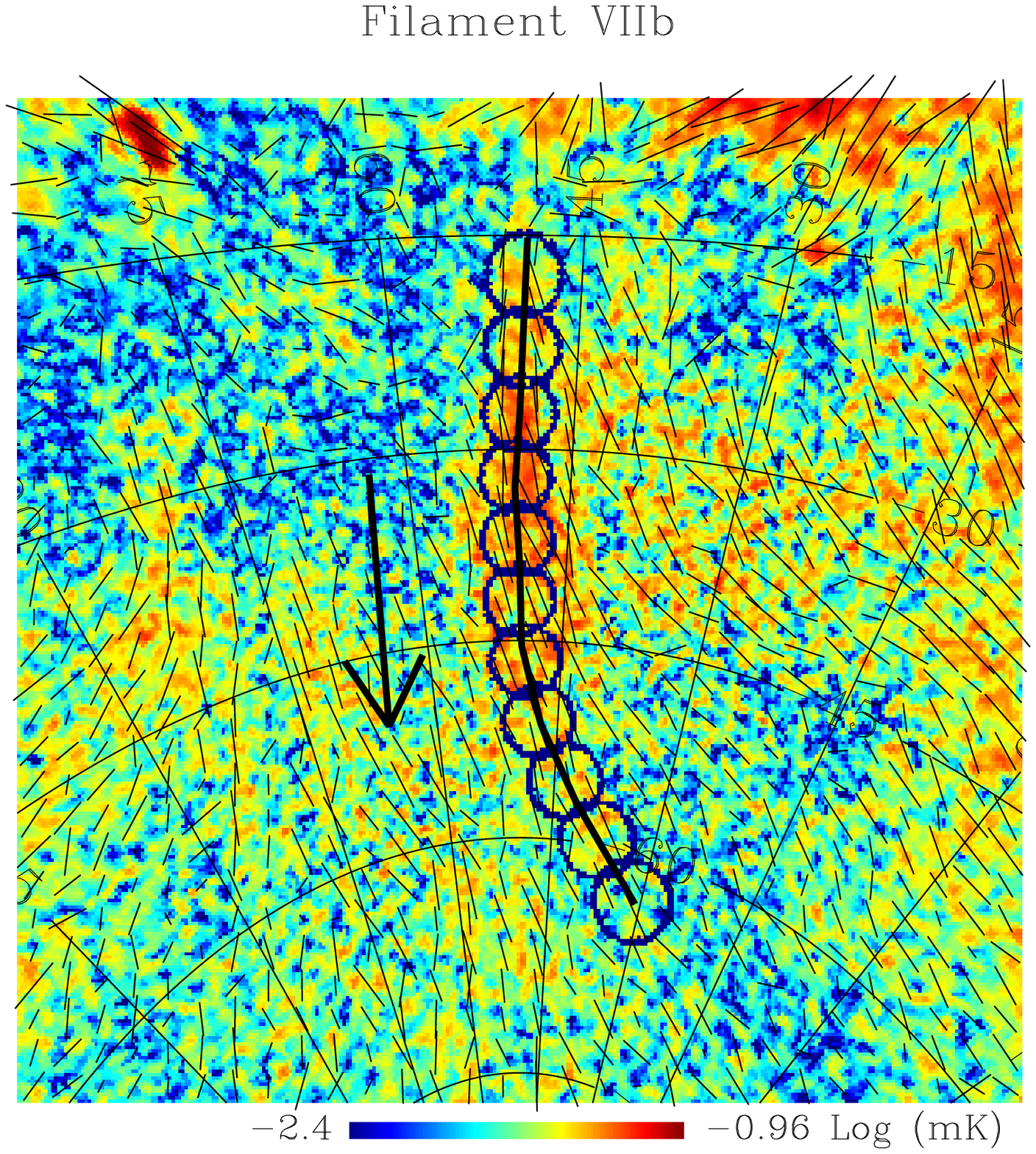}
  \includegraphics[angle=0,width=\widtha\textwidth]{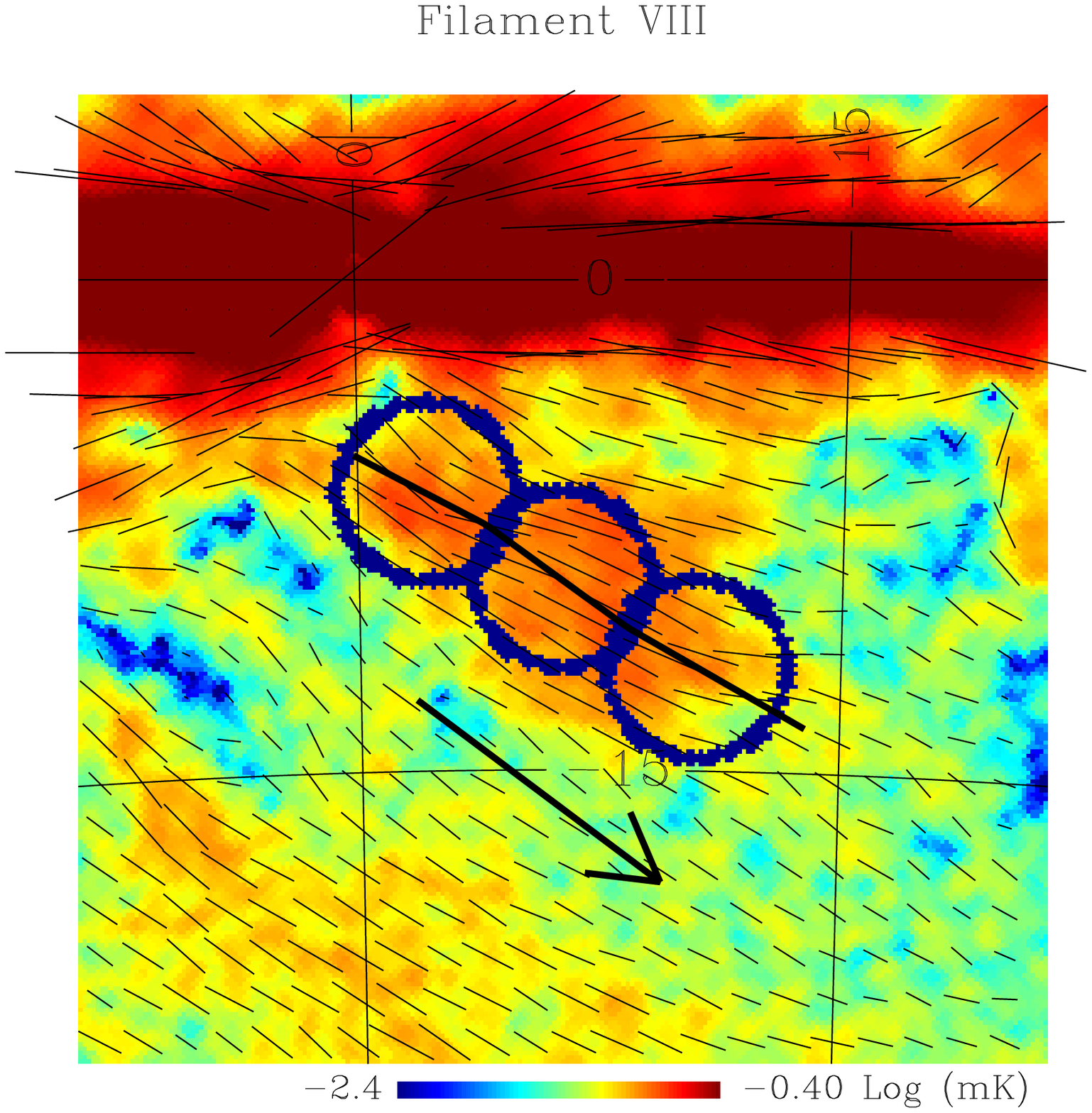}
  \includegraphics[angle=0,width=\widtha\textwidth]{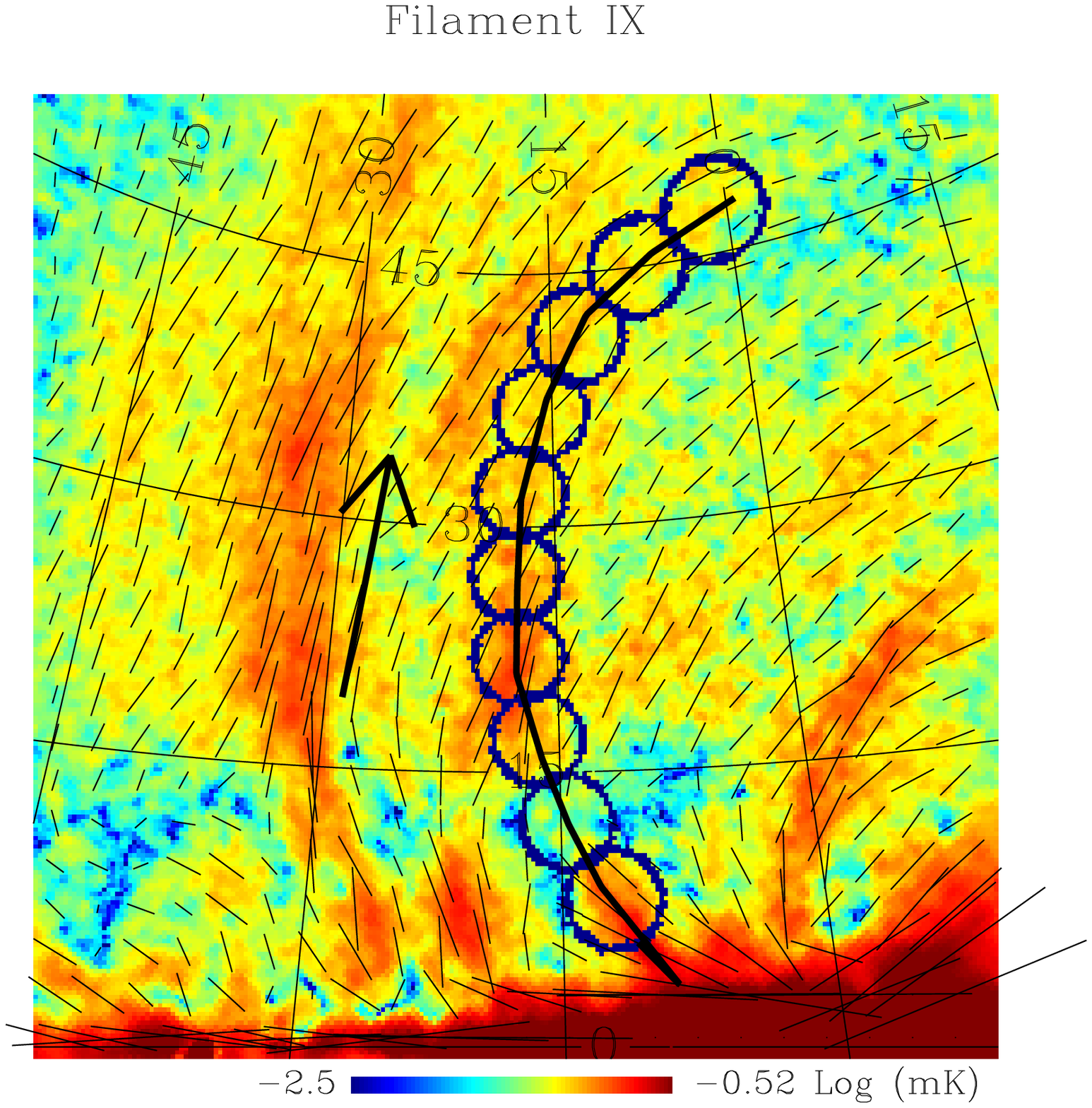}
  \includegraphics[angle=0,width=\widthb\textwidth]{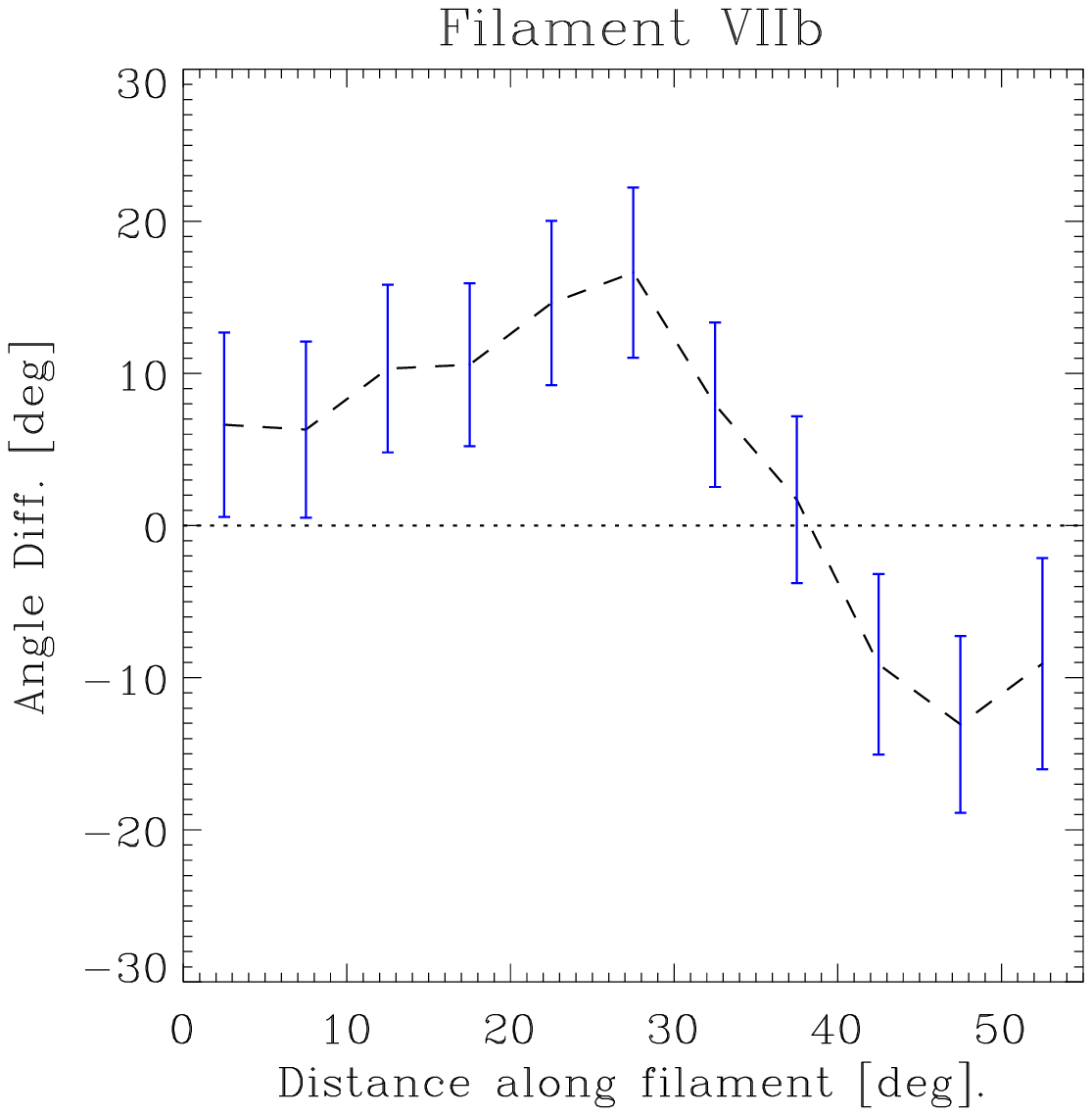}
  \includegraphics[angle=0,width=\widthb\textwidth]{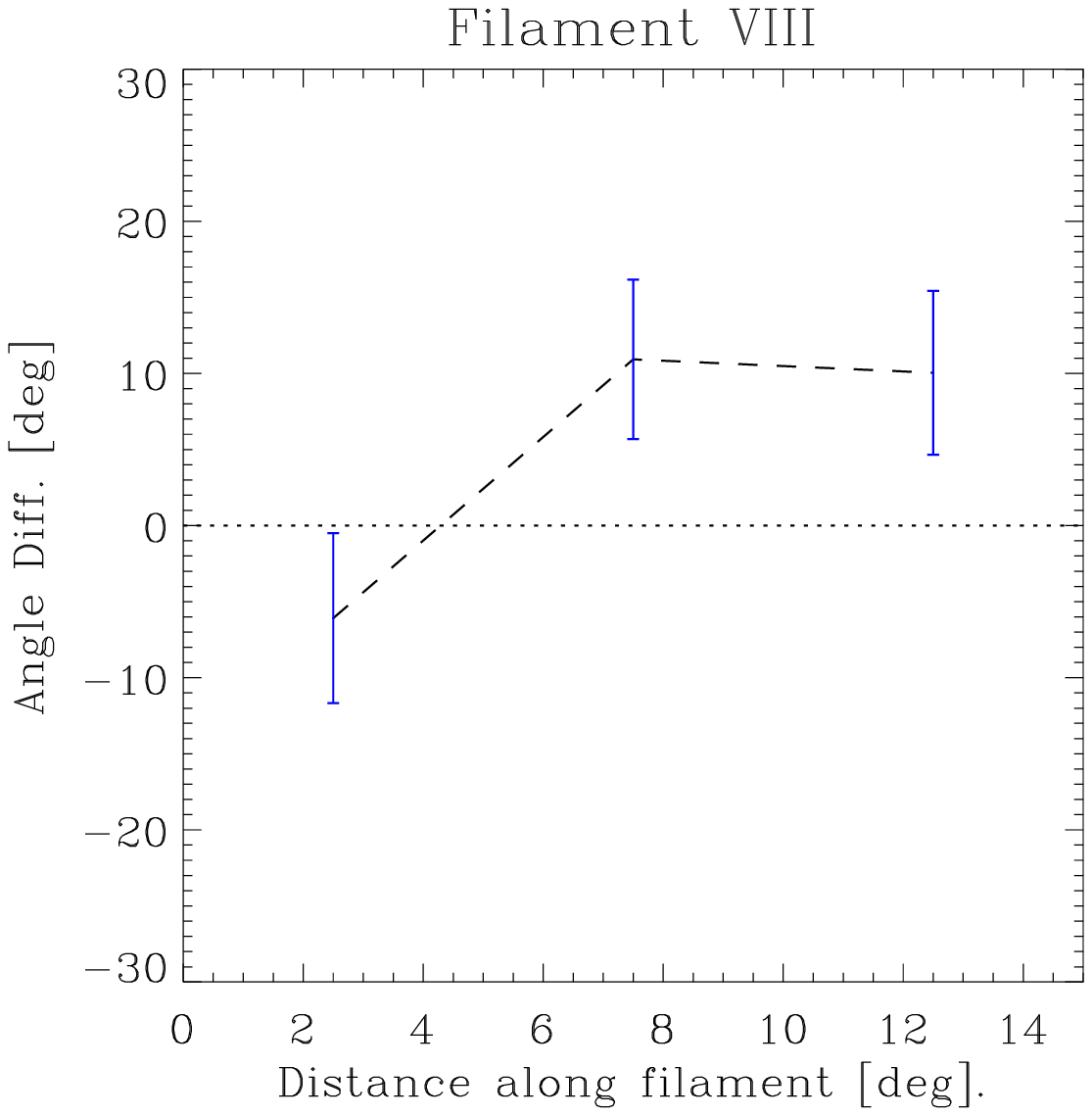}
  \includegraphics[angle=0,width=\widthb\textwidth]{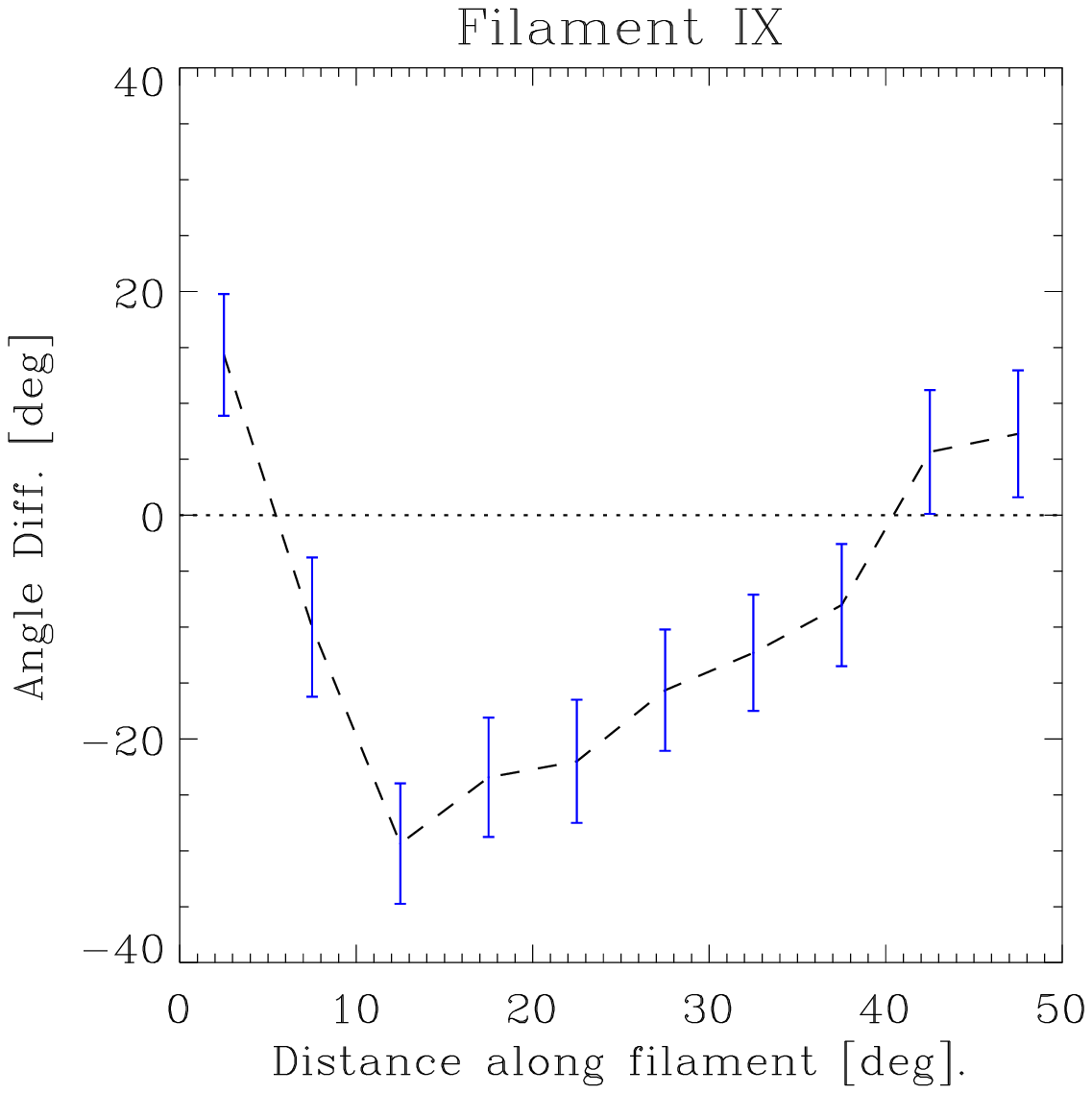}
  \includegraphics[angle=0,width=\widtha\textwidth]{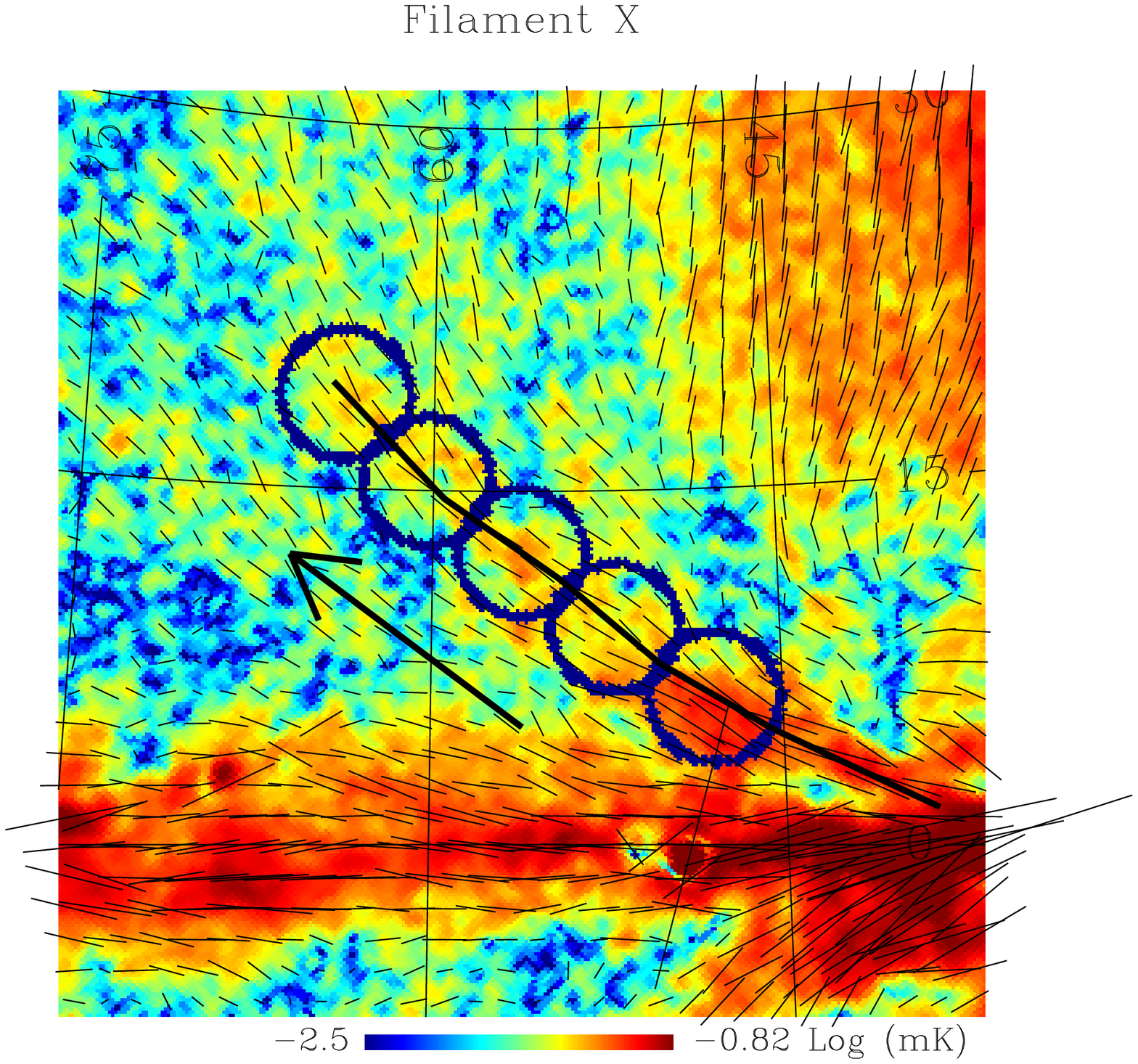}
  \includegraphics[angle=0,width=\widtha\textwidth]{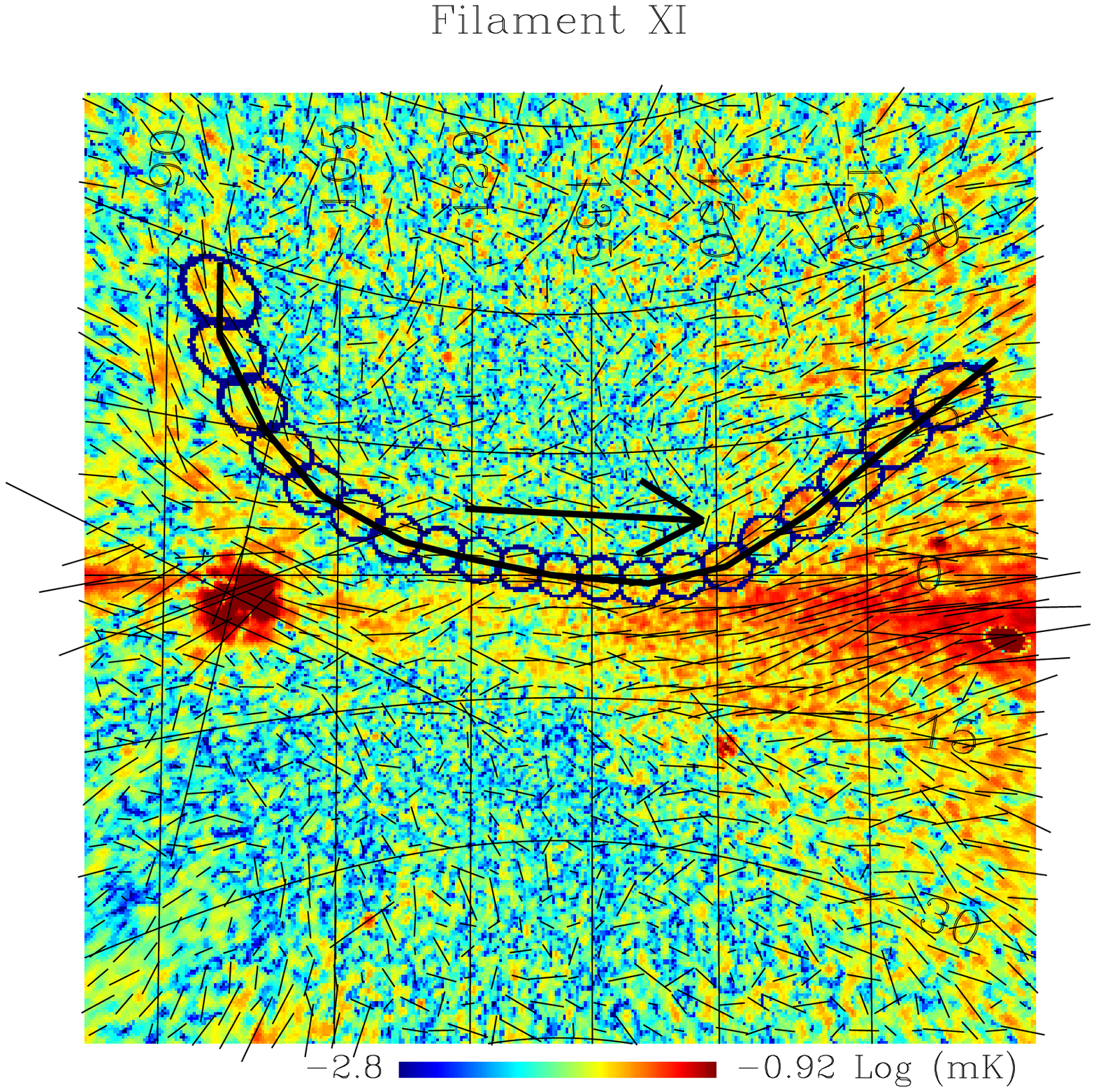}
  \includegraphics[angle=0,width=\widtha\textwidth]{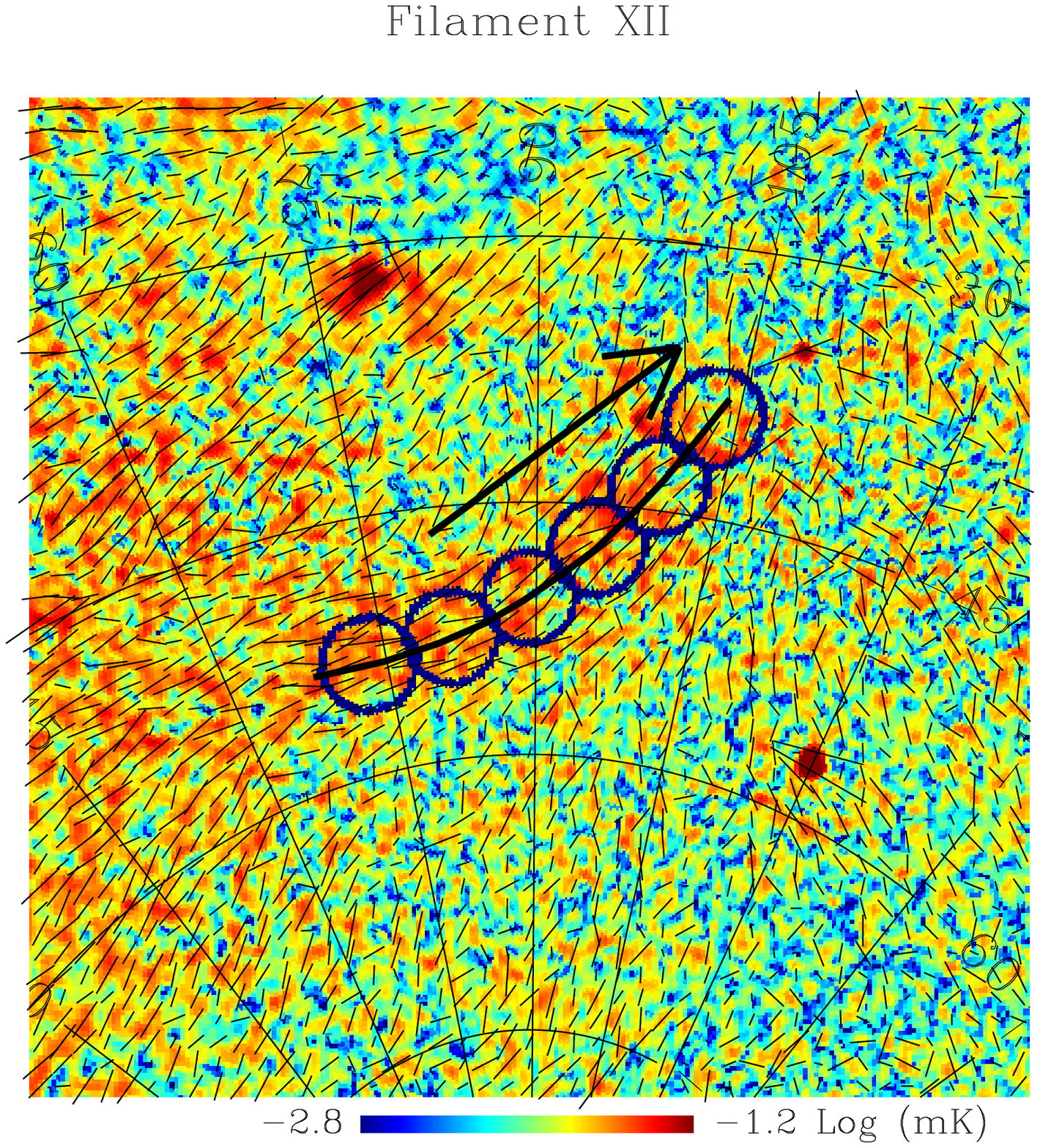}
  \includegraphics[angle=0,width=\widthb\textwidth]{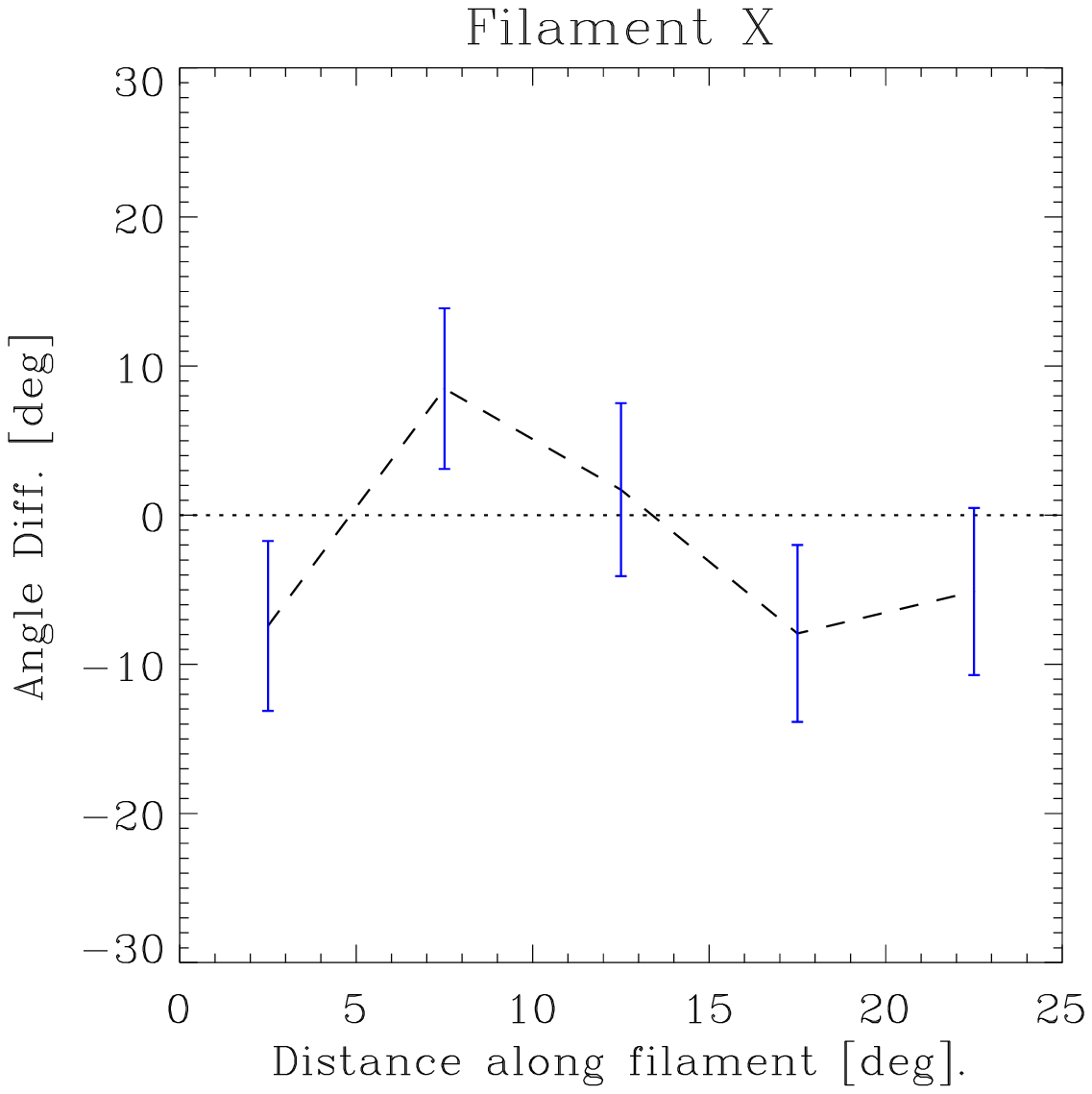}
  \includegraphics[angle=0,width=\widthb\textwidth]{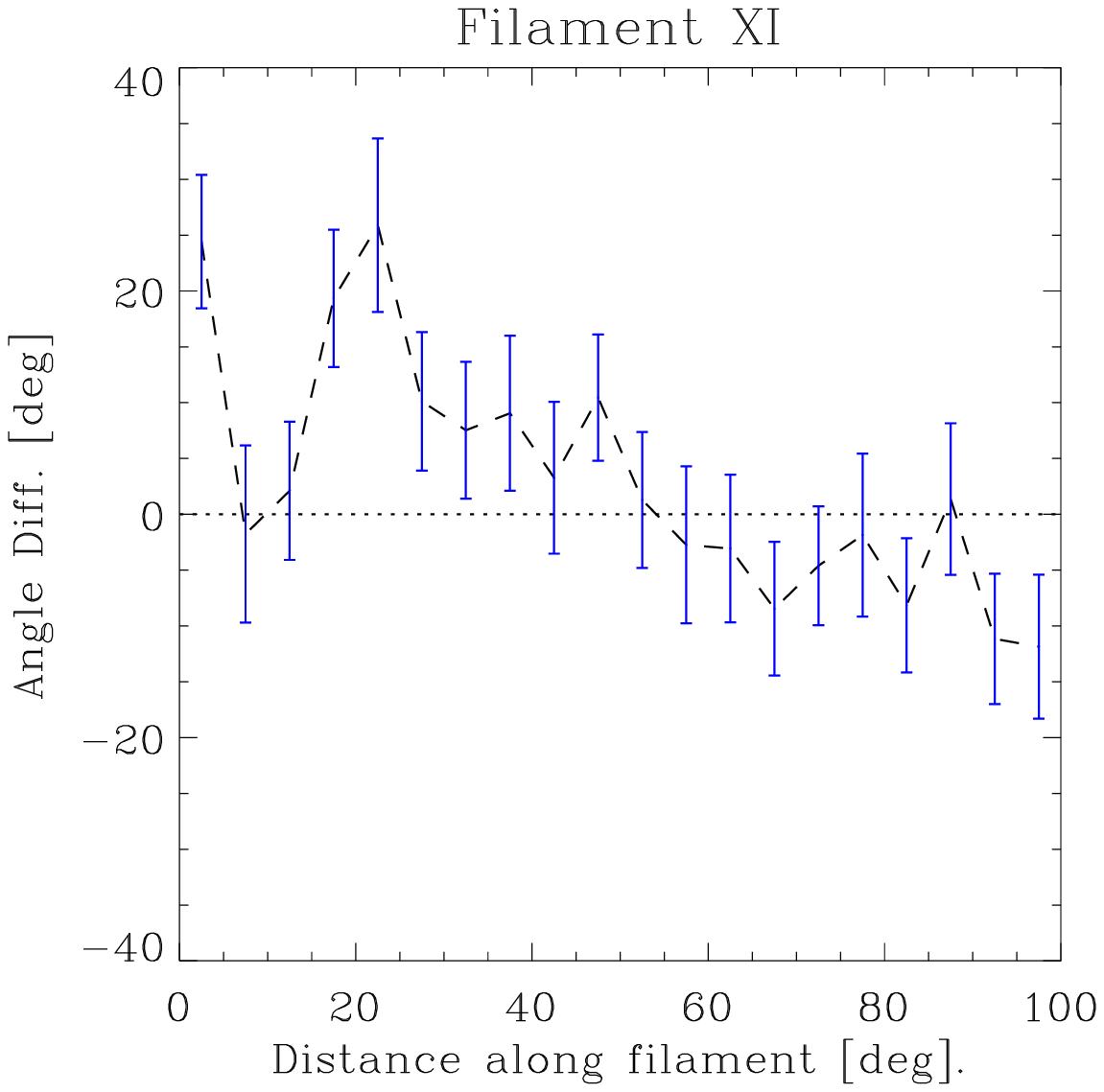}
  \includegraphics[angle=0,width=\widthb\textwidth]{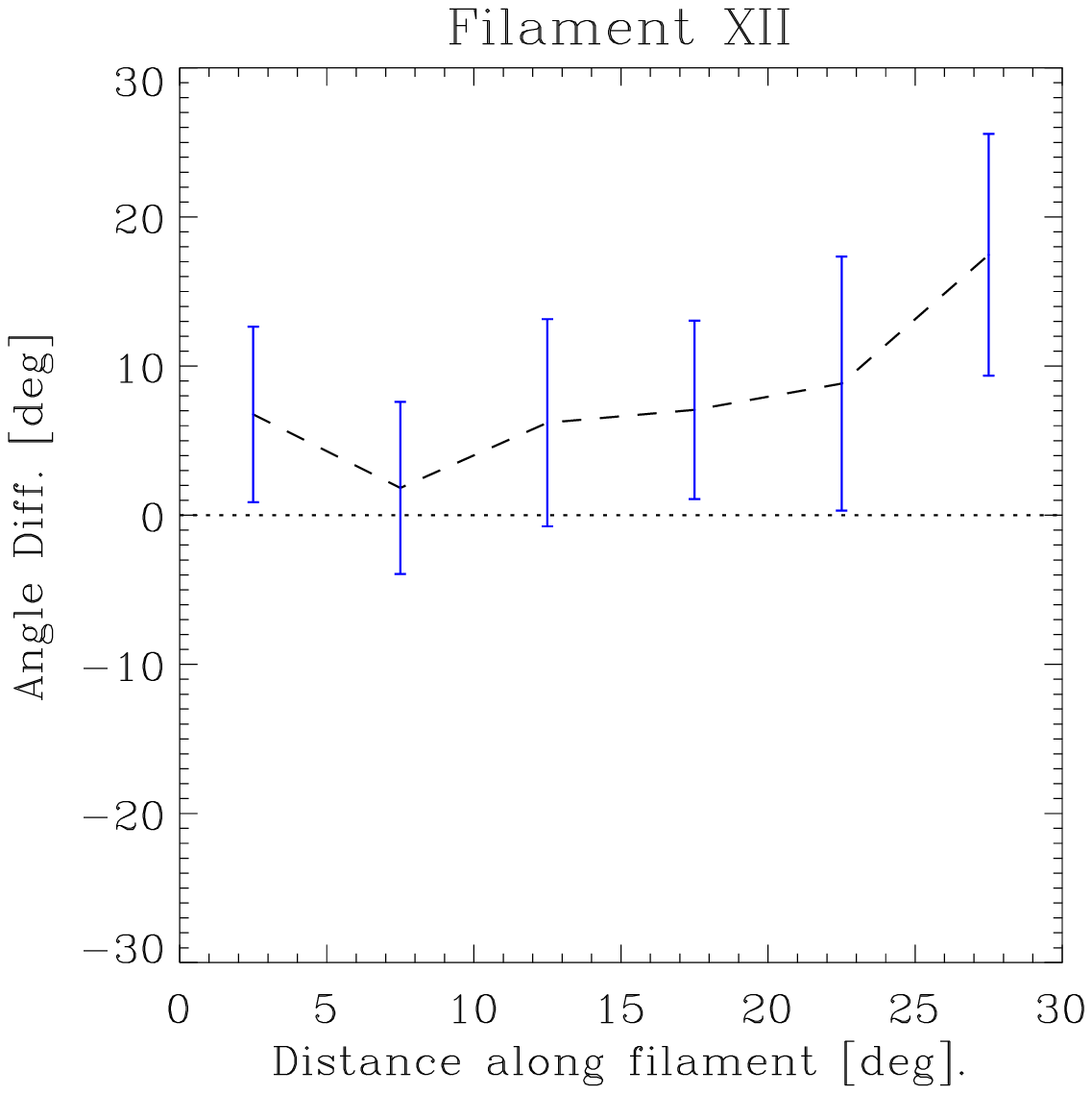}
   \caption{Continuation of Figure \ref{fig:diff_angle_1} }
  \label{fig:diff_angle_2}
\end{figure*}

\begin{figure*}
  \renewcommand{\widtha}{0.27}
  \renewcommand{\widthb}{0.27}
  \centering
  \includegraphics[angle=0,width=\widtha\textwidth]{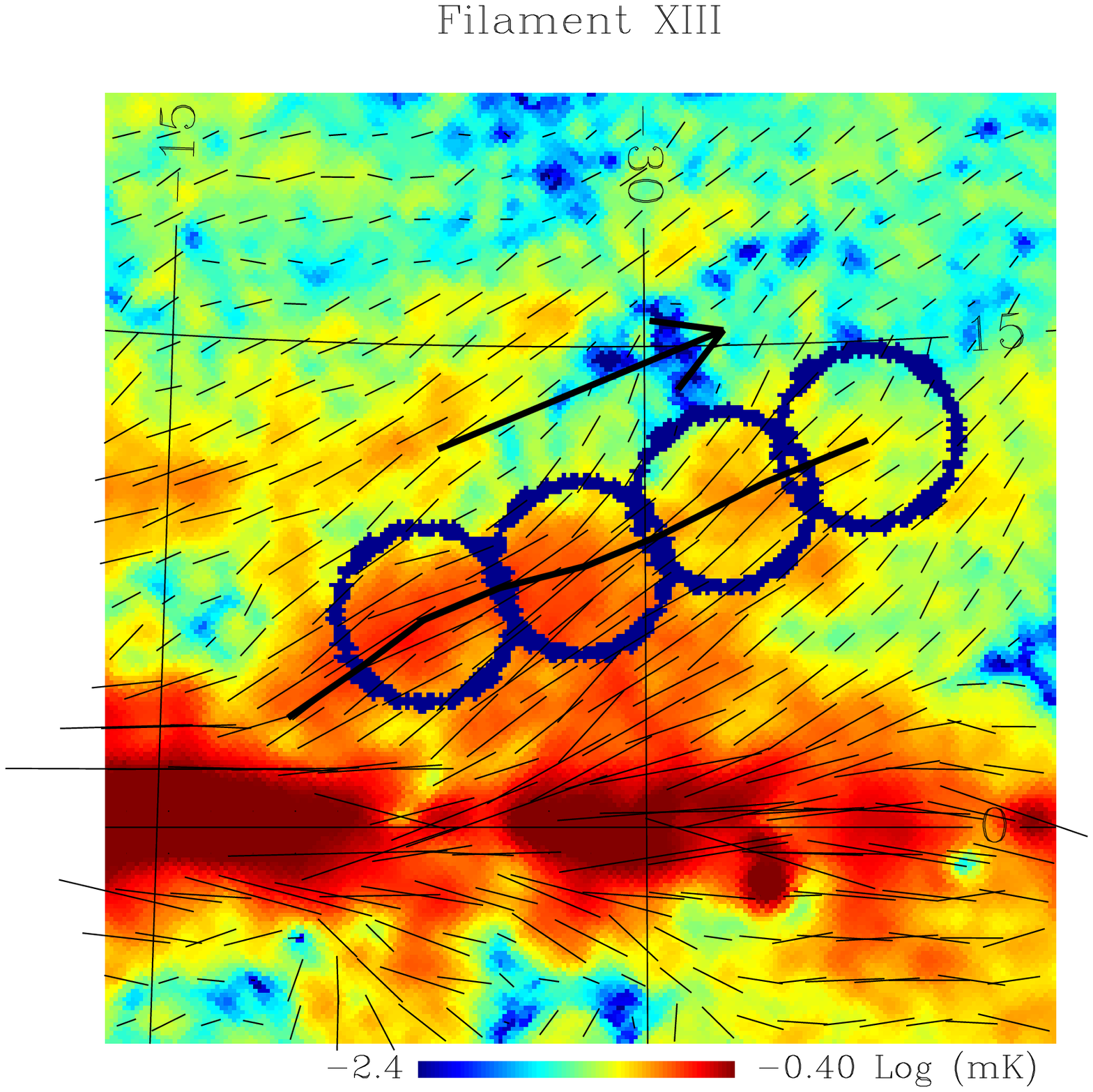}
  \includegraphics[angle=0,width=\widtha\textwidth]{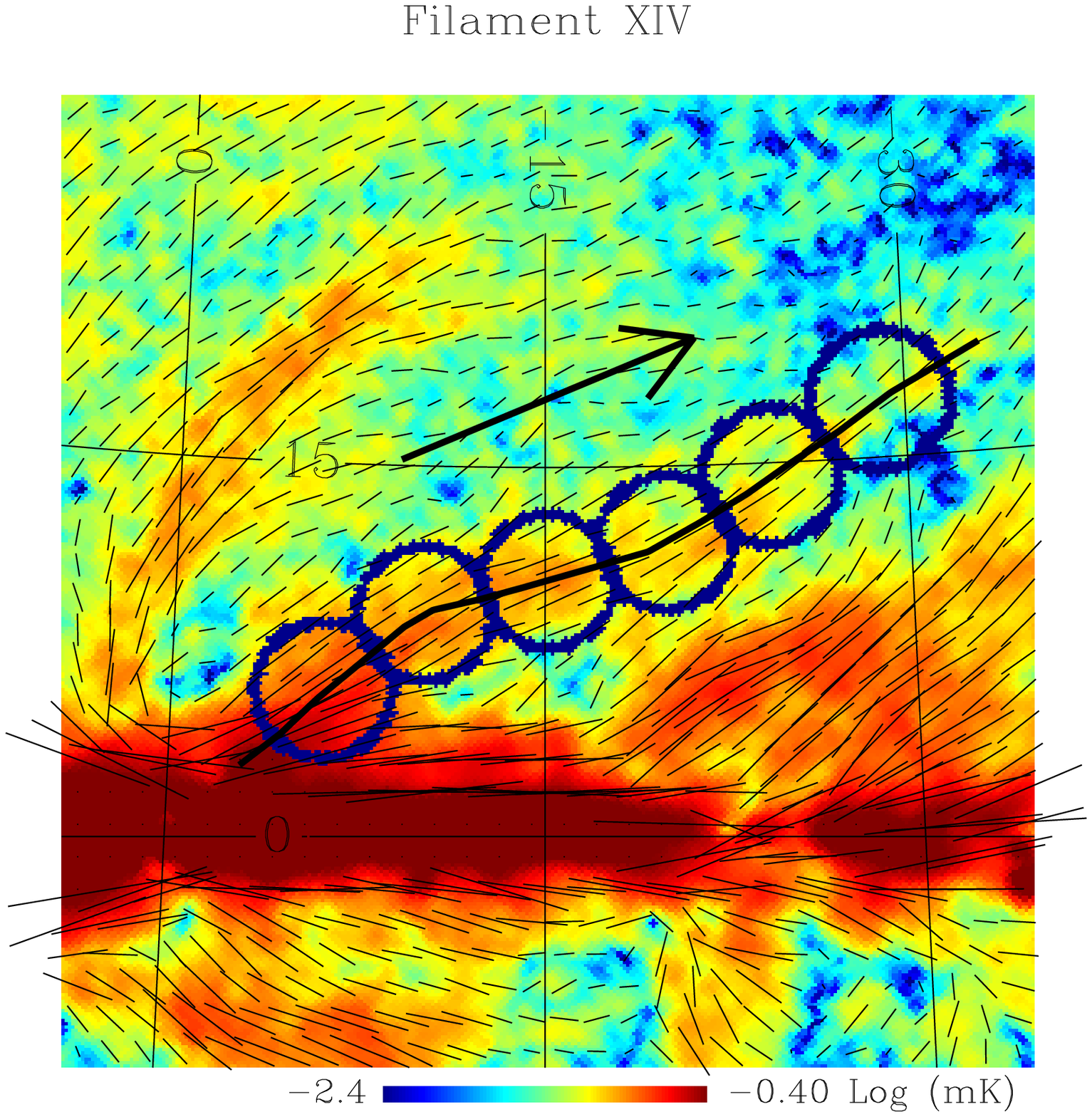}

  \includegraphics[angle=0,width=\widthb\textwidth]{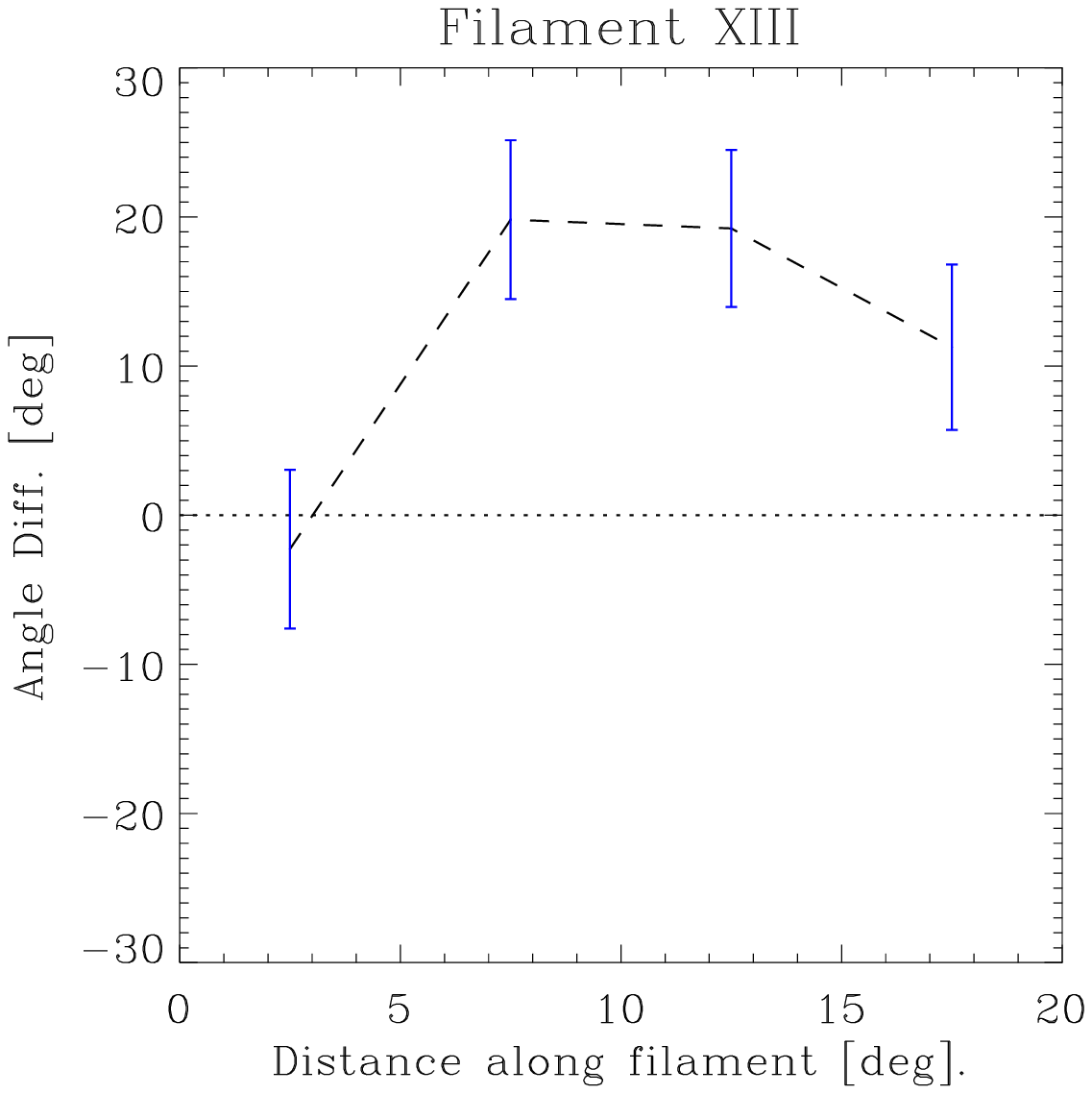}
  \includegraphics[angle=0,width=\widthb\textwidth]{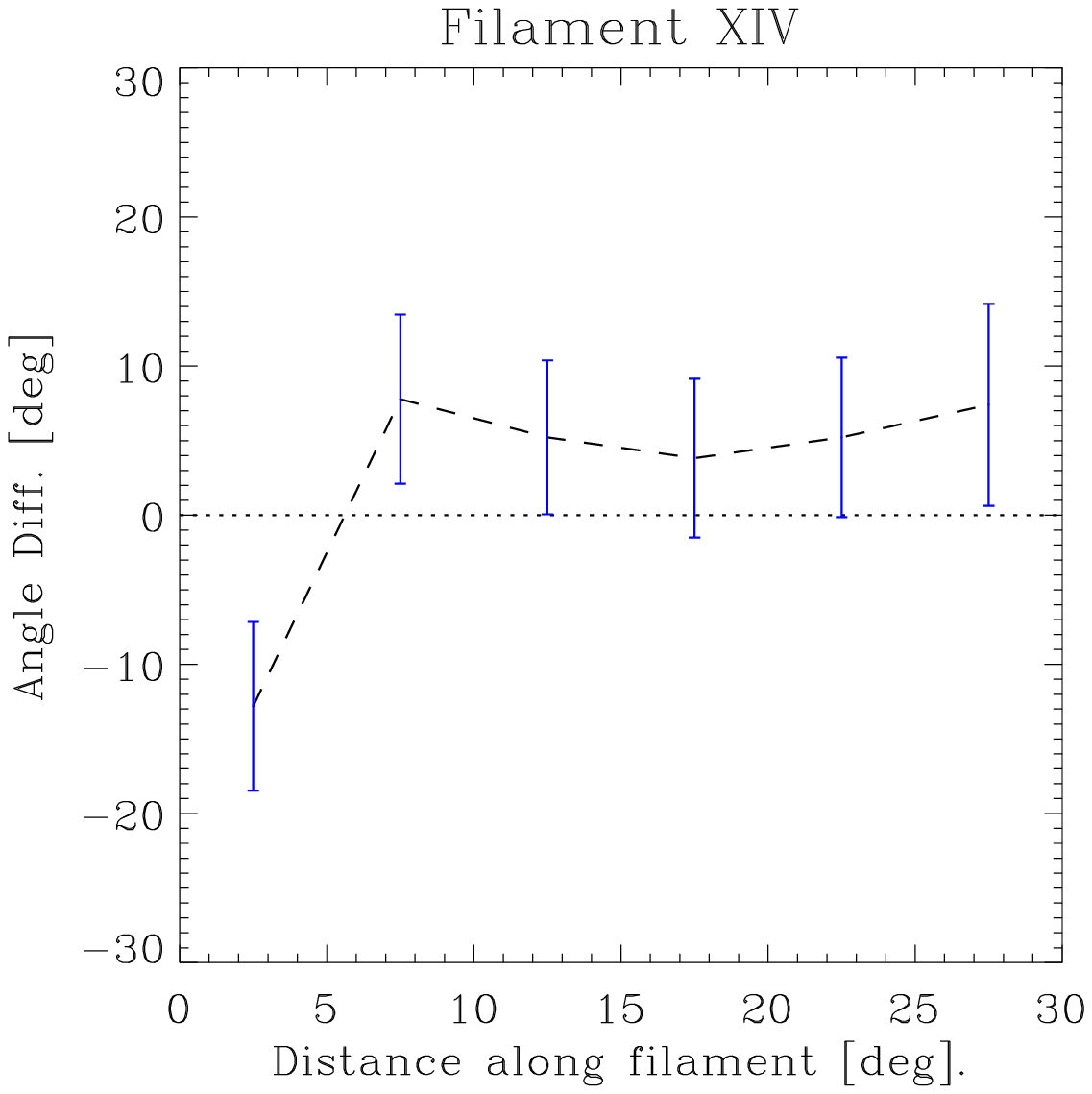}
  \caption{Continuation of Figure \ref{fig:diff_angle_1}}
  \label{fig:diff_angle_3}
\end{figure*}

\subsection{Polarisation angle along filaments}

The direction of the ordered component of the magnetic field can give
us information about the origin of the observed filaments. The
magnetic fields of old SN \citep[e.g. the DA~530
  supernova;][]{landecker:99} are tangential to the surface of the
shell. This is traditionally explained by the compression of the field
in radiative shocks with large radius \citep[see e.g. the review
  by][]{reynolds:12}. 

Looking at the de-biased \wmap \kband~polarisation map from
Fig. \ref{fig:wmap_3deg_all}, we see that most of the emission at high
latitudes comes from individual filamentary structures. We also can
see that the magnetic field direction vectors are roughly parallel to
the direction of the filaments. We will quantify this observation by
comparing the polarisation angle $\chi$ with the direction of the
filament.

In the {\it top} panel of Fig. \ref{fig:unsharp_haslam_kpol}, there
are some features that are not well fitted by a circular arc. We
include some of these non-circular features in this analysis, and also
ignore some of the loops and arcs that are listed in Table
\ref{tbl:circ_loops} due to their low surface brightness. The
filaments that we use in this analysis are shown in
Fig. \ref{fig:unsharp_haslam_kpol}.

In order to define the direction $\alpha$ of the filaments that we
want to compare with the polarisation angle $\chi$, we connected
pairwise the points of maximum brightness across each filament shown
in the {\it top} panel of Fig.~\ref{fig:unsharp_haslam_kpol} using
great circle arcs. This created a map with a smooth line that follows
the direction of each filament.  Then, this maximum brightness
template for each filament was convolved with a Gaussian beam of a
width similar to the apparent width of the filament, typically 2\dg--
4\dg. In the bottom panel of Fig.~\ref{fig:unsharp_haslam_kpol} we can
see the resulting templates for all the filaments that we study here.

The direction at each point of the filament, $\alpha$, can be
calculated using the spatial gradient of the previously described
template map for the filaments, $T$. The gradient of this map points
to a direction that is perpendicular to the extension of the filament,
so that we will have,
\begin{equation}  
  \alpha=\arctan\left(\frac{\partial T/\partial\theta}{\partial
    T/\partial\phi/\sin{\theta}}\right),
\end{equation}
where $\theta$ and $\phi$ are the spherical coordinates. The
$\sin(\theta)$ factor appears because of the use of spherical
coordinates.

$\alpha$ defines the direction of steepest gradient moving away from
the filament axis, so $\alpha \pm 90$ defines the orientation of the
filament axis, defined in the same sense as the position angle of
polarization.


We measured the difference between the observed polarisation angle
$\chi=0.5 \arctan(U/Q)$ and the direction angle $\alpha$. We did this
in adjacent circular apertures along each filament, with a diameter
between 3\dg and 5\dg depending on the size of the filament.
Figs.~\ref{fig:diff_angle_1}, \ref{fig:diff_angle_2} and
\ref{fig:diff_angle_3} show maps of each filament with the circular
apertures where we calculate the angle difference. A thick black line
in the maps indicates the direction of the filament defined from the
template map and the gradient method described. Also are plotted, for
each filament, the difference between the observed polarisation angle
and the angle of the filament. The error bars incorporate both the
random fluctuations given by the dispersion of the measured angles in
each aperture and an additional 5\dg systematic uncertainty added to
account for small errors in the definition of the spatial direction of
each filament.
 
For the NPS (Filament I in our numbering), we also compared the
polarisation angle $\chi$ with the direction angle defined by the
geometry of the loop in total intensity, i.e. the circular fit that
traditionally is used to describe the NPS, listed in Table
\ref{tbl:circ_loops}. The green line in that figure represents the
angle that the circular loop defines. The difference between the
circular loop and our definition for the direction of the NPS using
the gradient technique is small compared with the uncertainties, so
there is consistency between the two definitions for the direction of
the filament.

In the Galactic Centre Spur (GCS), the difference in angle is small,
averaging less that 5\dg\!\!. The magnetic field vectors are very
close to parallel with the direction of the filament. In Filament
No. I, the NPS, there is an offset of $\sim 10$\dg in the angles. It
can be seen even by looking at the map on
Fig. \ref{fig:diff_angle_1}. A similar difference occurs in Filament
IX. Filament X is also well-aligned, with a difference very close to
zero within the errors. Filaments III, IIIs, VIIb, XI and XII are
those with less signal-to-noise ratio, and the polarisation angle on
them is not parallel to their extension, there are even inversions in
the sign of the difference of the angles. Filaments Is and VIII also
have a systematic difference in angles but in this case, the
difference is positive. Emission from Filament XIV is well aligned
along its extension. Finally, Filament XIII shows a systematic
positive difference of $\sim 15$\dg\!\!.

It is interesting that in the inner Galaxy, the deviations from
parallel in the direction of the angles are different on the north
Galactic hemisphere compared with the south (compare filaments I and
IX in the north with Filaments Is and VIII in the south). The
direction of the emission from these filaments has a smaller radius of
curvature than the one defined by the polarisation vectors.  We will
discuss these observations later in the context of the origin of the
filamentary emission (Section \ref{sec:model}).

We note that in each case when there is a change in the sign of the
$\chi - \alpha$, there is an overlapping feature. Examples of this can
be seen in Filaments numbers IX, XIV and XIII from
Fig.~\ref{fig:diff_angle_1}, where the first circular aperture closer
to the Galactic plane includes emission both from the filament and
from the plane, where the field is parallel to the plane. A similar
effect can be seen where a filament comes close to another polarised
feature away from the plane. See for example the borders of Filament
IIIs.


\section{Polarised spectral indices}
\label{sec:spectral_indices}

Temperature spectral indices are useful in identifying the emission
mechanism of the measured radiation, thereby, obtain physical
properties of the emitting region. Because of the combination of
different emitting sources along an arbitrary line-of-sight, it is
difficult to measure the spectral index of an individual physical
emission mechanism. \citet{davies:06} selected small regions of the
sky that were expected to be dominated by a single emission mechanism,
which allowed them to measure the spectral indices of free-free,
synchrotron and thermal dust radiation.

Where polarisation data are available, the synchrotron spectral index
can be measured unambiguously without the use of component separation
techniques in the frequency range below $\sim$50\,GHz, because the
polarisation of the other mechanisms is much smaller.  The
polarisation of the CMB varies between 1-10\%, showing the larger
values at small angular scales ($\ell\sim1000$). Free-free emission is
intrinsically unpolarised \citep[although Thompson scattering could
  yield a polarisation signal at the edges of H{\sc ii} regions but
  its contribution to the full sky is less than 1\%;][]{macellari:11}.
Anomalous microwave emission (AME), which is observed at \wmap
frequencies shows little polarisation. \citet{dickinson:11} set upper
limits of less than 2.6\% on the polarisation of two AME regions (see
\cite{rubino-martin:12} for a recent review of AME polarisation
measurements).

The measurement of precise synchrotron spectral indices has become
even more necessary after the discovery of the ``haze'' by
\citet{finkbeiner:04a}, a diffuse emission of unknown origin centred
at the Galactic centre. \citet{planck_haze:13} indicates the existence
of the haze with a spectral index of $-2.55\pm 0.05$, favouring a
hard-spectrum synchrotron origin for this diffuse emission. This shows
that significant synchrotron spectral index variations can occur and
they should therefore be measured with care.  We note that the
foreground-cleaned CMB maps which the \wmap team provide are produced
on the basis of a fixed spectral index for synchrotron
emission. \citet{fuskeland:14} found variations in the polarisation
spectral index between 23\,GHz and 33\,GHz using \wmap data in large
regions on the sky.

In this section we measure the spectral index of the bias-corrected
polarised intensity between \wmap K, Ka and Q bands in a number of
regions across the sky using maps smoothed to 3\dg\!\!. We use the T-T
plot approach, in which we assume a power-law relationship between the
polarised intensities at different frequencies:
\begin{equation}
  \frac{T_{\nu_1}}{T_{\nu_2}}=\left(\frac{\nu_1}{\nu_2}\right)^{\beta}.
\end{equation}
This method has the advantage of being independent of any zero level
offset or large scale artefacts present in the maps.\footnote{\wmap
  polarisation maps are affected by poorly constrained modes on very
  large angular scales \citep{jarosik:11}.}. Here, the spectral index
$\beta$ is calculated by fitting a straight line $y=mx+c$ to the
measured brightness temperatures $T_{\nu_1}$, $T_{\nu_2}$:
\begin{equation}
  \beta=\frac{\log{m}}{\log(\nu_1/\nu_2)},
\end{equation}
where $m$ is the slope of the fit. We took into account the error in
both coordinates for the linear fit. Also, an absolute calibration
error of 0.2\% \citep{bennett:13} has been added in quadrature to the
random uncertainty (this corresponds to an uncertainty in $\beta$ of
0.0076).

\begin{figure*}
  \centering
  \includegraphics[angle=90,width=0.99\textwidth]{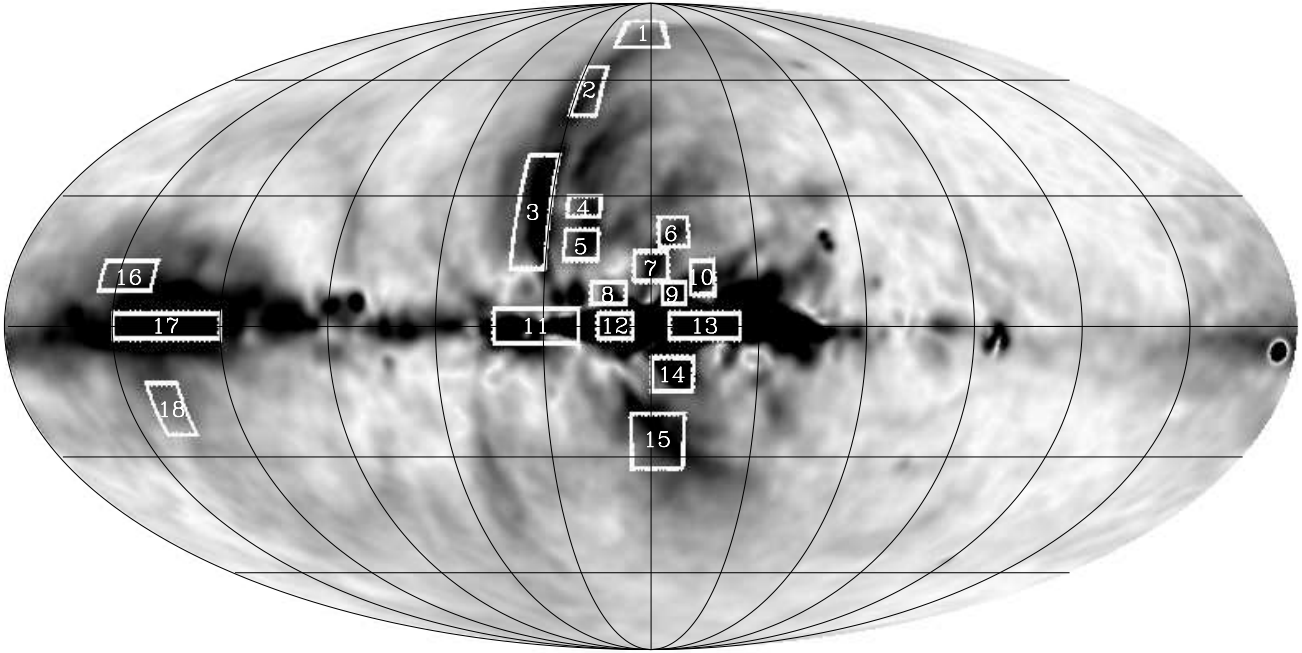}
  \caption{\wmap \kband~polarisation intensity map showing the 18
    regions chosen to measure spectral indices. }
  \label{fig:box_regions}
\end{figure*}
\begin{figure*}
  \centering
  \includegraphics[angle=90,width=1\textwidth]{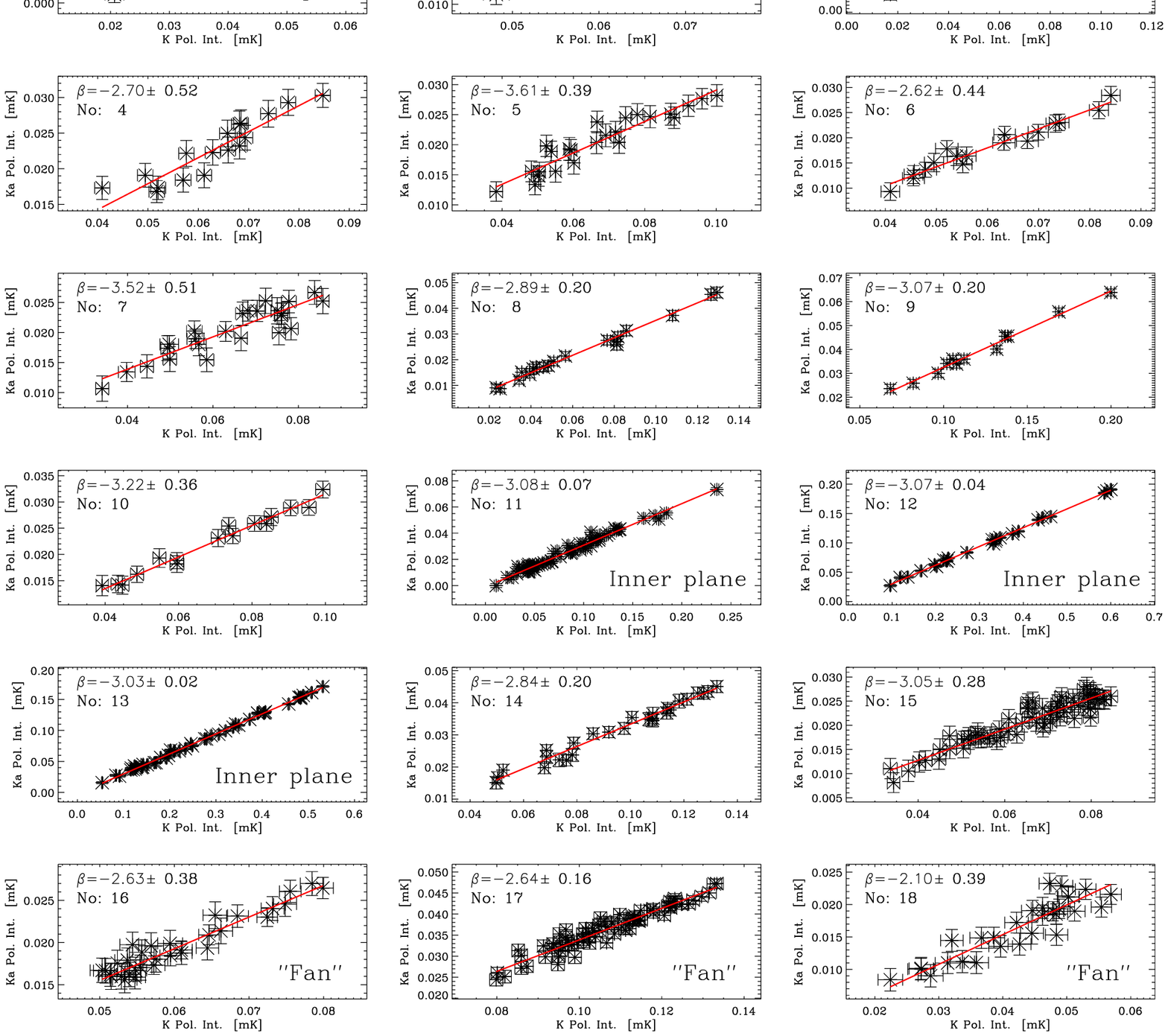}
  \caption{T-T plots of polarisation intensity between 23 and 33\,GHz
    of the eighteen regions defined in Fig. \ref{fig:box_regions}. The
    straight line represents the best linear fit. The error bars show
    the statistical fluctuations of each point.}
  \label{fig:ttplots_boxes_K_Ka}
\end{figure*}

In Fig. \ref{fig:ttplots_boxes_K_Ka} we show the T-T plots between 23
and 33\,GHz for each region. Table \ref{tab:spec_ind} lists the
location and size of each region, the measured spectral indices, the
reduced $\chi^2$ value and the $\chi^2$ probability, $q$. The T-T
plots between K and Ka bands, where the SNR is larger, allow us to
measure $\beta$ with small uncertainties. This occurs mainly on the
Galactic plane, where the signal is stronger. At higher latitudes, the
fractional uncertainties on the data are larger so the constraints on
$\beta$ are less tight.

\begin{table*}
\centering
\caption{Spectral indices in different regions between \wmap K--Ka,
  K--Q and Ka--Q bands. Also listed are the reduced $\chi^2$ of the
  fit and the $\chi^2$ probability or $q$-value, for each T-T plot. }
\begin{tabular}{lKKKK|HKK|HKK|HKK}
\toprule
 \textit{Region}
 & \multicolumn{1}{c}{$l_0$}
 & \multicolumn{1}{c}{$b_0$}
 & \multicolumn{1}{c}{$l_{\text{size}}$}
 & \multicolumn{1}{c|}{$b_{\text{size}}$}
 & \multicolumn{1}{c}{$  \beta_{K-Ka} $}
 & \multicolumn{1}{c}{ $\chi_{\text{red}}^2$}
 & \multicolumn{1}{c|}{ $q$} 
 & \multicolumn{1}{c}{$  \beta_{K-Q} $}
 & \multicolumn{1}{c}{ $\chi_{\text{red}}^2$}
 & \multicolumn{1}{c|}{ $q$} 
 & \multicolumn{1}{c}{$  \beta_{Ka-Q} $}
 & \multicolumn{1}{c}{ $\chi_{\text{red}}^2$}
 & \multicolumn{1}{c}{ $q$} \\
 \midrule   
 1 &   5.0 &  75.0 &  30.0 &  10.0 & -2.29 , 0.33 &  0.66 &  0.90 & -3.13 , 0.47 &  0.68 &  0.88& -4.74 , 1.44 &  0.80 &  0.74 \\ 
 2 &  25.0 &  57.0 &  10.0 &  14.0 & -2.12 , 0.48 &  1.24 &  0.20 & -3.49 , 0.92 &  0.41 &  0.99& -7.43 , 3.34 &  0.61 &  0.92 \\ 
 3 &  35.0 &  26.5 &  10.0 &  27.0 & -3.30 , 0.15 &  0.59 &  1.00 & -3.15 , 0.16 &  0.91 &  0.69& -2.80 , 0.50 &  0.80 &  0.88 \\ 
 4 &  20.0 &  27.5 &  10.0 &   5.0 & -2.70 , 0.52 &  0.87 &  0.60 & -2.26 , 0.45 &  1.31 &  0.19& -1.10 , 1.40 &  0.66 &  0.82 \\ 
 5 &  20.0 &  18.5 &  10.0 &   7.0 & -3.61 , 0.39 &  0.75 &  0.79 & -2.61 , 0.30 &  0.94 &  0.54& -0.75 , 1.06 &  0.87 &  0.64 \\ 
 6 &  -6.5 &  21.5 &   9.0 &   7.0 & -2.62 , 0.44 &  0.43 &  0.97 & -3.04 , 0.55 &  1.89 &  0.02& -3.03 , 1.52 &  1.31 &  0.19 \\ 
 7 &   0.0 &  13.5 &  10.0 &   7.0 & -3.52 , 0.51 &  0.88 &  0.62 & -2.34 , 0.32 &  1.86 &  0.01&  0.34 , 1.27 &  1.55 &  0.05 \\ 
 8 &  12.0 &   7.5 &  10.0 &   5.0 & -2.89 , 0.20 &  0.41 &  0.98 & -3.05 , 0.24 &  0.45 &  0.97& -3.38 , 0.73 &  0.60 &  0.90 \\ 
 9 &  -6.5 &   7.5 &   7.0 &   5.0 & -3.07 , 0.20 &  0.69 &  0.73 & -3.13 , 0.23 &  2.40 &  0.01& -3.10 , 0.68 &  1.33 &  0.21 \\ 
10 & -14.5 &  11.0 &   7.0 &   8.0 & -3.22 , 0.36 &  0.29 &  0.99 & -3.03 , 0.36 &  0.18 &  1.00& -2.75 , 1.16 &  0.28 &  1.00 \\ 
11$^*$ &  32.0 &   0.0 &  24.0 &   8.0 & -3.08 , 0.07 &  1.90 &  0.00 & -3.06 , 0.07 &  1.08 &  0.32& -3.06 , 0.24 &  1.16 &  0.19 \\ 
12$^*$ &  10.0 &   0.0 &  10.0 &   6.0 & -3.07 , 0.04 &  2.10 &  0.00 & -3.06 , 0.04 &  1.08 &  0.36& -3.07 , 0.12 &  1.05 &  0.40 \\ 
13$^*$ & 345.0 &   0.0 &  20.0 &   6.0 & -3.03 , 0.02 &  2.00 &  0.00 & -3.09 , 0.03 &  1.12 &  0.27& -3.20 , 0.09 &  0.80 &  0.83 \\ 
14 & 354.0 & -11.0 &  12.0 &   8.0 & -2.84 , 0.20 &  0.58 &  0.95 & -3.03 , 0.25 &  1.49 &  0.06& -3.17 , 0.76 &  1.11 &  0.32 \\ 
15 &  -2.0 & -26.5 &  16.0 &  13.0 & -3.05 , 0.28 &  0.69 &  0.97 & -4.00 , 0.52 &  0.83 &  0.82& -5.89 , 1.63 &  0.90 &  0.71 \\ 
16 & 147.5 &  11.5 &  15.0 &   7.0 & -2.63 , 0.38 &  0.44 &  1.00 & -2.66 , 0.41 &  0.69 &  0.90& -2.43 , 1.22 &  0.41 &  1.00 \\ 
17$^*$ & 135.0 &   0.0 &  30.0 &   6.0 & -2.64 , 0.16 &  1.10 &  0.26 & -2.67 , 0.16 &  0.99 &  0.50& -2.79 , 0.49 &  0.93 &  0.65 \\ 
18 & 138.5 & -19.0 &   9.0 &  12.0 & -2.10 , 0.39 &  1.15 &  0.26 & -2.37 , 0.41 &  1.23 &  0.19& -3.00 , 1.29 &  1.40 &  0.08 \\ 
\hline
\end{tabular}
\begin{flushleft}
      $^*$: These regions are on the Galactic plane. \\
 \end{flushleft}
\label{tab:spec_ind}
\end{table*}

\label{sec:ttplots_regions}

We selected the brightest regions from the filaments and also some
areas of interest that do not necessarily belong to a filament,
including regions on the Galactic plane. Fig. \ref{fig:box_regions}
shows the regions that were chosen. Most of them lie in the inner
Galaxy and three are in the Fan region, around $l=140$\dg\!\!,
$b=0$\dg\!\!.

There are a few regions that show a flatter spectral index than the
commonly accepted --3.0 value. Regions 1 and 2, at the top of the NPS
have $\beta_{K-Ka} > -3.0$ at a $\sim 2$-$\sigma$ significance
level. Also, in the Fan region, regions No. 16, No. 17 and No. 18 show
a flatter spectral index in at least two pairs of the frequencies
used. No polarised sources in these regions are listed in the
\citet{lopez-caniego:09} \wmap catalogue. On the plane, in the inner
Galaxy (regions 11, 12 and 13), the averaged spectral indices are
$\beta_{K-Ka} =3.04\pm0.02$, $\beta_{K-Q} =3.08\pm0.02$ and
$\beta_{Ka-Q} =3.15\pm0.07$. These values are consistent with the ones
measured by \citet{fuskeland:14} in a larger area on the Galactic
plane.

The flatter spectral index in regions No. 1 and 2 between K and Ka
bands is interesting as these regions belong to the upper end of the
NPS. The flat spectra however, are not observed in $\beta_{K-Q}$,
which is consistent with $-3.0$. These two regions may be interpreted
as either an excess of emission in the Ka-band (e.g. from a peaked
spinning dust contribution) or as a change in the underlying
synchrotron spectral shape, as we can see by the steep values of
$\beta_{Ka-Q}$. The low significance of the spectral indices measured
with the noisier Ka and Q bands do not allow us to draw strong
conclusions about these regions at these frequencies. Region number 3,
which includes the brighter emission from the NPS has a marginally
steeper spectral index $\beta_{K-Ka}=-3.30\pm0.15$ than the one
measured in total intensity by \citet{davies:06} of
$-3.07^{+0.09}_{-0.13}$ with an angular resolution of 1\dg\!\!\!. This
might indicate the presence of some AME not traced by the templates in
the area used in the total intensity analysis.

Regions 4 and 5, which represent the middle and lower region of
Filament IX show different spectral indices, with region 5 steeper
than region 4 in both $\beta_{K-Ka}$ and $\beta_{K-Q}$.  Regions 6 and
7, which represent the middle and lower sections of the GCS also show
a similar behaviour, the lower part of the spur has a $\beta_{K-Ka}$
steeper than the upper section.

The regions closer to the Galactic plane in the inner Galaxy (regions
8, 9, 10, 11, 12, 13 and 14) have values of $\beta_{K-Ka}$ consistent
with --3.0. These regions represent the area of the sky with the
largest amount of polarised emission. A box of 90\dg degrees in
longitude by 20\dg in latitude centred at $(l,b) =
0^{\circ},0^{\circ}$, which represents only $4.4 \%$ of the total area
of the sky encompasses more than $20\%$ of the polarised emission at K
band. Hence, this region is particularly significant for the
estimation of a full-sky average value of the synchrotron spectral
index.

Table~\ref{tab:spec_ind_summ} lists the weighted mean of the spectral
indices from all the regions. There is a hint (close to 2-$\sigma$) of
steepening of the spectral index with higher frequencies. The
dispersion quoted in Table \ref{tab:spec_ind_summ} is large.

\begin{figure*}
  \centering
  \includegraphics[angle=90,width=0.49\textwidth]{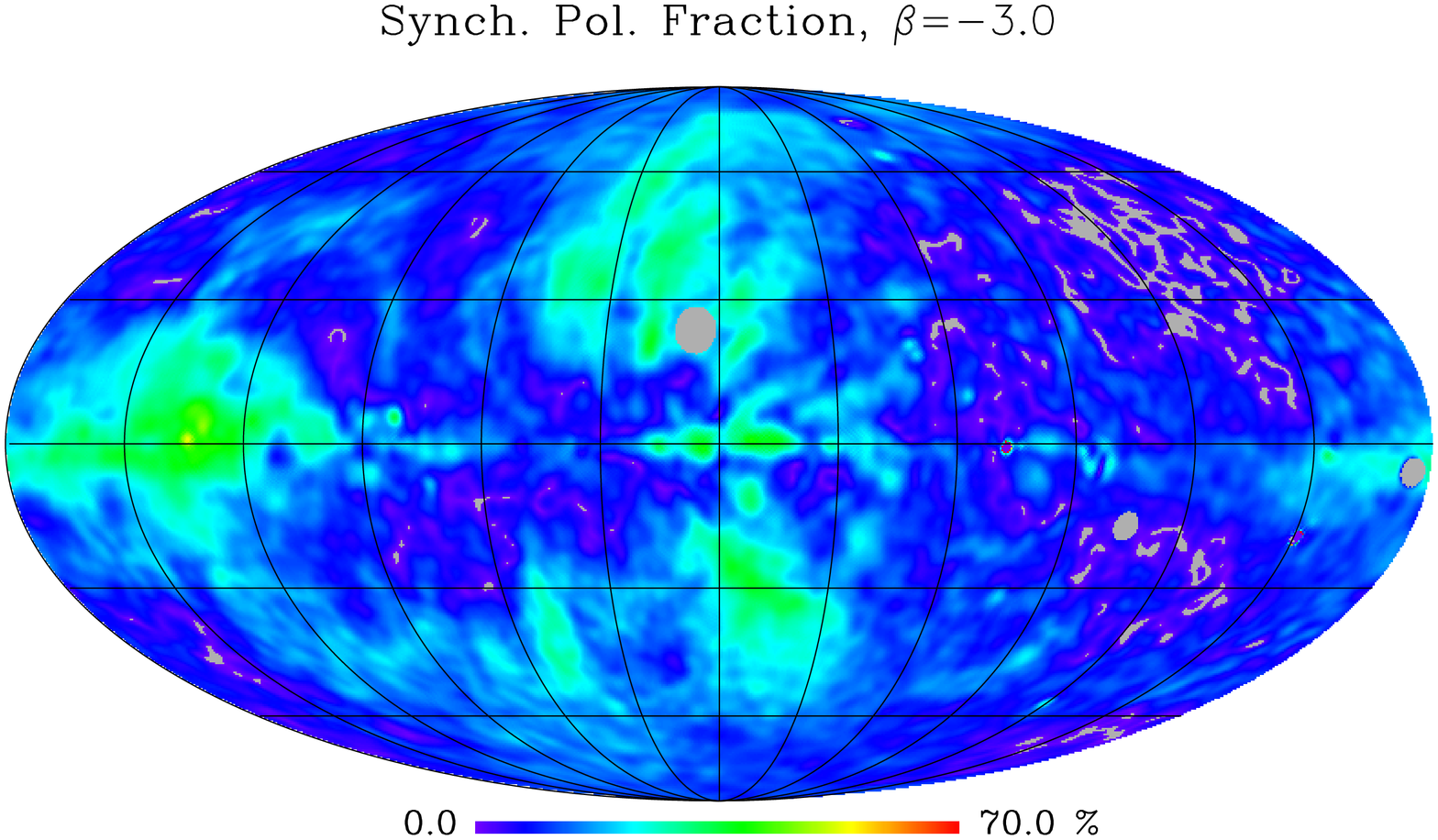}
  \includegraphics[angle=90,width=0.49\textwidth]{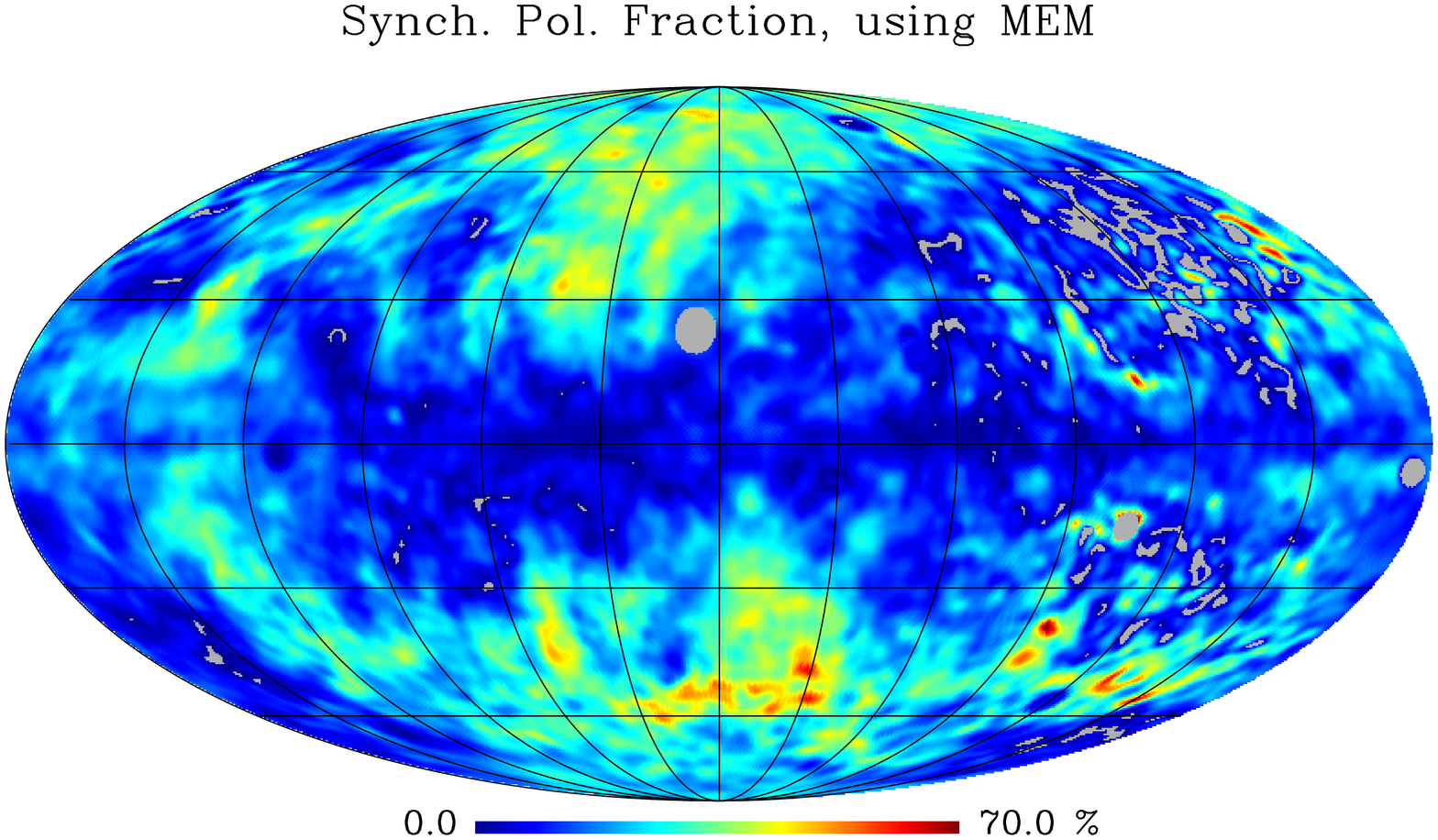}
  \includegraphics[angle=90,width=0.49\textwidth]{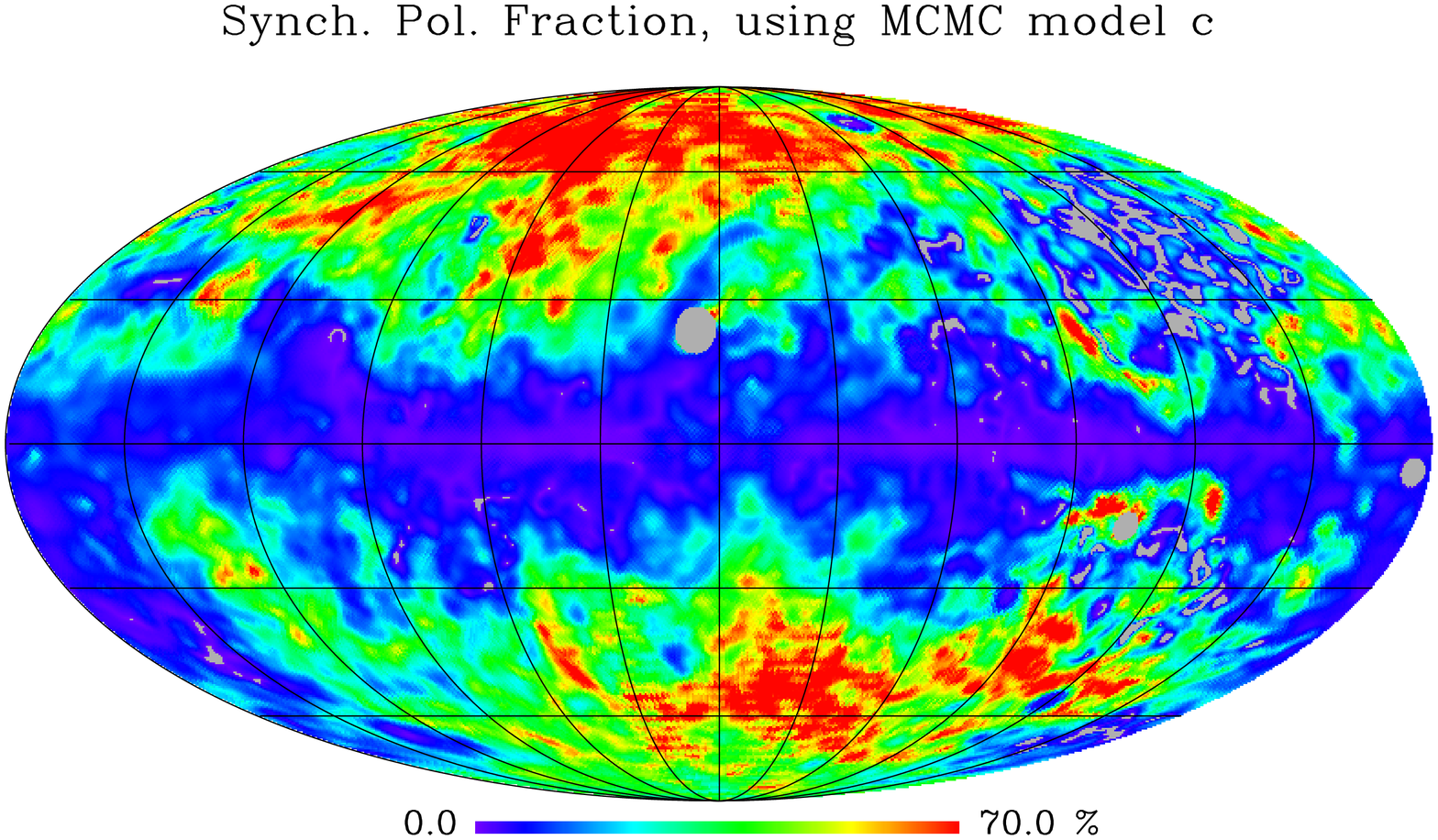}
  \includegraphics[angle=90,width=0.49\textwidth]{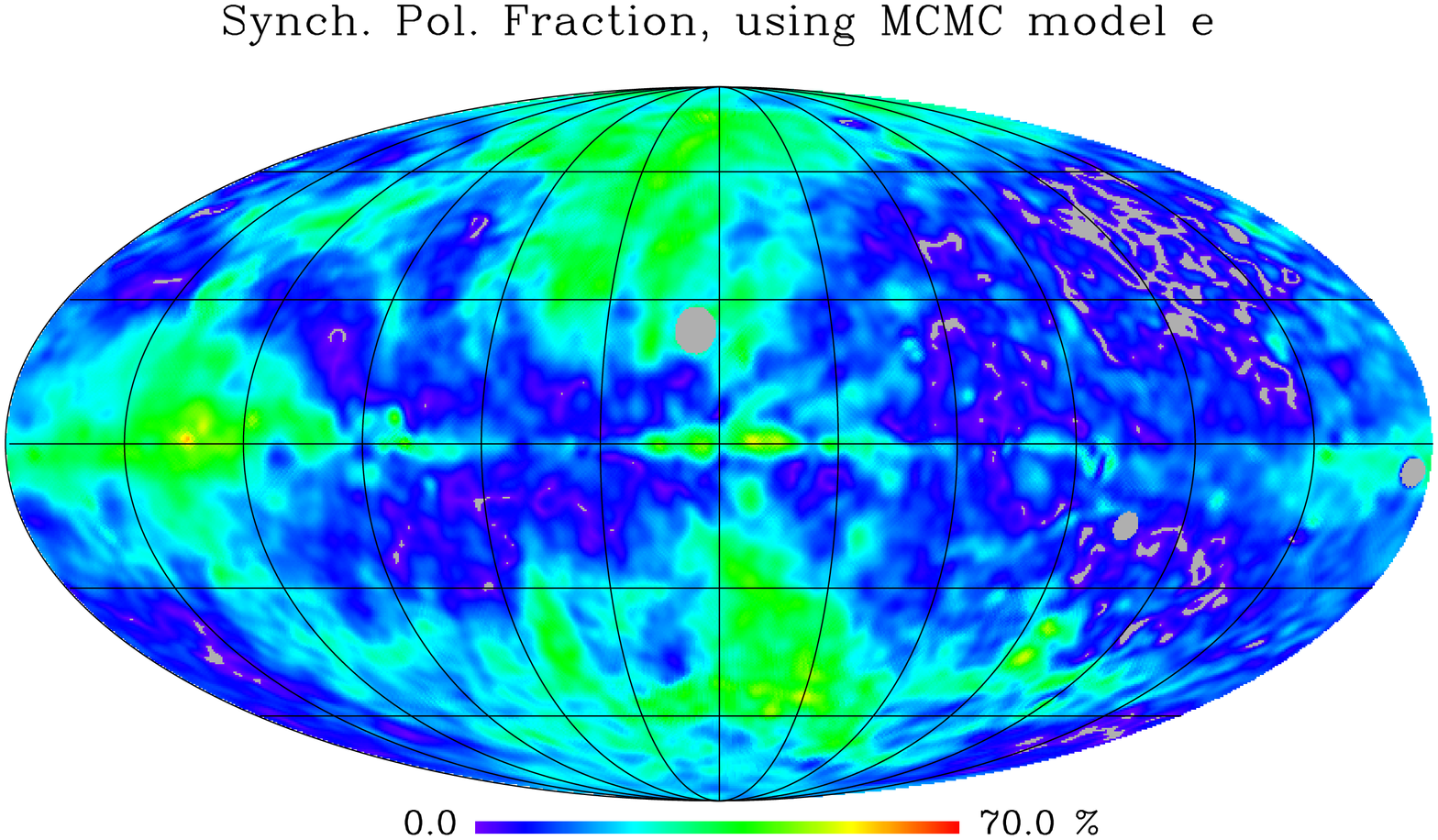}
  \caption[Synchrotron polarisation fraction maps at 23\,GHz using
    different templates for the synchrotron intensity.]{Synchrotron
    polarisation fraction maps at 23\,GHz using different templates
    for the synchrotron intensity (see text). The colour scale is
    linear, and it ranges from 0--75\%. Our preferred $\Pi$ map is
    derived from the \wmap\ MCMC model-e \protect\citep{gold:11}.}
  \label{fig:pol_frac}
\end{figure*}

\section{Polarisation fraction}
\label{sec:pol_fraction}

Calculating the polarisation fraction, $\Pi$, of synchrotron emission is
a good way to measure the degree of order of the magnetic field
perpendicular to the line-of-sight. This calculation is not trivial
because of the difficulty in obtaining the synchrotron total intensity
at GHz frequencies. Free-free and AME contribute to the total emission
so component separation methods are necessary. Here we compare the
polarisation fraction at \kband~using different templates for the
synchrotron intensity over the full sky.

\label{sec:pol_fraction_fs}
To obtain the polarisation fraction, we measured the ratio, $\Pi$,
between our debiased polarisation map at \kband~(produced using the
Wardle \& Kronberg estimator from Eq. \ref{eq:pwk_chi}) and a template
for the synchrotron intensity at this frequency,
\begin{equation}
  \Pi = \frac{P_{K}}{I_K}~.
  \label{eq:pol_frac}
\end{equation}

As a first step, we used the total intensity map at \kband~only, which
gives a lower limit in $\Pi$. We subtracted the contribution from the
CMB using the internal linear combination (ILC) map provided by the
\wmap team \citep{bennett:13}.  The median value over the sky is 9.5\%
with some regions along the NPS and other filaments showing a value
$\Pi \sim 40\%$. This is consistent with a substantial amount of
emission from other components beside synchrotron present in the
\kband\ total intensity map.

\begin{table}
\centering
\caption[Average spectral indices and the standard deviation of  $\beta$ for the 18 regions. ]{Weighted average spectral indices and the  standard deviation of $\beta$ for the 18 regions. }
\begin{tabular}{lcc}
\toprule
 & $\bar{\beta}$ & $\sigma_{\beta}$\\
\midrule
$K-Ka$ & $-3.03 \pm 0.02$ &  0.43 \\ 
$K-Q$ & $-3.07 \pm 0.02$ &  0.40 \\ 
$Ka-Q$ & $-3.17 \pm 0.06$ &  1.66 \\ 
\bottomrule
\end{tabular}
\label{tab:spec_ind_summ}
\end{table}

A way of estimating the synchrotron intensity is by extrapolating the
408\,MHz map up to 23\,GHz. At 408\,MHz there is some free-free coming
mainly from star forming regions on the Galactic plane. We subtracted
this component by scaling the \wmap MEM map free-free template to
408\,MHz using a free-free spectral index $\beta_{\rm ff}=-2.13$,
valid for the diffuse gas \citep{davies:06}. Then, we scaled the
resulting pure synchrotron map from 408\,MHz to 23\,GHz, using a fixed
synchrotron spectral index of $\beta_{s}=-3.0$. The largest source of
uncertainty comes from the fixed synchrotron spectral index between
0.408\,GHz to 23\,GHz. A variation in the synchrotron spectral from
$\beta_{s}=-3.1$ to $\beta_{s}=-2.9$ decreases the measured
polarisation fraction by a factor $\sim 2$. Therefore, accurate
measurements of the total intensity spectral index are critical in
order to find accurate polarisation fractions using this method.

A second approach to obtain the synchrotron polarisation fraction is
based on the modelling of the different emission mechanisms in total
intensity. Closer to the plane, there is a large free-free
contribution from H{\sc ii} regions so it is necessary to subtract it in
order to obtain the synchrotron emission. The \wmap team provides
all-sky foreground templates at each of the five \wmap frequencies,
generated using two procedures: Maximum Entropy Method (MEM) and Monte
Carlo Markov Chains (MCMC) fits \citep[for a detailed description of
  the methods, see][]{gold:11}. The MEM technique generates
synchrotron, free-free, spinning dust, and thermal dust templates and
assumes that the spectral indices of these foregrounds are constant
over the sky. The MCMC method produces four different synchrotron
templates, one for each sets of parameters of the prior
model. \citep[see][for a description of the models
  investigated.]{bennett:13} All the methods used to calculate the
synchrotron total intensity are listed in Table~\ref{tab:synch_models}.

We masked three regions on the synchrotron templates that are
problematic for calculating the polarisation fraction. First, Tau A
($l=184^{\circ}\!.3,~b=-5^{\circ}\!.8$), where the synchrotron
spectral index is flatter ($\beta_s\approx -2.3$) than that observed
in the diffuse gas; the $\zeta$\,Oph H{\sc ii} region
($l=6^{\circ}\!.3,~b=-23^{\circ}\!.6$), where the synchrotron
contribution is clearly underestimated in some MCMC models; and an
area from the Gum nebula ($l=254^{\circ}\!.5,~b=-17^{\circ}\!.0$),
where the same underestimation of synchrotron occurs. See Figure
\ref{fig:pol_frac}

\begin{table*}
  \centering
  \caption[Methods used to estimate the synchrotron polarisation
    fraction at 23\,GHz.]{Methods used to estimate the synchrotron
    total intensity at 23\,GHz. Also listed are the median uncertainty
    over the sky of the polarisation fraction derived form each
    method, $\Pi_{\mathrm{median}}$; the standard deviation along the
    polarisation fraction maps, $\Pi_{\mathrm{stdev}}$; the median of
    the polarisation fraction at high latitudes ($b>|30^{\circ}|$),
    $\Pi_{\mathrm{med}}^{b>|30^{\circ}|}$ and the median of the
    polarisation fraction at low latitudes ($b<|30^{\circ}|$),
    $\Pi_{\mathrm{med}}^{b<|30^{\circ}|}$. }
  \begin{tabular}{llccccc}
    \toprule
    Method &  Note & $\Pi_{\mathrm{median}}$   & $\Pi_{\mathrm{stdev}}$ & $\Pi_{\mathrm{med}}^{b>|30^{\circ}|}$ & $\Pi_{\mathrm{med}}^{b<|30^{\circ}|}$ \\
    & & [\%]  & [\%]  &[\%]  &[\%]   \\
    \midrule
    $(408\,\mathrm{MHz} - ff) \rightarrow 23$\,GHz  & Assumes fixed $\beta_{\mathrm{syn}}-3.0$.                  & 16.9  & 10.1 & 22.0 & 12.9 \\
    \wmap MEM                              & Includes AME. Fixed $\beta_{\mathrm{syn}}=-3.0$.           & 11.6  & 7.6  & 11.5 & 11.8\\
    \wmap MCMC model c                     & No AME component. $\beta_{\mathrm{syn}}$ can vary.         & 20.5  & 20.6 & 36.7 & 9.9 \\
    \wmap MCMC model e                     & Includes AME. Fixed $\beta_{\mathrm{syn}}=-3.0$            & 16.2  & 9.2  & 18.6 & 14.0\\
    \wmap MCMC model f                     & Includes AME.  $\beta_{\mathrm{syn}}$ can vary.            & 21.0  & 21.6 & 38.9 & 9.6\\
    \wmap MCMC model g                     & Includes AME.  $\beta_{\mathrm{syn}}$ varies as described  & 22.5  & 22.6 & 41.1 & 10.8\\
    & in \citet{strong:11} \\            
    \bottomrule
  \end{tabular}
  \label{tab:synch_models}
\end{table*}

Fig. \ref{fig:pol_frac} shows four out of the six different
polarisation fraction maps produced. The differences in these maps are
due only to the different synchrotron intensity template used. The
extrapolation of the free-free corrected Haslam et al. map
(Fig. \ref{fig:pol_frac}, top-left) produces a $\Pi$ map which is
morphologically very similar to the polarisation intensity map
(Fig. \ref{fig:wmap_3deg_all}, top-left), where the structures are
filamentary. Regions with high polarisation fraction are individual
filaments or specific areas on the plane (e.g. the fan region at $l
\sim 140^{\circ}$). The $\Pi$ map created using the free-free MEM model
(Fig. \ref{fig:pol_frac}, top-right) has a higher polarisation
fraction and also shows a bigger difference between mid and low
Galactic latitudes, where the polarisation fraction is near to zero on
the plane.  The output with the MCMC model e is very similar to the
one using the MEM model at high latitudes and to the Haslam map
extrapolated close to the plane. It was created with a fixed
 synchrotron spectral index of $-3.0$. MCMC models $c$, $f$ and $g$
produce very similar and incorrect values for $\Pi$, as they result in
polarisation fractions $\sim$100\% over large areas of the sky at high
Galactic latitudes. We therefore believe that these maps are biased
over large areas of the sky and we will not discuss further.

Qualitatively, we believe that the best model for the synchrotron
total intensity is the MCMC-e model of \cite{gold:11}. This model
predicts similar values on the inner Galactic plane and at
intermediate latitudes than those calculated extrapolating the Haslam
et al. map using $\beta=-3.0$, value similar to the one that we have
measured in the same region in Sec. \ref{sec:spectral_indices}.  Some
regions are highly polarised ($\Pi \approx 30\%$) regardless of the
synchrotron model used. Examples are the brighter region of the NPS,
the peak of Filament VIIb, the diffuse region in the southern part of
Filament Is and Filament IIIs.

\section{Faraday rotation}
\label{sec:far_rot}
A magnetic field with a component parallel to the line-of-sight will
induce the effect of Faraday rotation, in which the plane of
polarisation of a travelling photon rotates as it travels through the
ISM. The amplitude of the effect depends on the strength of the
magnetic field along the line of sight and on the electron
density. The observable, the rotation measure RM, is a change in the
polarisation angle, $\chi$ which scales with the square of the
wavelength $\lambda$. It is defined as follows,
\begin{equation}
  \mathrm{RM} = \frac{\mathrm{d}\chi}{\mathrm{d}\lambda^2}.
\label{eq:rot_m}
\end{equation}

If there is a polarised source in the line-of-sight with no intrinsic
Faraday rotation, then RM is equal to the Faraday depth, which is
defined as \citep{burn:66}
\begin{equation}
  \phi = \frac{e^3}{2\pi m_e^2c^4}\int_0^d n_e(s) B_{||}(s)
  \;\mathrm{d}s,
\end{equation}
where $e$ and $m_e$ are the electron charge and mass, $c$ the speed of
light and the integral is performed over the electron density $n_e$
and the line-of-sight component of the magnetic field $B_{||}$.

The Faraday rotation at \wmap frequencies is expected to be very small
on average. A 1\dg difference in polarisation angle between K and Ka
bands correspond to a rotation measurement of 190\,rad\,m$^{-2}$.  We
measured $\Delta\chi$ between K and Ka over most of the sky using the
1\dg smoothed polarisation maps.

\begin{figure}
  \centering
  \includegraphics[angle=0,width=0.49\textwidth]{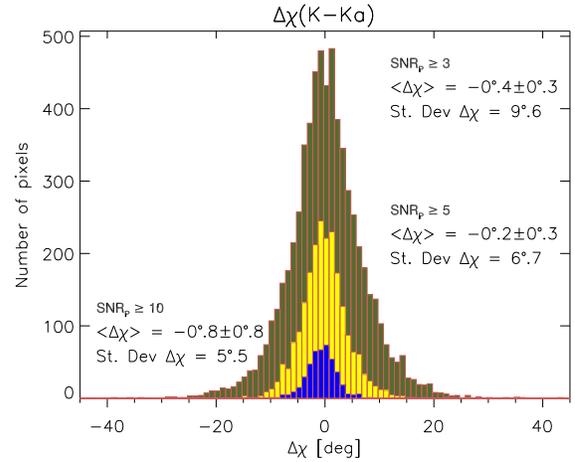}
  \caption[Histograms of the variation in the $\chi$ between K-Ka ]{
    Histograms of the variation in the $\chi$ between K-Ka bands using
    three SNR cuts. The olive histogram is made with the pixels that
    have SNR$_{\rm P}\ge3$, the yellow histogram with the pixels in
    which SNR$_{\rm P}\ge5$ and the blue histogram includes pixels
    that have SNR$_{\rm P}\ge10$. The mean values, uncertainties, and
    the standard deviations are quoted in the figure for each
    distribution.}
  \label{fig:dchi_hist_k_ka}
\end{figure}

We calculated the polarisation angle rotation for three different
signal-to-noise cutoffs (${\rm SNR_{cut}}$). Figure
\ref{fig:dchi_hist_k_ka} shows the three histograms of the difference
between the polarisation angle measured in \wmap--K and Ka bands for
three different $\rm SNR_{cut}$ of 3, 5 and 10. The three
distributions are centred at zero within the uncertainties.  The mean
Faraday rotation over the region studied is less than 1\dg\!\!. This
value is consistent with the prediction using the Faraday depth map
from \citet{oppermann:12}. In the high SNR selection (SNR$>10$, blue
histogram in Fig.  \ref{fig:dchi_hist_k_ka}), most of the pixels come
from the inner Galactic plane. It is here where the Faraday rotation
should have its larger effect as the column density is the
largest. Nevertheless, on average, the difference in angle measured is
consistent with zero. This shows consistency also in the \wmap
polarisation angle calibration between K and Ka bands.

\begin{figure}
  \centering
  \includegraphics[angle=0,width=0.49\textwidth]{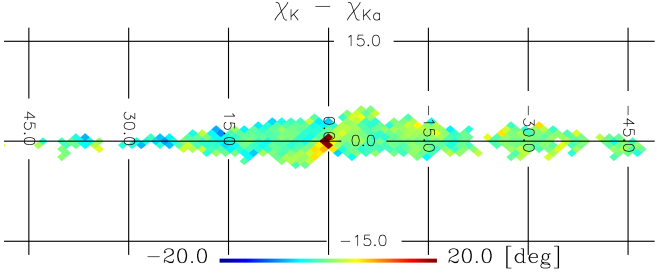}
  \caption{ Difference in polarisation angle between \wmap K and Ka
    bands in the inner Galaxy. Only the pixels where the SNR in the
    polarisation angle is greater than 10 are shown. The central
    pixels shows a peak value of $\Delta \chi_{\rm K-Ka}=$34\dg\!\!. }
  \label{fig:dchi_k_ka}
\end{figure}

The spread in the histograms of Figure \ref{fig:dchi_hist_k_ka} is
consistent with what is expected for a random noise component plus a
significant residual. A ${\rm SNR_{cut}} = 3$ is equivalent to an
uncertainty in the polarisation angle\footnote{Ignoring QU
  correlations, $\sigma_{\chi}=\sqrt{Q^2\sigma^2_Q +U^2\sigma^2_U}/P =
  (2\rm{SNR_{P}})^{-1} $.} of 9$^{\circ}\!\!.5$, while the measured
value using this ${\rm SNR_{cut}}$ is 9$^{\circ}\!\!.6$. Table
\ref{tab:far_rot_summ} lists the expected width of the histograms for
each SNR cut for a noise only component. It also shows the observed
width of the histograms and the difference between these two values,
which can be attributed to Faraday rotation. We note that this
difference is actually a slight underestimate of the true dispersion
because some pixels will have SNR larger than the cutoff. The region
with the higher SNR cut (pixels in the blue histogram of
Fig.~\ref{fig:dchi_hist_k_ka}) has the widest residual distribution of
4$^{\circ}\!\!.7$, which corresponds to a rotation measures of $\pm
893$\,rad\,m$^{-2}$. However, this region represents only 0.9\% of the
area of the sky. Figure \ref{fig:dchi_k_ka} shows the difference in
polarisation angle between K and Ka bands in the inner Galaxy. Here,
the Galactic centre shows significant amount of Faraday rotation with
a peak value of $\Delta \chi_{\rm K-Ka}=$34\dg at the central pixel.

\begin{table}
\centering
\caption{For three different SNR cut values, we list the fraction
  of the sky which is used, the width of the distribution expected for
  a noise only contribution $\sigma_{\chi-{\rm noise}}$, the observed
  width of each distribution, shown in Figure \ref{fig:dchi_hist_k_ka}
  and the difference between these two widths, $\sigma_{\rm Faraday}^2
  = \sigma_{\Delta\chi}^2 - \sigma_{\chi-{\rm noise}}^2 $, which can
  be attributed to Faraday rotation. }
\begin{tabular}{ccccc}
\toprule
 SNR$_{\rm P}$  & Sky fraction [\%] & $\sigma_{\chi-{\rm noise}}$ & $\sigma_{\Delta\chi}$ & $\sigma_{\rm Faraday}$ \\
\midrule                    
3        & 13.1              & $ 9^{\circ}\!\!.5 $ & $9^{\circ}\!\!.6$ & $1^{\circ}\!\!.4$  \\ 
5        &  4.3              & $ 5^{\circ}\!\!.7 $ & $6^{\circ}\!\!.7$ & $3^{\circ}\!\!.5$  \\ 
10       &  0.9              & $ 2^{\circ}\!\!.8 $ & $5^{\circ}\!\!.5$ & $4^{\circ}\!\!.7$  \\ 
\bottomrule
\end{tabular}
\label{tab:far_rot_summ}
\end{table}


\section{Discussion}
\label{sec:discussion} 
We have presented a range of observational parameters that can drawn
from the \wmap~polarisation data. We have seen that an important part
of the polarised emission at microwaves comes from filamentary
structures at high Galactic latitudes. What is the origin of these
filaments? How do they relate with the rest of the ISM, the neutral
gas for example?  We start this discussion by looking at H{\sc i}
data, which traces the neutral gas of the Galaxy. We then go through
the hypothesis for the origin of these filaments and use \wmap data to
test the predictions of one model.

\subsection{H{\sc i} morphology}
\label{sec:hi} 
\begin{figure}
  \hspace{0.62cm}
  \includegraphics[angle=90,width=0.435\textwidth]{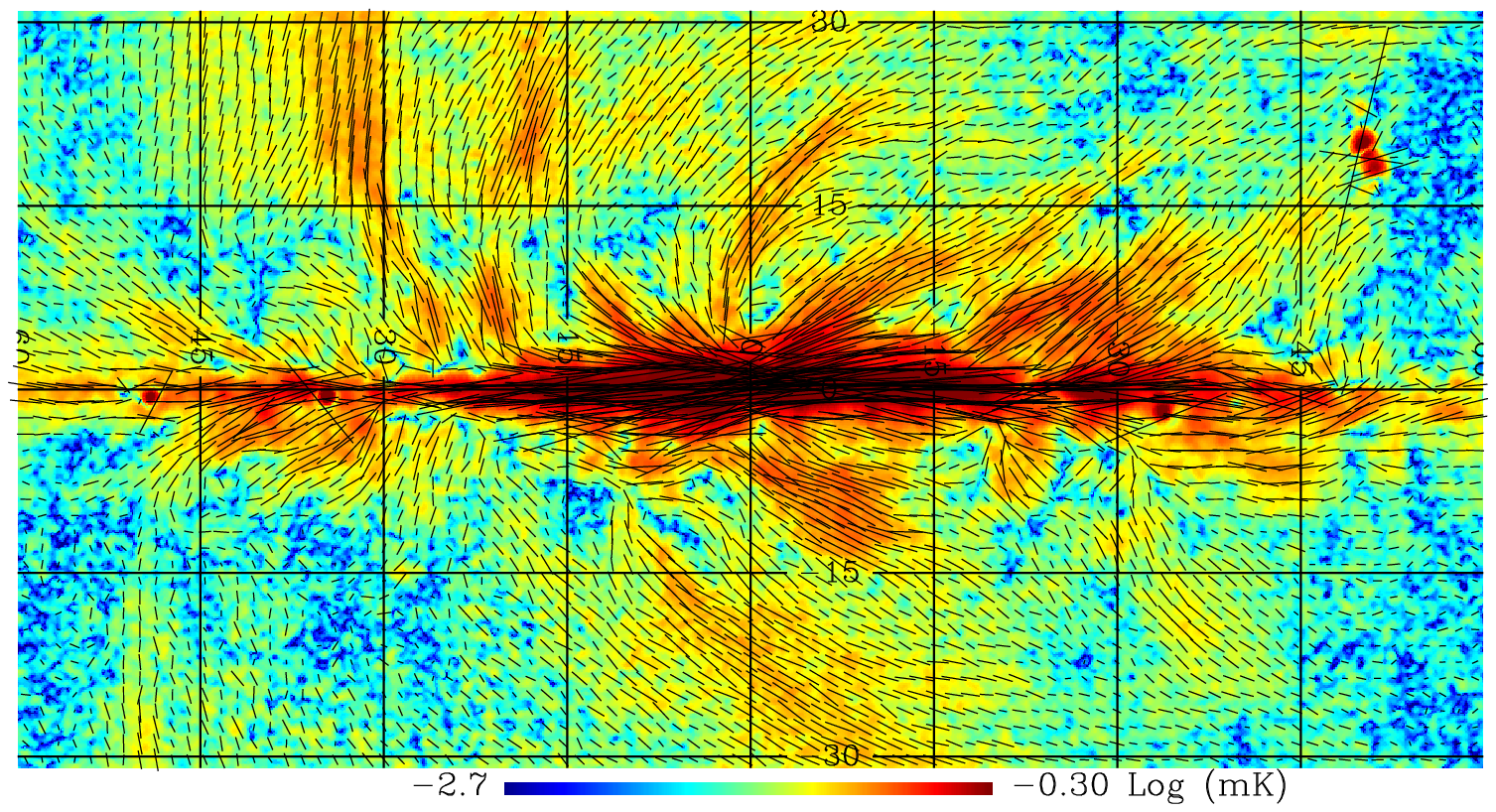}
  \vspace{-0.5cm}
  \hspace{-0.2cm}
  \includegraphics[angle=0,width=0.5\textwidth]{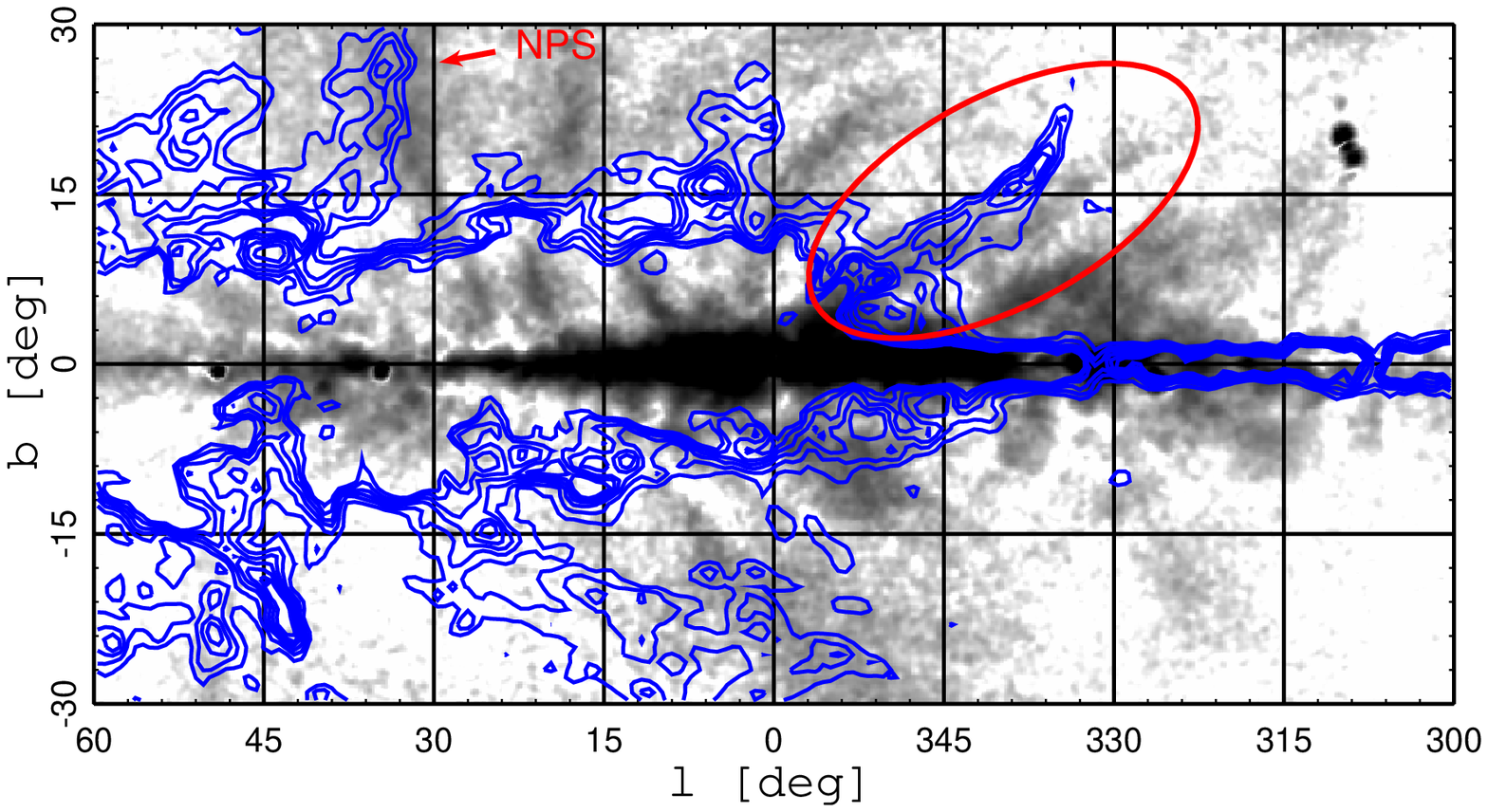}
  \caption{{\it Top:} \wmap \kband~polarisation intensity map showing
    the inner Galaxy between $l =\pm60$\dg. The polarisation vectors
    are rotated by 90\dg to indicate the direction of the magnetic
    field.  {\it Bottom:} Same as the map on {\it Top:} but using a
    linear colour scale. The contours of H{\sc i} 21\,cm emission at
    $v=+10$\kms are shown in blue. There is a possible association
    between the H{\sc i} and the polarised intensity on filament
    XIV. The angular resolution is 1\dg and the graticule has 15\dg
    spacing in both Galactic coordinates. Both maps are centred at
    $(l,b) = (0^{\circ},0^{\circ}$).}
  \label{fig:inner_galaxy_fils}
\end{figure}

Most of the filamentary features that extend roughly perpendicular to
the Galactic plane lie in the inner Galaxy, with $|l| \lesssim
60^{\circ}$.  In these structures, the polarisation angle is roughly
aligned along the extension of the filaments.  The top panel of
Fig. \ref{fig:inner_galaxy_fils} shows the inner region of the Galaxy
between $|b| < 30^{\circ}$ visible in \kband~polarisation.

These structures appear similar to the H{\sc i} ``worms'' described by
\citet{heiles:84,koo:92}. Heiles proposed that these worms observed in
the inner Galaxy are shells with open tops. Stellar winds and
supernova explosions create shells in the ISM and if the expansion
energy is large enough, hot gas will flow out through chimneys to the
Galactic halo. Chimney walls are vertically supported by magnetic
fields \citep{breitschwerdt:91}, which can glow in radio continuum if
cosmic rays diffuse through them.

\begin{figure}
  \centering
  \newcommand{\widthfig}{0.495}
  \includegraphics[angle=90,width=\widthfig\textwidth]{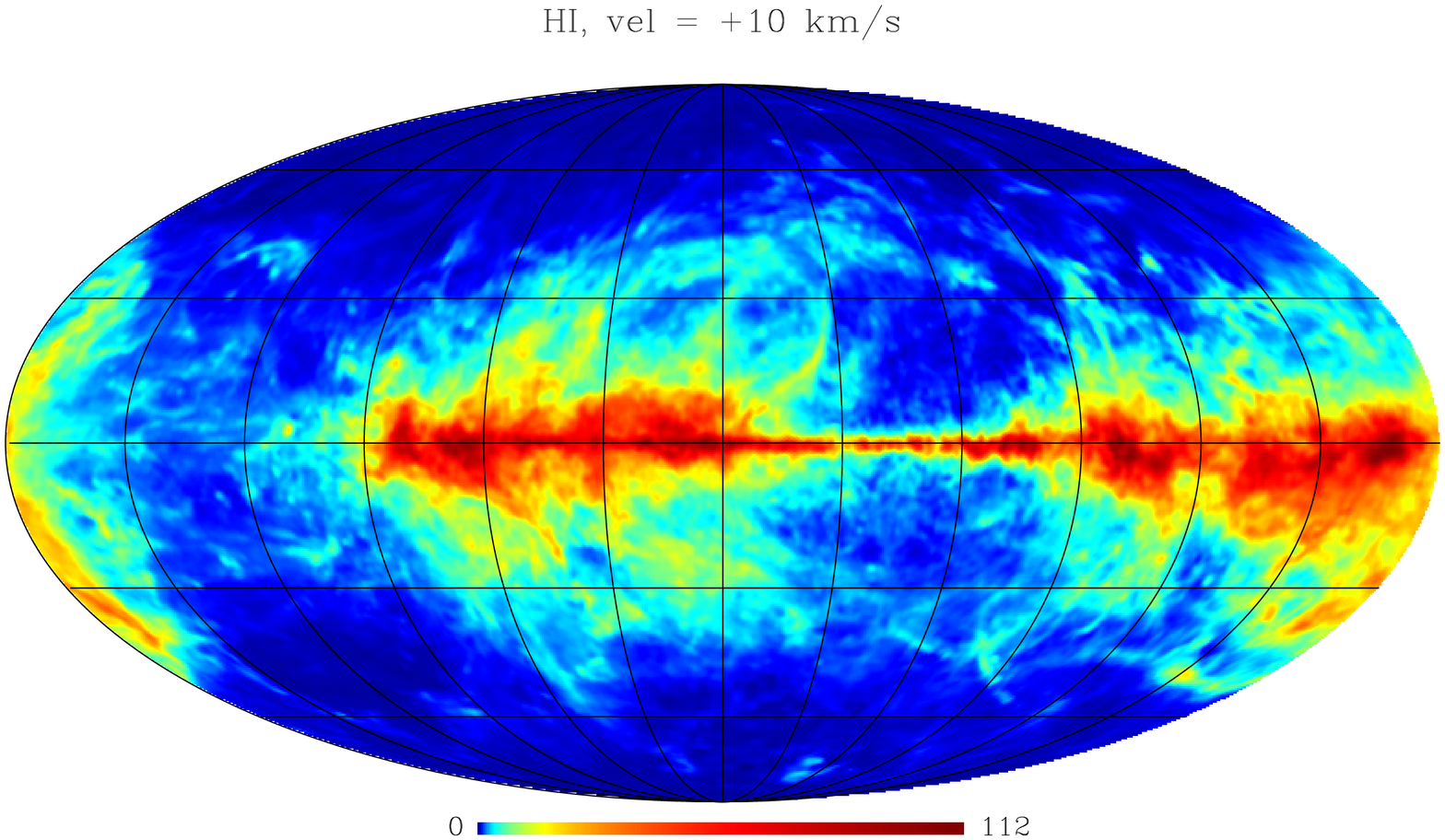}
  \includegraphics[angle=90,width=\widthfig\textwidth]{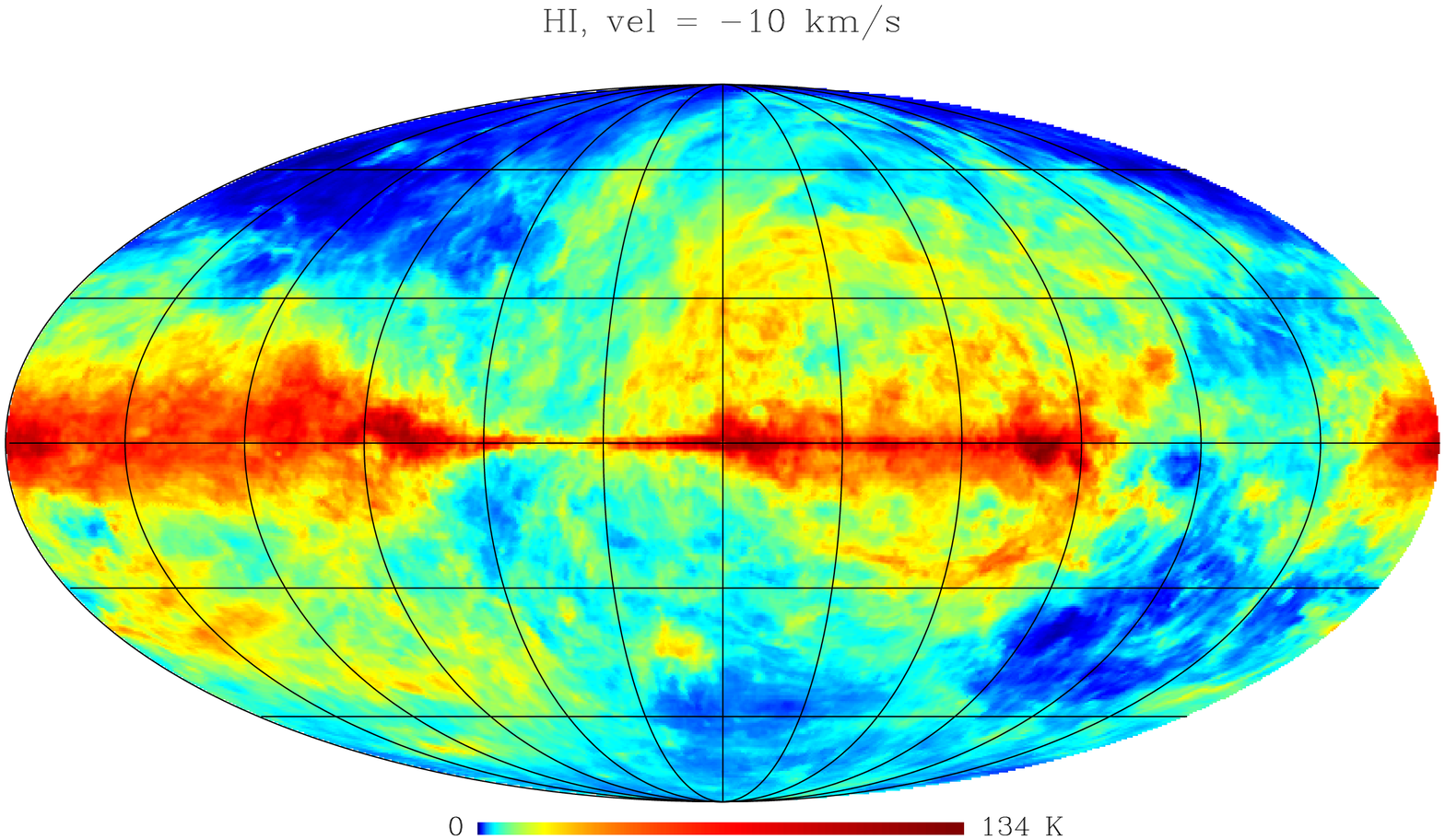}
  \caption{H{\sc i} maps of the full sky at two line-of-sight
    velocities from the LAB survey \citep{kalberla:05}:
    $+10$\,km\,s$^{-1}$ ({\it top}) and $-10$\,km\,s$^{-1}$ ({\it
      bottom}). The graticule has 30\dg spacing in both Galactic
    coordinates and the maps are plotted using a histogram-equalised
    scale.}
  \label{fig:HI_full_sky}
\end{figure}

\citet{sofue:88} describes a number of radio spurs present in the
408\,MHz map of Haslam et al., and most of these spurs lie in the
inner Galaxy. He did not find definitive correlations between H{\sc i}
and the radio continuum structures in this region. Here, we examine
the more sensitive H{\sc i} 21\,cm maps from the Leiden/Argentine/Bonn
(LAB) survey \citet{kalberla:05}, which maps the full sky with an
angular resolution of 36\,arc\,min FWHM in the velocity range from
--450\,\kms~to +400\,\kms\!, at a resolution of 1.3\kms\!.  We
searched for morphological correlations between the polarisation spurs
and the H{\sc i} velocity maps.  We did not find obvious correlations
either. The only features that have H{\sc i} counterpart are the NPS
and the polarised filament No. XIV in our
nomenclature. Fig.~\ref{fig:inner_galaxy_fils} ({\it bottom}) shows the
polarisation map with the H{\sc i} contours of the velocity range that
shows the correlation. From the figure, we can see that the
correlation is better near the base of the filament and with
increasing latitude, there is a spatial separation in the H{\sc i}
emission from the polarised synchrotron. This H{\sc i} filament in
particular is part of a larger shell, probably the limb-brightened
periphery of a H{\sc i} bubble. This is visible in the {\it top} panel
of Fig. \ref{fig:HI_full_sky} where we show the full sky H{\sc i} map
with a velocity of +10\,km\,s$^{-1}$, where there is a shell-like
structure of $\sim 30$\dg in diameter centred around
$(l,b)=(350^{\circ},20^{\circ})$.

If the polarised emission that comes form filament No. XIV is indeed
physically correlated with the H{\sc i} emission from this bubble,
this would resemble the \citet{heiles:84} picture previously
described. Here, an enhancement in the magnetic field, produced by the
compression of the ambient field by an expanding shock, can induce the
emission of polarised synchrotron radiation along the field lines.

We now draw attention to the structure centred at $l \sim 320^{\circ}$
on the plane, the large shell that is most visible at --10\kms ({\it
  bottom} panel of Fig.~\ref{fig:HI_full_sky}). This shell is known
as the Scorpius-Centaurus super-shell, a cavity in the local ISM which
is thought to be the expanding remnants of a number of SNe explosions
\citep[e.g.][]{weaver:79}. Inside this cavity lies the Sco-Cen OB
association, the nearest group to the Sun containing massive stars,
where the distances of its members lie between 118 and 145\,pc and the
oldest groups of stars are $\sim 15$\,Myr old \citep{preibisch:08}. We
note that the H{\sc i} counterpart of the NPS lies on the periphery of
the Sco-Cen shell. We will discuss this connection in more detail in
the next section.

\subsection{Large scale polarised loops origin: towards a possible model} 
\subsubsection{Background}

Since the early attempts to explain the origin of the radio continuum
loops, SN remnants expanding in the Galactic magnetic field have been
a preferred
alternative. \citet{spoelstra:71,spoelstra:72,spoelstra:73} applies
the model of a sphere expanding in the ISM by \citet{vanderlaan:62a}
to Loops I, II, III and IV. In this model, the compression of the
magnetic field due to the shock wave from the SN will act as a
synchrotron source. The emission will be more easily observable on the
periphery of the shell due to limb-brightening, which explains the
loop-type shape of the remnants.

\citet{sofue:74} discuss a number of difficulties that this SNR
hypothesis faces to explain diverse observational data. They present a
new model in which the spurs are the result of the tangential view of
a region of shocked gas produced at the spiral arms that extends above
and along the arms. This hypothesis explains the origin of Loops II and
III and many other spurs according to the authors. The NPS on the
other hand, can be explained according to \citet{sofue:77} in terms of
a magnetohydrodynamical phenomenon associated with the Galactic
centre. A similar scenario connected to the Galactic centre was more
recently proposed by \citet{bland-hawthorn:03}, based on diffuse
filamentary infra-red emission observed on both sides of the Galactic
plane, which is thought to be formed by a central outburst with an
energy scale $\sim10^{55}$\,ergs. This interpretation implies that the
spur has a kpc scale, in comparison with the hundred-pc scale that is
expected from the SNR hypothesis.

It is clear that a distance determination for the spurs was necessary
to settle which interpretation was correct. \citet{bingham:67}, using
optical polarisation data, which is correlated with the polarised
emission from the NPS, determined a distance of $100\pm20$\,pc. The
discovery of H{\sc i} emission from the periphery of the NPS in the maps of
\citet{heiles:74} and \citet{colomb:80} has also helped in this
matter.  \citet{puspitarini:12} have shown that H{\sc i} gas that belongs to
the top region ($b \ge +70^{\circ}$) of Loop I is located at a
distance of $98\pm6$\,pc and the material at intermediate latitudes
($+55^{\circ}\le b \le +70^{\circ}$) is around 95--157\,pc. This is strong
evidence that favours a nearby shell as the origin of Loop I.

In this nearby scenario, a problem with a single SN event as the
origin for Loop I is the discrepancy on the expected age for such SN
event. The low expanding velocity of the H{\sc i} gas, $|v| \le
20$\,km\,s$^{-1}$ \citep{sofue:74,weaver:79,kalberla:05}, implies an
SN older than several $10^{6}$\,yr. On the other hand, soft X-ray
emission detected from the interior of the radio loop by
\citet{bunner:72} suggest an age 10 times younger. Moreover, the
initial SN energy in either case is $\sim 10^{52}$\,ergs, an order of
magnitude larger than the standard $10^{51}$\,ergs \citep{egger:95a}.

A related scenario in which instead of a single SN event, the Loop I
cavity is a super-bubble, the result of stellar winds and consecutive
supernovae in the Sco-Cen OB association is more attractive. This idea
has been suggested by a number of authors, \citep[see
  e.g.][]{weaver:77,heiles:80}. \citet{egger:95b} present a model in
which a recent SN inside the Sco-Cen super-shell shocks the inner
walls of the bubble, giving rise to the NPS emission. 

An expanding super-shell in a magnetised medium has been modelled by
\citet{tomisaka:92}. In their model, during the early stages of the
shell (a few Myr), the expansion is spherically symmetric. In a later
phase ($\sim 50$\,Myr), the magnetic field stops the expansion
perpendicular to the field lines producing a highly elongated bubble.
Cosmic rays accelerated by the shock will emit polarised synchrotron
travelling through the distorted field lines produced by the shell. 

\subsubsection{A proposed model}
\label{sec:model}
Assuming that the unperturbed magnetic field lines in the vicinity of
the Sun are parallel to the Galactic plane, an expanding super-shell
will bend the lines in a simple manner. The originally parallel field
lines will follow lines of constant longitude on the surface of the
expanding sphere. The observed pattern from Earth however is not
trivial and it will depend on the viewing angle of the
shell. Following what \citet{heiles:98} showed using starlight
polarisation, we can compare the direction of the field lines in this
scenario, produced by a shell centred at the location of the Loop I
bubble, with the polarisation angle of the emission seen by
\wmap\!\!. This type of modelling was also used by
\citet{wolleben:07}, who placed two overlapping shells that deforms
the magnetic field medium to reproduce the polarisation directions
observed at 1.4\,GHz. These data show a depolarisation band confined
at $|b|\approx 30^{\circ}$, so their fit is based only on the high
latitude data. The low Faraday rotation of the \wmap data allow us to
compare directly the angles in the $|b| \le 30^{\circ}$ band, where
there is strong emission from the spurs.

\begin{figure}
  \centering
  \includegraphics[angle=0,width=0.49\textwidth]{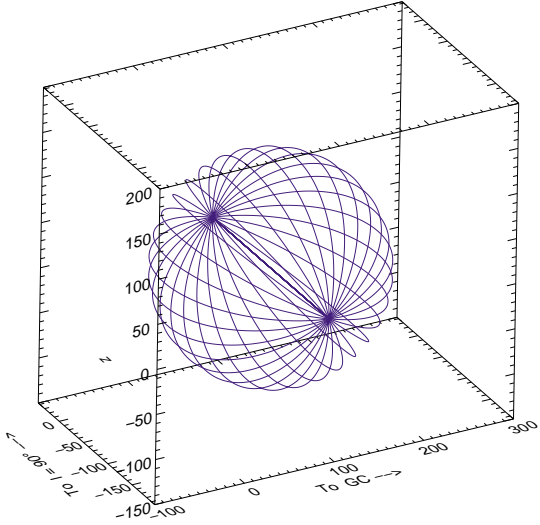}
  \caption[Model of magnetic field lines on a spherical
    shell.]{Direction of the magnetic field lines on the surface of a
    spherical shell expanding in a magnetised medium with an initial
    uniform field parallel to the Galactic plane. The field lines will
    follow ``meridian'' lines on the surface of the shell as it is
    shown in the figure. The Sun is located at the origin of the three
    axes and the units of are parsecs. The radius of the shell is
    120\,pc and it is located at 120\,pc from the Sun in the direction
    $(l,b)=(320^{\circ}\!\!,5^{\circ})$.  }
  \label{fig:model_field}
\end{figure}

We locate the centre of the bubble at 120\,pc, in the direction of the
Sco-Cen supper shell as seen in H{\sc i} at
$(l,b)=(320^{\circ}\!\!,5^{\circ})$ \citep{heiles:98}. We note that
this is not the centre of the Loop I as measured in radio data (table
\ref{tbl:circ_loops}) Fig. \ref{fig:model_field} shows a 3D projection
of the field lines on the surface of the shell. From the figure, it is
clear that the vantage point will define the appearance of the field
lines for any observer.

\begin{figure}
  \centering
   \includegraphics[angle=180,width=0.49\textwidth]{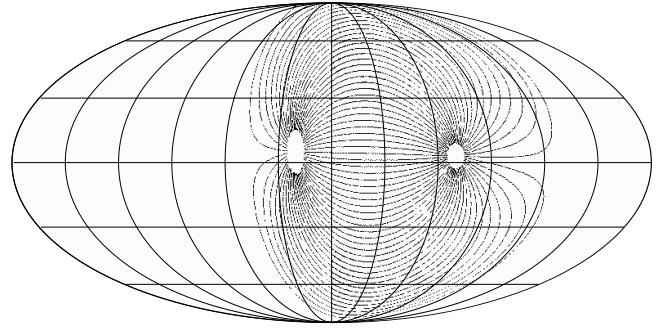}
  \caption[Magnetic field lines of shell over the sky.]{Projection on
    the sky of the magnetic field lines directions of a spherical
    shell of 120\,pc of radius located at 120\,pc in the direction.
    The grid spacing is 30\dg in both $l$ and $b$.}
  \label{fig:field_lines}
\end{figure}

\begin{figure*}
  \centering
 \includegraphics[angle=0,width=1\textwidth]{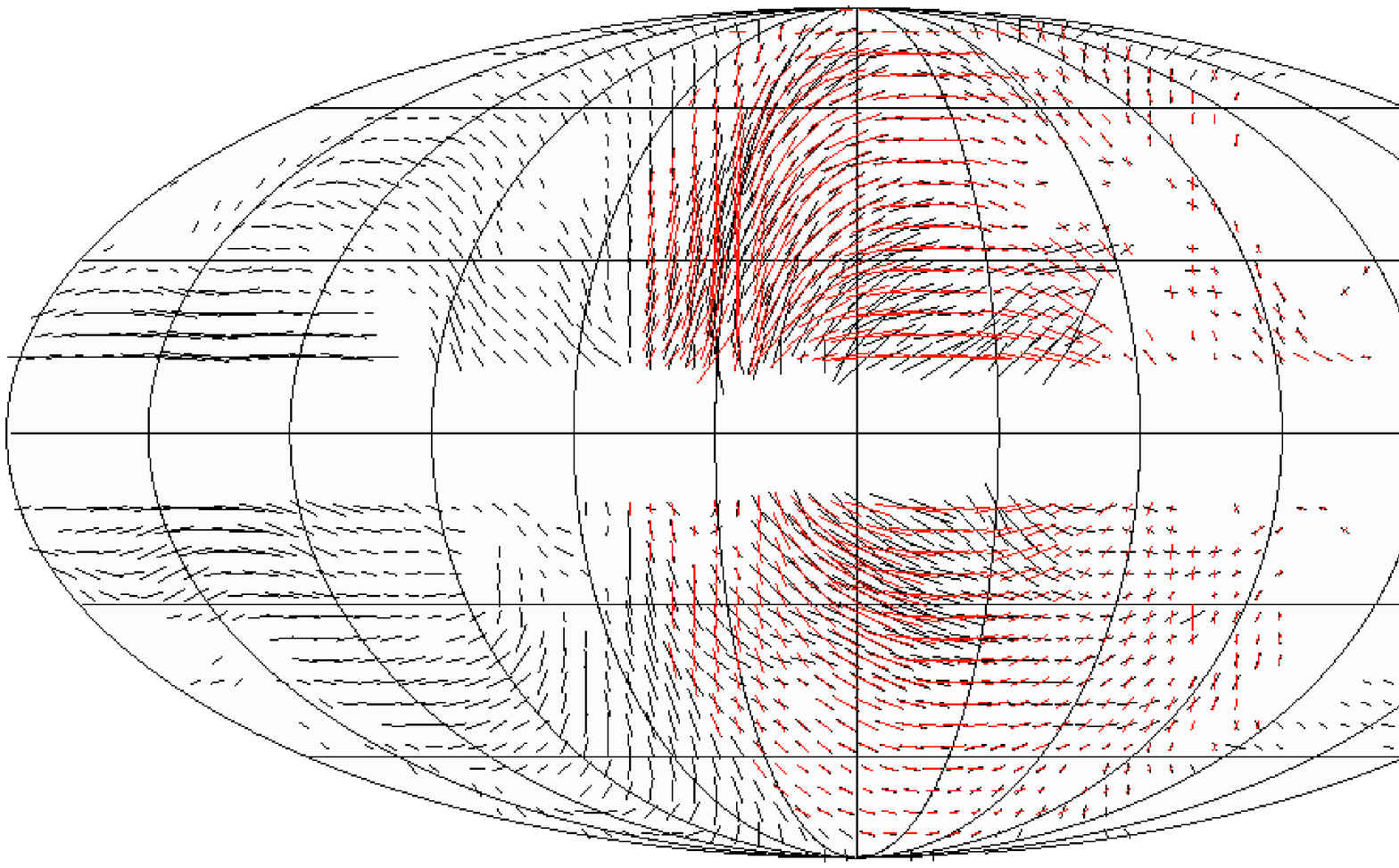}
 \includegraphics[angle=90,width=1\textwidth]{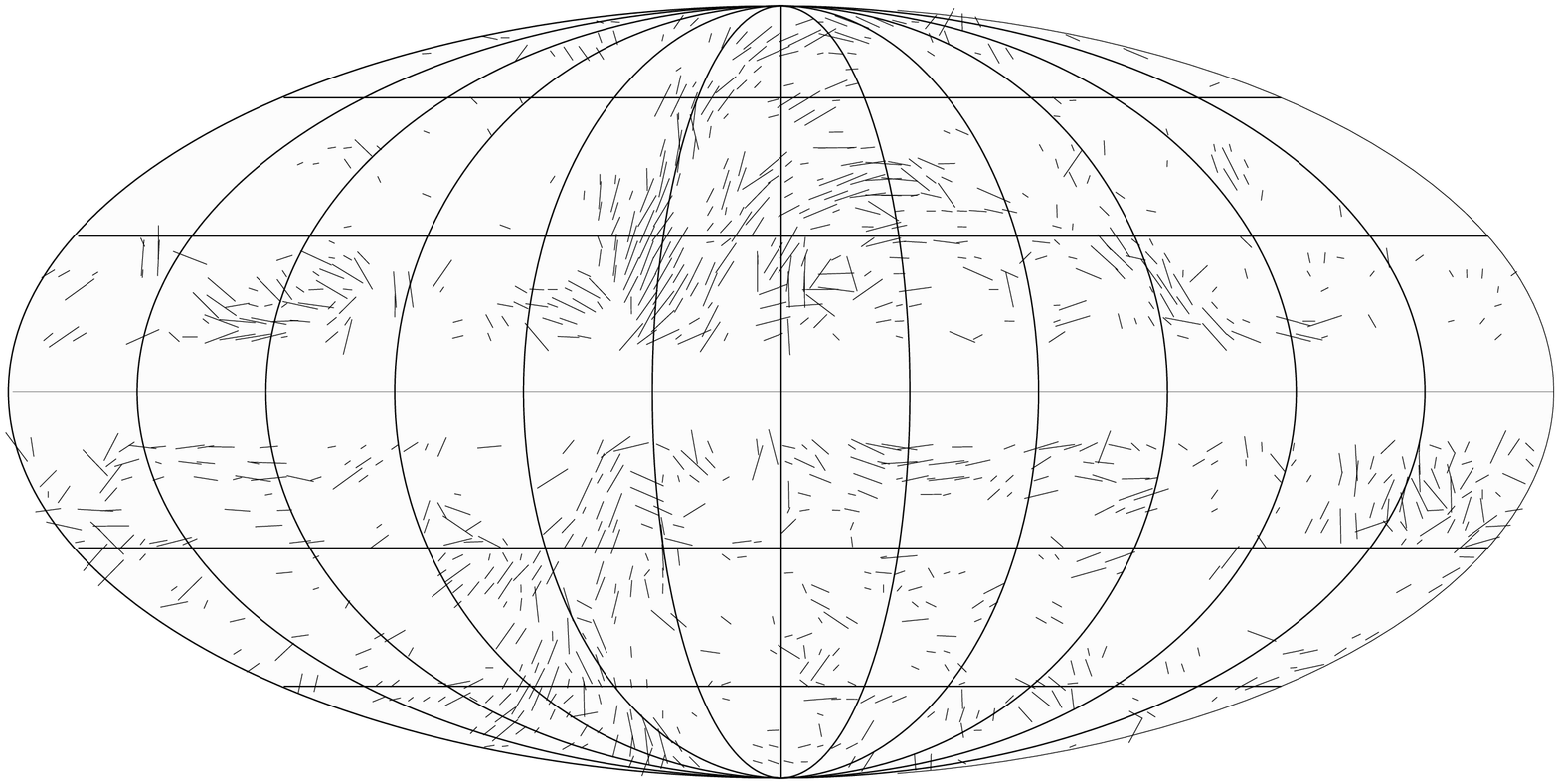}
  \caption{{\em Top:} Comparison between the predicted polarisation
    vectors using the field lines shown in Figure
    \ref{fig:field_lines} in red and the the polarisation vectors
    observed by \wmap at \kband\ in black.  The red vectors have been
    scaled to have the same amplitude as the polarisation observed at
    \kband.  {\em Bottom:} Pseudovectors representing the polarisation
    angle from starlight polarisation measurements, taken from the
    \citet{Heiles2000} compilation. The grid spacing is 30\dg in both
    $l$ and $b$.}
  \label{fig:vectors}
\end{figure*}

A Mollweide projection of the view of the field lines onto the sky is
shown in Fig. \ref{fig:field_lines}. The lines closer to the ``poles''
of the sphere have been masked to avoid crowdedness. In Figure
\ref{fig:vectors} we compare the predicted polarisation angle
direction vectors (in red) on top of the \wmap \kband polarisation
vectors.  It is remarkable that this very simple model fits very well
the direction of the largest loops, above the Galactic plane.  The
bottom panel of Fig. \ref{fig:vectors} shows the pseudovectors
from starlight polarisation measurements, taken from the
\citet{Heiles2000} compilation. We only selected the stars with
distances less than 300\,pc. There is also very good agreement between
the structures seen in starlight polarisation with the polarisation
vectors from \wmap and our model (top panel of the
Fig. \ref{fig:vectors}), particularly at the Northern Galactic
hemisphere, at the interior of Loop I. This is good evidence that the
common structures between the two maps are nearby structures, as we
only used stars within a 300\,pc radius to create the map.

In this scenario, where the emission comes from the compression of the
interstellar field lines, one would expect emission coming from most
of the surface of the shell. This is clearly not the case. The NPS and
the other large loops are the only places where there is a significant
amount of synchrotron.  The reason for this difference in synchrotron
emission on different regions might be connected with the density of
the ISM at different sides of the shell. The interaction of the shell
with a denser medium might trap cosmic rays more efficiently than in a
less dense environment. \citet{lallement:03} presented a 3D map of the
nearby ($d\lesssim500$\,pc) ISM by mapping the absorption of Na lines
by neutral interstellar gas using 1000 lines-of-sight. The section of
the shell that has higher emission, corresponding to the NPS, is close
to the denser Ophiuchi clouds, while the rest of the shell is
expanding into a tenuous low density cavity. The difference in local
density might be responsible for the different synchrotron emissivity
along the shell.
 
\citet{mertsch:13} analysed the angular power spectrum of the diffuse
synchrotron emission as observed in the Haslam et al. map. They
compared the observed power spectrum with a modelled one, which
includes ${O}(1000)$ SN shells distributed in the Galaxy with scales
$\sim 100$\,pc.  The inclusion of these SN shells greatly improves the
fit of the power spectrum. If the results from their modelling are
correct, these shell-like structures are quite common in the Galaxy,
supporting the idea of an SN remnant origin for the diffuse
synchrotron radiation.

We think that the model discussed here, which implies a nearby origin
for Loop I and the largest filaments, describes the data better than
the models connecting these large filaments with activity from the
Galactic centre. One of the arguments that \citet{bland-hawthorn:03}
use to connect the NPS with nuclear activity is that the radio
continuum emission from the NPS is thermal, originated by a
shock-induced UV field which ionized the gas.  The spectral indices
measured in this work show the opposite, a steep spectral index
consistent with diffuse synchrotron emission.  Also, the polarisation
observed is the signature of synchrotron and not free-free
emission. In another work, \citet{jones:12} claim that the magnetic
field is perpendicular to the extension of the GCS and hence is
consistent with a toroidal magnetic field along the spur. This is an
error, as we have shown that the field is actually parallel.  We have
recreated their plot of the \wmap data by replacing $Q,U$ by $-Q,U$;
but to show the magnetic field direction $-Q,-U$ must be
used.\footnote{The same error would occur if \wmap $-Q,-U$ were
  plotted in a package assuming the normal (IAU) sign convention for Stokes
  $U$\!, since \wmap uses the cosmological convention with the opposite
  sign.}  As the magnetic field lines are parallel to the extension of
the filament, a toroidal field along the spur can be ruled out.
\citet{carretti:13}, with polarisation data at 2.3\,GHz, detected two
polarised radio lobes that encompass around 60\dg in the inner
Galaxy. They claim that the origin of these lobes is connected with a
large star formation activity in the central 200\,pc of the Galaxy
that can transport $\sim 10^{55}$\,ergs of energy into the Galactic
halo. This interpretation includes the emission from the GCS, NPS-S,
filaments VIII and IX from our analysis. We remark that this
hypothesis is not incompatible with the existence of the local shell
that we have discussed here. 

If the large-scale polarisation pattern is due to local features, as
it seems to be the case, it cannot be used to test global models of
the Galactic magnetic field. Large scale modelling of Faraday rotation
and other polarisation data has been attempted in a number of works
\citet{page:07,Han2009,Taylor2009,Mao2012}.  But, as shown here, the
locally measured high-latitude polarizations and Faraday rotations
result from local structures, not from the large-scale global Galactic
magnetic field structure. This is a very important consideration that
has to be taken into account in further modelling of the Galactic
field structure of the Galaxy.

We note with interest that \citet{Clark2014} have recently found much
smaller-scale ($10'$) filaments in the atomic hydrogen distribution,
which also align with the local magnetic field. They show that most of
their filaments lie outside the local cavity, at distances larger than
100\,pc. This is in agreement with the model that we studied here, as
he fibres would be aligned by the same mechanism, where the local
magnetic field is distorted by an expanding shell.

In summary, we have used a simple model, based on the idea from
\citet{heiles:98} where a expanding shell is deforming the local
magnetic field to explain the large scale polarisation features seen
by \wmap. There is a very good agreement between the polarisation
direction predicted from the model with the \wmap data, as can be seen
in Fig. \ref{fig:vectors}, and also with the polarisation orientation
as measured by starlight observations. We remark that this good
correlation with the starlight only occurs when the selected stars are
nearby, i.e. $d \leq 300$\,pc. This is good evidence that the magnetic
structures that give rise to the polarisation observed by \wmap is of
local origin. We also note that a large number of the filaments
described in Sec. \ref{sec:fil_angles} are represented by this
model. This means that some of the filaments, such as I, Is, IX, CGS,
CIV, XIII, VIII, XII, XII and XI from
Fig. \ref{fig:unsharp_haslam_kpol} are likely to be part of the same
local structure. In order to get a more clear picture, the \planck
polarisation is very valuable as it provides polarisation data sets
bot at low frequency, tracing the synchrotron emission and also, more
importantly, full sky polarisation maps of the dust emission, which
could be related with H{\sc i} data. This analysis however is outside
the scope of this paper.


\subsection{CMB foreground contribution}
 
With the aim of quantifying the contribution of the polarised
filaments in the context of foreground to the CMB, we have calculated
the power spectra of the polarised synchrotron sky. Using masks which
include/remove the observed filaments, we calculate the spectra and
then compare these with the expected E and B-modes components of the
CMB, to asses the level of contamination that the filaments induce at
different angular scales. Recently, \citet{liu:14} suggested that the
radio Loops can also be traced at high frequencies by polarised dust
emission so there might be significant CMB foreground contamination at
their location.

\begin{figure}
  \centering
  \includegraphics[angle=90,width=0.49\textwidth]{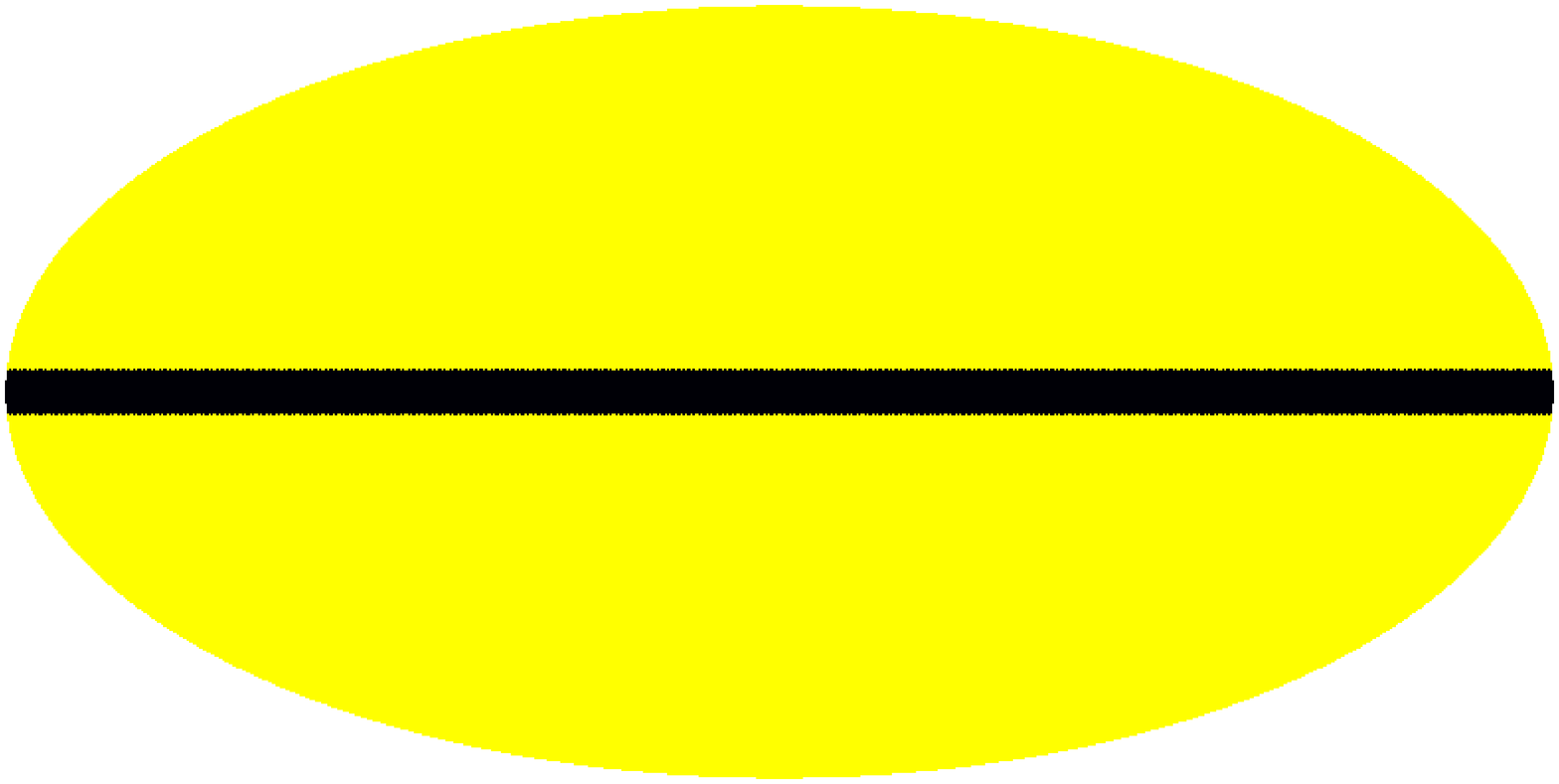}
  \includegraphics[angle=90,width=0.49\textwidth]{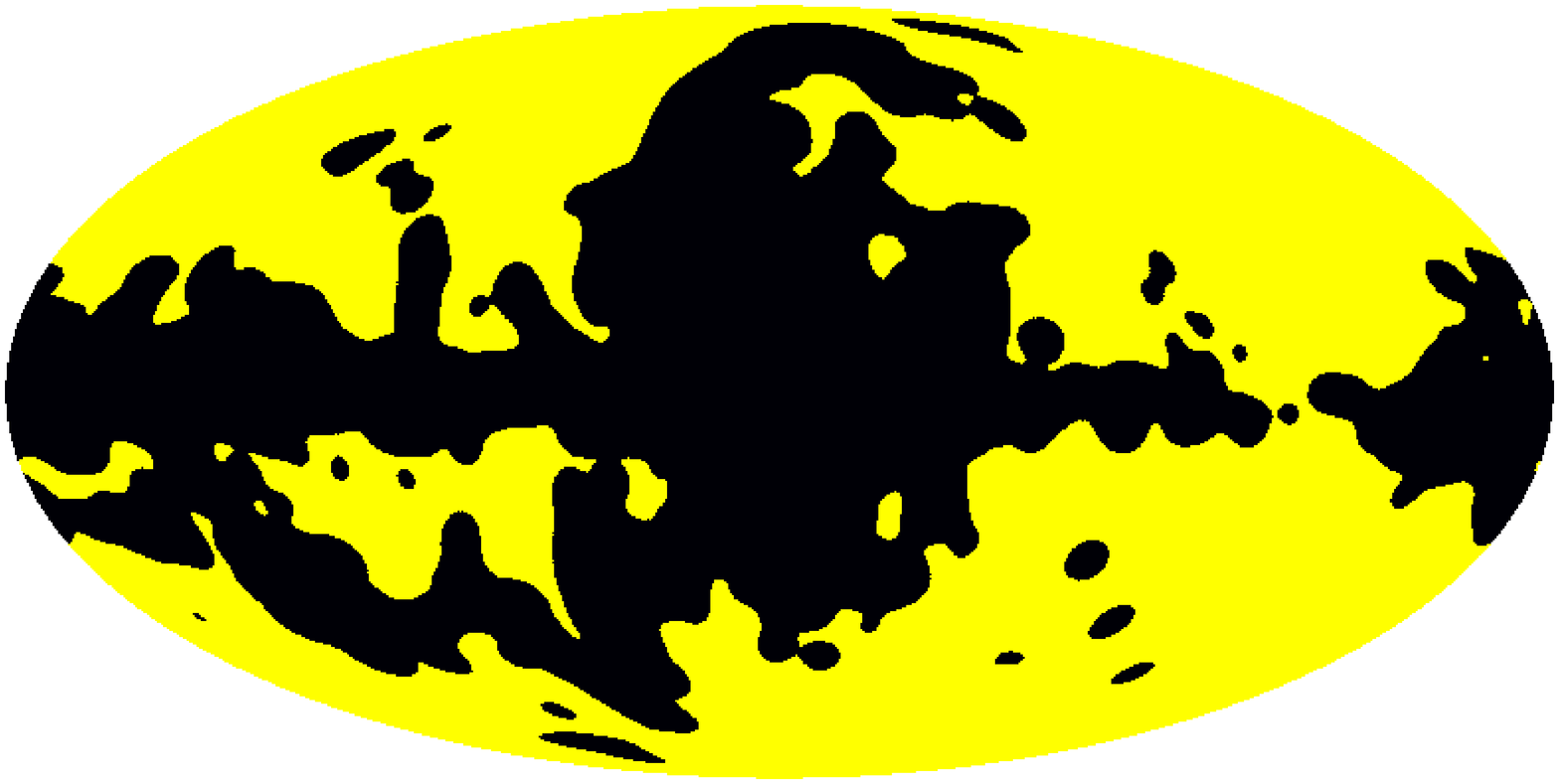}
  \caption[Masks of the Galactic plane and of the filaments]{The two
    mask used in the analysis. On the {\it top} is the one that masks
    the Galactic plane, all the pixels where $|b| \le 5$\dg and it
    covers $\approx 7\%$ of the sky. On the {\it bottom} is the one
    that we used to mask the emission from the filaments. This masks
    covers $\approx 27\%$ of the sky. }
  \label{fig:ps_masks}
\end{figure}

In order to test the contribution of the filaments to the power
spectrum of the full sky, we calculated the BB spectrum of the \wmap
\kband\ polarisation maps using two different masks: one that
suppresses all the pixels on a 10\dg strip on the Galactic plane and a
second mask that masks out the polarised features at high latitudes.
These masks can be seen in Fig. \ref{fig:ps_masks}.

The power spectrum was computed using the publicly available {\tt PolSpice}
package \citep{chon:04}. This software measures the the angular auto-
and cross- power spectra $C(l)$ of Stokes $I$, $Q$ and $U$. It is well
suited for our applications because it can correct for the effects of
the masks, taking into account inhomogeneous weights given to the
pixels of the map. The mixing of the E and B modes due to the masked
sky and pixel weights is corrected for, so an unbiased estimate of the
B-mode spectrum is given.

We also calculated a theoretical CMB power spectrum using the CAMB
package \citep{lewis:00}. For this, we used the cosmological
parameters from the \wmap 9-year release, listed in
\citet{hinshaw:13}.  We used a tensor-to-scalar ratio $r=0.2$ for the
B-mode spectrum calculation based on the recent measurement by the
BICEP2 collaboration\footnote{We note that there has been considerable
  debate about whether this detection is contaminated by foregrounds
  \protect\citep{flauger:14,mortonson:14}, which may result in a much
  lower value for $r$.} \citep{bicep2:14}. We calculated the power
spectrum for two different frequencies. First, at 23\,GHz, where the
polarised synchrotron is expected to dominate and also at 150\,GHz,
where its contribution is significantly reduced. Because we aim to
estimate the contribution of the synchrotron filaments, we do not
include any dust contribution at these higher frequencies. We used a
polarised spectral index of $\beta=-3.0$ to scale the emission from 23
to 150\,GHz. This value is an average for the high latitude filaments
as we have derived previously. We note that $\beta_{23-150}$ might be
steeper than the --3.0 value that we used due to a possible frequency
steepening of the spectrum. If this occurs, the contribution that we
show here at 150\,GHz will be an overestimate for the power of the
synchrotron filaments at 150\,GHz.

\begin{figure}
  \centering
  \includegraphics[angle=0,width=0.49\textwidth]{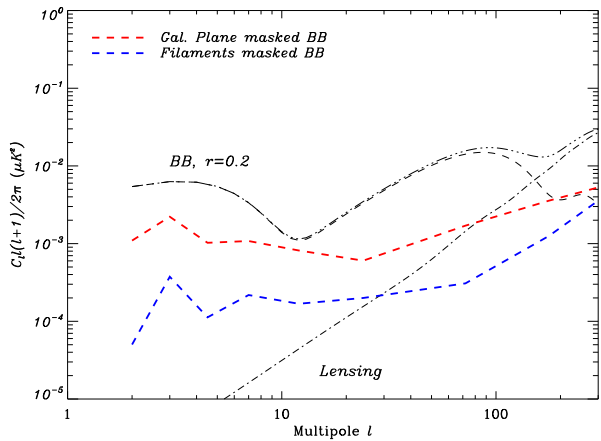}
  \caption{B-modes power spectra for the polarised synchrotron
    component at 150\,GHz compared to the theoretical CMB B-modes
    spectrum. In black are the CMB spectra where the B-modes were
    calculated using a tensor-to-scalar ratio value of $r = 0.2$. In
    colour are the B-mode spectra of the synchrotron polarisation map
    calculated using the two masks shown in
    Fig. \ref{fig:ps_masks}. The dot-dashed line shows the
    contribution from lensing to the cosmological B-mode spectrum.}
  \label{fig:ps_pol}
\end{figure}

In Fig.~\ref{fig:ps_pol} we show the resulting polarisation power
spectrum at 150\,GHz. The black lines shows the expected B-mode,
including the primordial and a lensing component. The red line show
the B-modes spectrum of the polarised synchrotron map computed using
only the Galactic plane mask, while the blue lines show the spectrum
computed using the filaments mask. The difference between the spectra
calculated using these two masks is more pronounced at the large
angular scales (low multipole value), as expected given the extension
of the filaments. The B-mode power calculated using the filament mask
is $\sim 10$ times smaller for $\ell =5$ than the power estimated
using only the Galactic plane mask. With increasing multipole value,
the difference of the spectra calculated with the different masks is
smaller. The large difference at low-$\ell$ on the B-modes spectrum is
important because the ``low-$\ell$ bump'' of the CMB B-modes spectrum
will be a target for the experiments that aim to detect the
cosmological B-mode signal \citep[see e.g.][]{katayama:11}.  However,
at 150\,GHz the synchrotron power is expected to be small compared
with the B-modes spectrum if $r=0.2$. We note that the last bins in
the power spectrum are noise dominated.

The poorly-constrained modes in the \wmap polarisation data will give
excess power at $\ell \lesssim 8$ \citep{jarosik:11}. In the \wmap
analysis, these were explicitly downweighted in the power spectrum
estimation, but we have not done this. This implies that our power
spectrum is slightly high in this multipole range.

\section{Conclusions}
\label{sec:conclusions}

The polarised sky at 23\,GHz is dominated by synchrotron emission and,
away from the Galactic plane, it originates mostly from filamentary
structures with well-ordered magnetic fields. Some of these structures
have been known for decades in radio continuum maps: the ``radio
loops'', with the North Polar Spur being the most studied. The origin
of these filaments is not clear and there are many filaments that are
visible for the first time in these polarisation data. We have
identified 11 filaments, including three of the well known radio
continuum loops. Five of these filaments are only visible in the
polarisation data. The geometry of these filaments can be described
using circular arcs. We fitted for the centres and radii of these
``loops''. We also compared polarisation angles along the filaments
with the direction defined by their extension. The polarisation angle
is well aligned along the filaments, being typically tangential to
their direction. We found however some systematic differences between
the polarisation angle $\chi$ and the direction defined by the
extension of the filament.

We measured the polarisation spectral indices of 18 small regions in
the sky and we found significant variations in $\beta$ over the
sky. Some of the regions show a spectral index flatter than the usual
$\beta=-3.0$, for example the Fan region, around Galactic longitude
$l=140$, which shows $\beta_{K-Ka}=-2.68\pm0.16$. On average, no
significant steepening with frequency of the spectral index is
observed between $\beta_{K-Ka}$ and $\beta_{Ka-Q}$, although some
individual regions do show some steepening, such as some regions on
the Galactic plane. The average spectral index in all the 18 regions
considered is $\beta_{K-Ka}=-3.04\pm0.02$.

We quantified the polarisation fraction of the synchrotron emission
over the entire sky, using different templates to estimate the
synchrotron total intensity at 23\,GHz. The results depend critically
on the model used for the synchrotron total intensity. Nevertheless,
some of the polarised filaments show a large polarisation fraction
($\Pi \approx 30\%$) regardless of the synchrotron total intensity
template used.

We measured the Faraday rotation between K and Ka bands at an angular
resolution of 1\dg. We only find signs of Faraday rotation on the
inner Galactic plane. Here, the change of polarisation angle between K
and Ka bands has an rms value of 4$^{\circ}$\!\!.7, corresponding to a
rotation measure of 890\,rad\,m$^{-2}$. This region however, covers
only 0.9\% of the total area of the sky. Over most of the sky, the
difference in polarisation angle between K and Ka bands is less than
1\dg\!\! (RM = 190\,rad\,m$^{-2}$).

To explain the large-scale observed polarisation pattern, we invoke a
model originally proposed by \citet{heiles:98}, in which an expanding
shell, located at a distance of 120\,pc compresses the magnetic field
in the local ISM. Under the assumption that the unperturbed magnetic
field lines are parallel to the Galactic plane, an expanding spherical
shell will bend the lines in a simple manner. We calculated how these
field lines would appear from our vantage point within the Galaxy. The
predicted direction of the magnetic field lines is in good agreement
with the observed polarisation angles at 23\,GHz for most of the
relevant area of the sky. We highlight that this model includes
emission from many of the observed filaments, therefore connecting
them as part of a big local structure.  This result suggests that a
substantial part of the filamentary and diffuse emission seen by \wmap
is local. This has to be taken into account when trying to model the
global Galactic polarisation emission and magnetic field.

Finally, we estimated the level of contamination that the filaments
add to the CMB E- and B-mode power spectra. We compared the B-mode
power spectrum of the sky at 23\,GHz using two different masks, one
that covers only the Galactic plane and a second one that masks-out
the diffuse filaments. The power measured at $\ell=3$ using the
Galactic plane masks is $\sim 10$ times larger than the power measured
using the filaments mask. This implies that a careful subtraction is
required to precisely measure the CMB B-mode spectrum at the largest
angular scales.

\section*{Acknowledgements}

We thank the referee, Carl Heiles, for his comments and suggestions
that have greatly improve the quality of the paper. We thank Anthony
Banday and Michael Keith for some very useful comments on this
work. MV acknowledges the funding from Becas Chile. CD acknowledges an
STFC Advanced Fellowship, an EU Marie-Curie IRG grant under the FP7
and ERC Starting Grant (no. 307209).  We acknowledge the use of the
Legacy Archive for Microwave Background Data Analysis
(LAMBDA). Support for LAMBDA is provided by the NASA Office of Space
Science. Some of the work of this paper was done using routines from
the IDL Astronomy User's
Library\footnote{http://idlastro.gsfc.nasa.gov/}. Some of the results
in this paper have been derived using the HEALPix \citep{gorsky:05}
package.

\label{lastpage}

\bibliographystyle{mn2e}
\bibliography{refs} 

\bsp

\end{document}